\documentstyle[aps,prb,twocolumn,epsf]{revtex}

\begin{document}
\draft
\title
{Calculations of the Knight Shift Anomalies in Heavy Electron Materials}

\author{E. Kim and D. L. Cox$^1$}
\address{Department of Physics, Ohio State University, Columbus, OH 43210}
\date{\today}
\maketitle
\setlength{\baselineskip}{12pt}
\widetext

\begin{abstract}

We have studied the Knight shift $K(\vec r, T)$ and  magnetic 
susceptibility $\chi(T)$ of  heavy electron materials,
modeled by the infinite $U$ Anderson model with the NCA method.
A systematic study of   $K(\vec r, T)$  and $\chi(T)$ for different
Kondo temperatures $T_0$ (which depends on the hybridization width
$\Gamma$) shows a low temperature anomaly (nonlinear relation between $K$ and
$\chi$)  which increases as the Kondo temperature $T_0$  and distance 
$r$  increase .  We carried out an incoherent lattice sum by adding 
the $K(\vec r)$ of a
  few hundred shells of rare earth atoms around a nucleus and   compare
  the
numerically calculated
results with the experimental results.
For CeSn$_3$,  which is a concentrated heavy electron material, 
both the $^{119}$Sn NMR Knight shift and positive
muon Knight shift are studied.  
Also, lattice coherence effects by conduction electron scattering 
at every rare earth site are included using the average-T matrix
approximation.
The calculated magnetic susceptibility and $^{119}$Sn NMR Knight shift
show excellent agreement
with experimental results for both incoherent and coherent calculations.
The  positive muon  Knight shifts are calculated for both possible
positions of muon (center of the cubic unit cell and middle of
Ce-Ce bond axis). 
 Our numerical results show a low temperature anomaly for the muons
  of the correct
 magnitude but we can only find agreement with experiment if we take 
 a weighted average of the two sites in a calculation with lattice
 coherence present.
 For YbCuAl, the  measured $^{27}$Al NMR Knight shift shows an anomaly with
   opposite sign to CeSn$_3$ compound.   Our calculations agree 
   very well with the experiments. 
For the proposed quadrupolar Kondo alloy Y$_{0.8}$U$_{0.2}$Pd$_{3}$, our 
$^{89}$Y
NMR Knight shift  calculation doesn't show the observed Knight
shift anomaly.

\end {abstract}

\narrowtext
\section{Introduction}
\label{intro.sec}
Many heavy electron materials show  Knight shift anomalies, which are a
deviation from a linear relation of the Knight shift $K(T)$ to the magnetic 
susceptibility $\chi(T)$ below the Kondo temperature $T_0$.  
 The origin of the Knight shift anomalies  
has been a subject of great interest
in the condensed matter community over a period of nearly twenty five years
~\cite{heeger,bs74,gz74,i76,mv75,mb79,m85,cjk87,ph88,lm94}.
If the impurities in metals 
have local magnetic moments, they display 
interesting properties comparing to metals with  nonmagnetic impurities, 
such as a resistivity minimum and anomalies in specific heat and susceptibility. 
This Kondo effect is a consequence of interaction between the magnetic ion and
 conduction electron. 
The central physical concept is that the
many body screening
cloud surrounding a Kondo impurity site should give rise to an
anomalous temperature
dependent Knight shift at nuclear sites due to the coupling of the
local moment to the
nuclear spin through the screening
cloud~\cite{heeger,gz74,i76,cjk87,ph88}.
Such a ``non-linear Knight shift anomaly'' is to be distinguished from the 
non-linear susceptibility related to the field dependence of $\chi$.
Another way to describe this effect is to say  that in the absence of an
anomaly, the contribution $K(\vec r,T)$ from a local 
moment at
distance $\vec r$ from the nucleus can be written as $f(\vec r)\chi(T)$.
This factorization does not hold
if there is an 
\newpage\hfill\\*[8cm]
 anomaly (instead $K(\vec r,T) = f(\vec r,T)\chi(T)$ due to
the temperature dependent polarization cloud).
After Heeger~\cite{heeger} suggested that the anomalous spin cloud was detected
at low temperatures, the central question has been whether a  conduction 
electron spin cloud with
huge coherence length $\xi_K=\hbar v_F/k_BT_0$, where $v_F$ is the Fermi 
velocity and $T_0$ is the Kondo temperature exists.
The  main motivation of this paper is to help clear up  this conflict.
The  Knight shift calculations  presented here are  the first performed 
 using a realistic impurity model.

This paper is organized as follows. In Sec.~\ref{review.sec},
the Kondo effect and the Knight shift anomaly in heavy electron materials will
be reviewed. 
First, general characteristics of the Kondo effect are discussed. 
We will then review  the history of  Knight shift anomalies in 
heavy electron systems.
In Sec.~\ref{model.sec}, the model Hamiltonian is introduced for both Ce
and U compound. Also we review the methods we have  used to evaluate the Knight
 shift (the Non Crossing Approximation(NCA) and average-T matrix 
approximation (ATA)).
Our  formalisms for  numerical calculations are explained, and
a  detailed derivation of the Knight shift Feynman diagram will be given 
in the appendices.
In the next section, the numerical results for Ce  and Yb ions which are single
channel Kondo materials and  for U ions in a  proposed quadrupolar Kondo 
alloy will be examined and 
compared with the experimental results. The calculated NMR Knight shift of 
Ce and Yb compounds show low temperature
anomalies and agree well with the experimental results.
But, there is no calculated Knight shift anomaly  for the 
proposed quadrupolar Kondo U compound, in contrast to experiment.
The last section includes  conclusions and directions for future work.

\section{Review}
\label{review.sec}
\subsection{Kondo Effect}
\label{kondo.ssec}
The existence of localized moments in dilute alloys that couple to  
conduction electrons has important consequences for the electrical properties.
It has been known since 1930\cite{mv30} that
 the resistivity has a rather shallow minimum 
occurring at a low temperature that depends weakly on the concentration of
magnetic impurities instead of dropping monotonically with decreasing 
temperature like metals with 
nonmagnetic impurities. 
In 1963, Kondo~\cite{kondo} explained that this minimum arises from  some 
unexpected features of the
scattering of conduction electrons 
off a local magnetic moment, with the 
simplified model Hamiltonian 
\begin{equation}
{\cal H} = \sum_{\vec k \alpha}\epsilon_{\vec k}c^{\dagger}_{\vec k \alpha}
c_{\vec k \alpha} -\frac{J}{2N}\sum_{\vec k \vec k'}c^{\dagger}_{\vec k \alpha}
\vec {\sigma}_{\alpha\beta}c_{\vec k' \alpha}\cdot\vec S_I
\end{equation}
where $c_{\vec k \alpha}$ is the annihilation operator of the conduction 
electron, $\alpha$ is the spin index, $\vec {\sigma}$ are Pauli matrices,
$\vec S_I$ is the spin operator of the impurity, and $J < 0$ is the
exchange coupling.
 Kondo discovered that 
 the magnetic scattering cross section  is divergent in perturbation theory.
The anomalously high scattering probability of magnetic ions at low temperatures
is a consequence of the dynamic nature of the scattering induced by the exchange
coupling and the sharpness of the Fermi surface at low temperatures.
Subsequent analysis
by Kondo and others has shown that a nonperturbative treatment
removes the divergence, yielding instead a term in the impurity
contribution to the resistivity that increases with decreasing temperature.
In spite of the simple model Hamiltonian, a magnetic $S_I = 1/2$ local moment
interacting with the conduction electron gas, this result is an indication that
 the problem is explicitly a many body problem, meaning that the electron in 
state $\vec k$ which is being scattered is sensitive to the  occupation of all 
 other electron states $\vec q$.
 
For this single channel Kondo model, there is only one characteristic
energy scale, the Kondo temperature $T_0$, provided that the temperature
$T$ is much smaller than the conduction electron bandwidth $D$, and 
corrections of order $T/D$ are neglected. The Kondo temperature is given by
\begin{equation}
k_BT_0=D[N(0)J]^{1/2}exp(-1/N(0)J),
\end{equation}
 where $N(0)$ is the conduction electron density of states at the Fermi level.
Any physical quantities are universal functions of $T/T_0$. At low temperature,
with all material properties buried in $T_0$.

This Kondo model can explain the anomalies in the transport coefficients,
specific heat and magnetic susceptibility for some alloys with magnetic
impurities. 
 The Kondo effect is characterized  by the development of the Kondo resonance
peak with width of order of $T_0$.
At high temperatures $T > T_0$, the impurity resistivity increases 
logarithmically as the temperature decreases, and saturates to a finite
value at low temperatures below $T<T_0$. The magnetic susceptibility has a
Curie-Weiss form at high temperatures and shows Pauli paramagnetism at
low temperatures with $\chi(T=0)\sim 1/T_0$,concomitantly $C/T(T=0)\sim 1/T_0$.
The $\chi$ behavior is explained by  the fact that the magnetic moments which
 exist at high temperatures
are screened out by the conduction electron spin clouds  at low temperatures 
with the formation of a   singlet ground state. 
This conduction spin cloud and the Knight shift
anomaly will be discussed more in the next subsection.

\subsection{Review of the Knight Shift Anomalies of Heavy Electron Materials}
\label{anomaly.ssec}
There have been many theoretical and  experimental works about the Knight shift
 anomaly for
the heavy electron materials.  Whether there is  an observable
conduction electron spin cloud with huge coherence length  has been  an
issue in condensed matter physics more than 25 years and this is   
our main motivation to carry out this paper.

In the simplest approximation, the added electron is bound into a singlet with
the impurity~\cite{mattis}. Because of the falloff the amplitude of wave 
function with the
energy $\epsilon _k$, only states within roughly 
$\delta k \sim (k_BT_0/E_F)k_F $ 
of the Fermi surface are involved in the singlet. As a result, in coordinate
space, the singlet wave function extends to a very large distance of order 
$(\delta k)^{-1}$. 
 Heeger {\it et al.}~\cite{hee68} calculated the susceptibility at $T=0K$
using the Appelbaum-Kondo theory~\cite{ak67,ak68} and found
\begin{equation}
\chi =\chi_{\mbox{Pauli}}+\chi_{L}+\chi_{Q}  \label{eq:chi1}
\end{equation}
where
\begin{equation}
\chi_{L}=\chi_{Q}= |J|\rho[\mu^2/(9/8)k_BT_0] \label{eq:chi2}
\end{equation}
Equations (\ref{eq:chi1}) and  (\ref{eq:chi2}) give the very interesting result
that one half the excess susceptibility is localized on the impurity site
($\chi_{L}$) and one-half is associated with the partially polarized quasi 
particle($\chi_{Q}$). The associated spin polarization around the partially 
magnetized impurity is given by
\begin{equation}
\sigma(r)=\sigma_0 + \sigma_{RKKY}(r) + \sigma_{Q}(r)
\end{equation}
where $\sigma_0$ is the uniform polarization due to the external field,
$\sigma_{RKKY}(r)$ is the usual RKKY term~\cite{rk54,kasuya,yosida,vv62,bf56}
 which $r$ dependence is given by
\begin{equation}
(1/r^3)\cos(2k_Fr),
\label{eq:RKKY}
\end{equation}
and $\sigma_{Q}(0)$ is the quasiparticle term with
\begin{equation}
<\sigma_{Q}(r)> = <S_Z>\frac{3}{2}N\frac{k_BT_0}{E_F}(\frac{\sin k_Fr}
{k_Fr})^2 \ln^2(r/\xi).
\end{equation}
 where $\xi_K = (2E_F/k_BT_0)k_F^{-1}$. This expression is valid for $r <\xi_K$,
and at greater distances $\sigma_Q(r)$ rapidly approaches zero. In both the 
RKKY
term and $\sigma_Q(r)$ term the value $<S_Z>$ is not the free spin value but
is determined by the local susceptibility $\chi_L$. The existence of the RKKY
term  for $T < T_0$ was shown by Suhl~\cite{suhl}. 

 The technique of nuclear magnetic resonance has been of primary importance in 
the development of our current understanding of the localized moment problem.
The reasons are twofold. First, the nuclei in the host metal in the vicinity
of the impurity are sensitive to local perturbations in the spin density(via
the hyperfine interaction) and the charge density(via the nuclear quadrupolar
interaction). Moreover, the nuclei themselves are only weakly coupled to the 
electronic system and therefore act as passive "spies" into the phenomena
of interest. Secondly, the nuclear relaxation is sensitive to the low-lying
excitations of the electronic system and consequently can provide information
on the dynamical aspects of the impurity problem.
For the most part NMR experiments in heavy electron compounds are carried out
on the nuclei of the non-$f$ ions, so that coupling to the $f$ moments
occurs via indirect interactions such as transferred hyperfine and dipolar 
fields.

The NMR experiment of Boyce and Slichter~\cite{bs74} on Fe 
 impurities in Cu metals showed no evidence for a Knight shift anomaly 
at low temperature and was interpreted to indicate the absence of
this screening cloud or at least a screening cloud of size of the order of 
a lattice
spacing. These mixed results have led to theoretical discussions about
the size of any screening conduction electron spin cloud or even whether it 
exist~\cite{i76,cjk87,ph88,soaf95,soaf96,baaf96}.

 In contrast,
pronounced Knight shift
anomalies have been observed in the concentrated heavy
electron materials CeSn$_3$\cite{mv75,m85} and YbCuAl\cite{mb79,m85},
which have been described as
Kondo lattice
systems with $T_0\geq 400K$.  In view
of the Boyce-Slichter
result, the question is raised whether these anomalies represent a
coherent effect of the periodic
lattice rather than a single ion effect. However, recent experiments on
the
proposed quadrupolar Kondo alloy\cite{c87,yupd}
Y$_{1-x}$U$_x$Pd$_3$ demonstrate that
for concentrations
of 0.1-0.2 there are pronounced non-linearities in the Y Knight shift
for sufficiently large
distances away from the U ion\cite{lm94}.

Ishii~\cite{i76} calculated the field induced spin polarization 
for the degenerated  Anderson model  and
 confirmed that an anomalous spin cloud is formed outside
of the Kondo screening length $\xi_K$ at $T\rightarrow 0$.
The spin polarization $\sigma(r)$ for electron-hole symmetry case 
($\epsilon_d=-U/2$) 
is given by
\begin{equation}
\sigma(r)=\frac{\chi H}{g\mu_B}\frac{\cos 2k_Fr}{4\pi r^3}[1-\frac{2E_F}
{k_Fr\Delta } \overline{\gamma}].
\label{br}
\end{equation}
 where $\overline{\gamma}=\chi_{\uparrow\uparrow}$ for $U\rightarrow 0$ and
$3/2(2l+3)\chi_{\uparrow\uparrow}$ for $U\rightarrow\infty$, where $l$ is the
orbital angular momentum.
$\chi_{\uparrow\uparrow}$ changes from $(g\mu_B)^2/2\pi\Gamma$ for $U= 0$ 
to $(g\mu_B)^2S(S+1)/3k_BT_0$ for $s-d$ Exchange model ($U=\infty$).
Therefore the coherence length $\xi_K=2E_F/k_F\Delta\overline{\gamma}$ varies
from $2\hbar v_F/\Gamma$ at $U=0$ to $2\pi\hbar v_F/3k_BT_0$ in the
$s-d$ limit.
Also, for the strong $U/\pi\Gamma$ limit, spin polarization 
is calculated for  $r<\xi_K$. It is given by
\begin{equation}
\sigma(r)=\frac{4\Gamma}{\pi U}\frac{\chi H}{g\mu_B}\frac{\cos 2k_Fr}{4\pi r^3}.
\label{sr}
\end{equation}
 By the relation $4V^2/U=-J/2N$~\cite{sw}, the above spin polarization is just
the  RKKY contribution.
Comparing the equations (\ref{br}) and (\ref{sr}), the conduction electron spin
cloud which is formed outside the Kondo coherence length has the RKKY form but
with $2N/N(0)J$ times bigger amplitude than the spin cloud inside
the Kondo coherence length.

Chen {\it et al.}~\cite{cjk87} calculated the 
zero frequency response function $C(\vec r, T)$ around a magnetic 
impurity, using a perturbative
thermodynamic scaling procedure and nonperturbative renomalization
group method for $S=1/2$ Kondo  model.
The host nuclei near  magnetic impurities at positions $\vec r$ 
displays satellite resonances in the tail
of the main
magnetic resonance  signal, with a Knight shift given by
$K+\Delta K(\vec r)$, 
where $K=\rho_0g_e/2$ is the Knight shift of the pure host.
Then $\Delta K$ is
\begin{equation}
\Delta K= C(\vec r,T)-K
\end{equation}
Chen {\it et al.}~\cite{cjk87} showed that this Knight shift is factorized into 
a product of  temperature and spatial dependent functions, specifically
\begin{equation}
\Delta K=F(\vec r)\chi(T)
\end{equation}
where $\chi(T)$ is the magnetic susceptibility.
The computed $F(r)$ has the RKKY form.

The conduction electron spin polarization in the Anderson model
  has been studied with the NCA
previously by W. Pollwein {\it et al.}~\cite{ph88}. However, this study was
carried out only
for the spin 1/2 model with infinite Coulomb repulsion, and for a
limited
parameter regime (only results for very low $T_0$ values and short distances
$(r \ll \xi_K)$ have been numerically calculated). In consequence,
no strong evidence was found for a Knight shift anomaly in this
previous work.

Recently, E. S\o rensen and I. Affleck~\cite{soaf95,soaf96,baaf96}  showed that
 the Kondo coherence length
$\xi_K=\hbar v_F/k_BT_0$, varies when temperature
changes by combining a finite size scaling ansatz with density matrix 
renomalization group calculations.
They write the scaling hypothesis  for three dimensional susceptibility is
\begin{eqnarray}
\chi_{\mbox{imp}}(r)& =& \chi-\rho/2\nonumber \\
   &=& \frac{\cos (2k_Fr)}{8\pi^2 v_F r^2}f(rT/v_F,T/T_0) \nonumber \\ 
   & & (\mbox{for } r \gg 1/k_F;T,T_0 \ll E_F),
\end{eqnarray}
where $f$ is a real universal scaling function. $\rho/2$ is the standard
Pauli bulk susceptibility with $\rho$ the density of states per spin.
At higher temperatures $T> T_0$, the local susceptibility shows 
RKKY behavior and
at lower temperatures $T<T_0$ it has a local Fermi liquid form.
So the Knight shift has longer range at low temperatures where the conduction 
electron screening cloud has formed than at high temperatures where it has not.
Sorensen and Affleck tried 
to explain the experiment by Boyce and Slichter~\cite{bs74} by 
the possible
factorization of the scaling functions deep inside the screening cloud where
the experiment was done. 

\section{Model and Formalism}
\label{model.sec}

\subsection{Model Hamiltonian}
\label{model.ssec}

\begin{figure}
\vspace*{7pt}          
\begin{center}         
\leavevmode            
\epsfxsize=3.0in
\epsfbox{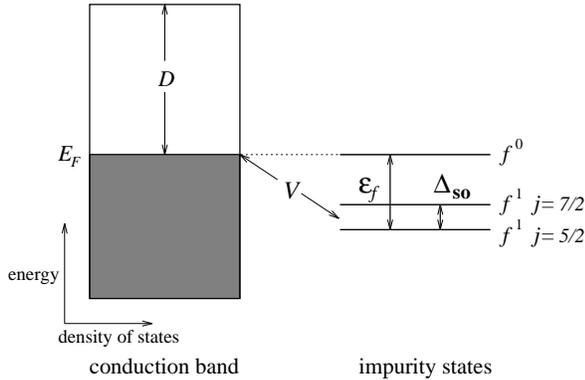}
\end{center}
\caption {Schematics of the  $U=\infty$ single impurity Anderson model for Ce
ions.   $D$ is the conduction band width and
$V$ is the single particle hybridization strength. The on site Coulomb interaction
 $U$  is infinity in our model.}
\label{anderson.fig}
\end{figure}

In our work, 
we use the  on-site Coulomb interaction $U=\infty$ single impurity Anderson 
model~\cite{a61}. 
The Anderson model can be canonically transformed by the Schrieffer-Wolff 
transformation to the Kondo model at $U=\infty$ limit~\cite{sw}, 
and is a good
model Hamiltonian to describe heavy electron materials with $f$ electrons.
Schematics of this model are shown   in  Fig~\ref{anderson.fig} for Ce ions.
This $U=\infty$ model  can be a good
approximation in the limit when the ratio of the virtual $4f$ level width
$\Gamma$ to the on site Coulomb repulsion $U$ is small. 
For  real materials this  interaction energy is of the
order of $U=5eV$~\cite{hww78,j79}, and  the hybridization width 
is of the order of $\Gamma=0.1eV$.     
 For Ce(Yb) ions, 
 we keep  $f^0$ and $f^1(f^{14}$ and $f^{13})$  configurations,
 and for U ions, which are proposed two channel quadrupolar Kondo alloys,
we keep  $f^2$ and $f^3$  configurations.
For Ce and Yb ions, spin orbit coupling is included and for U ions, spin orbit 
coupling and also crystal field splittings are  included.
The crystal electric field(CEF) will split the spin-orbit multiplet and  
 can mix two different angular momentum multiplets($j$, $j'$). But the  correction of
the mixing term between two different $j$'s by CEF is small, we only consider
the splitting effect. 
These crystal electric field effects in the Anderson model were first considered by Hirst~\cite{hirst}
 based on  group theory and it will be discussed more in the Appendix~\ref{cef.app}.

  We shall first discuss the
situation for Ce$^{3+}$ and Yb$^{3+}$ ions, and write down the model
only for the Ce case(the Yb$^{3+}$ ion has a lone $4f$ hole and our procedure
describes this with a simple particle hole transformation). 

  For a single Ce site at the origin, the model is
\begin{equation}
{\cal H} = {\cal H}_c + {\cal H}_f + {\cal H}_{cf} +{\cal H}_z
\end{equation}
with
\begin{equation}
{\cal H}_c = \sum_{\vec k\sigma} \epsilon_k c^{\dagger}_{\vec
k\sigma}c_{\vec k\sigma}
\end{equation}
the conduction band term for electrons with a broad featureless density
of states of width $D$, taken to be Lorentzian here for convenience,
with
\begin{equation}
{\cal H}_f = \sum_{jm_j} \epsilon_{fj} |f^1jm_j \rangle  \langle f^1jm_j|
\end{equation}
where $j=5/2,7/2$ indexes the angular momentum multiplets of the Ce ion
having azimuthal quantum numbers $m$, with
$\epsilon_{f5/2}=-2
eV$,$\epsilon_{f7/2}=\epsilon_{f5/2}+\Delta_{so}=-1.71eV$
(we take the $f^0$ configuration at
zero energy), with
\begin{equation}
{\cal H}_{cf} = {\sum_{\vec kjm\sigma}} [\;V_{\vec kj\sigma m_j}
c^{\dagger}_{\vec k\sigma}|f^00 \rangle  \langle f^1jmi_j| + h.c.\:]
\end{equation}
where $V_{\vec kj\sigma m_j} =VY_{3m_j-\sigma}(\hat
k) \langle 3m_j-\sigma,1/2\sigma|jm_j \rangle /\sqrt{N_s}$,
$V$ being the
one particle hybridization strength and $N_s$ the number of sites.
We can rewrite the hybridization Hamiltonian as
\begin{equation}
{\cal H}_{cf} = {\sum_{ kjm_j}}[\;V_kc^{\dagger}_{\vec kjm_j}|f^00\rangle
\langle f^1jm_j| + h.c.\:]
\end {equation}
where \begin{equation}
c^{\dagger}_{kjm_j} =\sum_{m\sigma}\langle 3m\frac{1}{2}\sigma|jm_j\rangle
c^{\dagger}_{km\sigma}
\end{equation}
with $ c_{km\sigma}=k\int d\hat{k} Y^{*}_{3m}(\hat{k})c_{\vec k \sigma}$.
For the Yb case, the hybridization Hamiltonian is given by 
\begin{equation}
{\cal H}_{cf} = {\sum_{ kjm_j}} [\;V_k
c^{\dagger}_{kjm_j}|f^{13}j-m_j \rangle  \langle f^{14}0| + h.c.\:].
\end{equation}
${\cal H}_z$, the Zeeman energy of the electronic system for a magnetic field
$H_z$ applied along the $z$-axis is given by
\begin{equation}
{\cal H}_z =- \mu_BH_z[2\sum_{\vec k\sigma} \sigma n_{\vec k,\sigma} -
\sum_{jm} g_jm|f^1jm \rangle  \langle f^1jm|\:].
\end{equation}
 In addition to this, we must add a term coupling the
nuclear spin system
to the conduction electrons, which we take to be of a simple contact
form $\sim \vec I(\vec r)\cdot
\vec S(\vec r)$ for each nuclear spin $\vec I(\vec r)$ at position
$\vec r$ with $\vec S(\vec
r)$ the conduction spin density at the nuclear site, and a nuclear
Zeeman term.
In terms of the parameters, the Kondo scale characterizing the low
energy physics is
given by
\begin{equation}
  k_BT_0= D(\frac{\Gamma}{\pi|\epsilon_f|})^{1/N_g}(\frac{D}{\Delta_{so}})^{N_{ex}
/N_g} \exp(\frac{\pi \epsilon_f}{N_g\Gamma})
\end{equation}
where the single particle hybridization width $\Gamma = \pi N(0)V^2$ with
$N(0)$ which is the density of the states at the Fermi energy. Other 
parameters are defined in  Table~\ref{ceyb}.

\begin{figure}
\vspace*{7pt}          
\begin{center}         
\leavevmode            
\epsfxsize=3.0in
\epsfbox{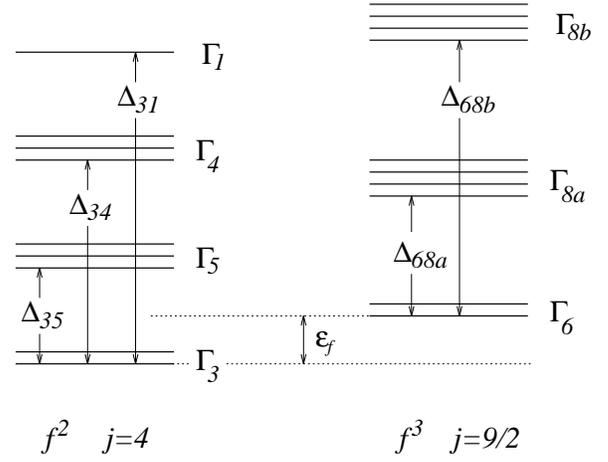}
\end{center}
\caption {Schematic  configuration diagram for the Uranium compound.
$f^2$, $j=4$ and $f^3$, $j=9/2$ spin-orbit states are split  by the crystal
electric field.
  All the notations are explained in the text and all the values are listed in
Table~\ref{uvalue}.}
\label{uconfig.fig}
\end{figure}

For the $Y_{1-x}U_{x}Pd_3$  which has the cubic AuCu$_3$ structure,  
the crystal 
field effect (CEF)  must   be  included. This  crystal electric
field effects lift the angular momentum degeneracy of U  ions and 
their spin-orbit multiplet decomposes into irreducible representation of
the cubic field.
The $f^2, j=4$ Hund's rule ground state of U compound  is split to a $\Gamma_3$ nonmagnetic
doublet, $\Gamma_5$ and $\Gamma_4$ magnetic triplet  and $\Gamma_1$
singlet states\cite{mooketal}. And $f^3, j=9/2$ spin-orbit multiplet is split
to $\Gamma_7$ doublet and two $\Gamma_8$ quartets.
In our calculation we choose $j=4$ $\Gamma_3$ for
the ground
 state for $f^2$ configuration and $j=9/2$ $\Gamma_6$ for the ground state 
of the  $f^3$ configuration.
Fig.~\ref{uconfig.fig} shows the schematic configuration diagram.
All parameter values are listed in  Table~\ref{uvalue} in the unit of $D$.
For an explicit derivation of Hamiltonian for  U ions, see the Appendix~\ref{u.app}.

\subsection{Non Crossing Approximation}
\label{nca.ssec}
We treat the Anderson Hamiltonian with the non-crossing approximation
(NCA), a self-
consistent diagrammatic perturbation theory discussed at length in the
paper of Bickers
{\it et al.}\cite{bcw}.
 This is  useful because this method 
provides ways of calculating the dynamic response functions, such as the 
one electron Green's functions and dynamic susceptibility and
 it makes possible a more extensive
comparison between the theoretical predictions and experimental results. 

In the NCA, we do the  $1/N$ expansions  with the
new variable $N$, the large orbital degeneracy of the ground state of $f$
electrons. 
  These simple approximation schemes work very well for
values of $N$ that are of interest in applications to rare earth impurities.
For example, the lowest spin-orbit split multiplet for Ce 4$f^1$ has $j=5/2$,
corresponding to $N=6$, and for Yb $4f^{13}$, $j=7/2$, 
 corresponding to $N=8$. Even for $N=2$ one can get  good semi-quantitative results,
which can be accurate to within a few percent for some quantities. 

It was noted that, in a perturbative $1/N$ expansion, the $f$-electron spectral
density exhibits a singularity at $\omega=T_0$, with
$T_0 \simeq D\exp(\pi\epsilon_f/N\Gamma)$.
This singularity remains order by order, preventing a complete description of 
$f$ photoemission and electronic transport. In order to remove this singularity,
it is necessary to perform an infinite-order resummation in $1/N$
~\cite{bcw,kk71,grewe,kura,col,zhanglee,ina,keiczy,kurkoj,zero,bc85,mk85a,mk85b}. 

\begin{figure}
\vspace*{7pt}         
\begin{center}       
\leavevmode           
\epsfxsize=4.0in
\epsfbox{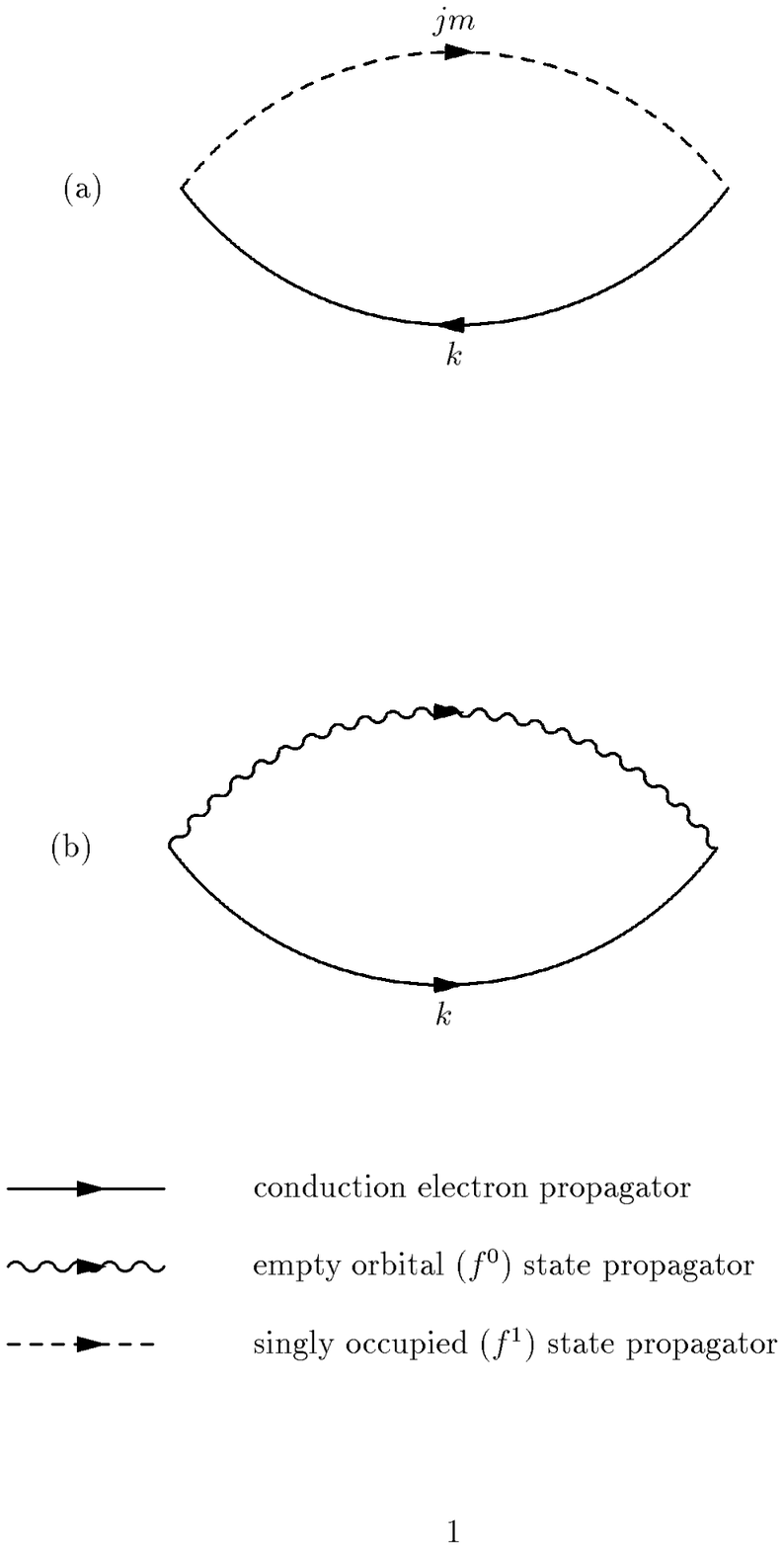}
\end{center}
\vspace{0.5cm}
\caption
{ Leading order Feynman diagrams for self energies.
 (a) pseudoboson self energy, $\Sigma_0(z)$r,
 (b) pseudo fermion($f^1$,$j$ multiplet) self energy
 $\Sigma_{jm}(z)$.}
\label{self.dia}
\end{figure}

In the NCA , our starting basis is the conduction   band plus the atomic
Hamiltonian projected to the atomic electron Fock space and treat the 
hybridization
between the conduction band and the atomic orbital as a perturbation.
The strength of this approach is that the strong on-site Coulomb interaction
for atomic electrons is included at the outset.  The conventional Feynman diagram
technique which uses Wick's theorem can not be applied for  strongly
correlated problems with restricted Hilbert spaces. Pseudo particle
Green's functions are introduced for     each atomic electron occupation state
which is neither fermionic nor bosonic
(i.e., $f^1$ $j=5/2$, $7/2$ and $f^0$ in the
present model for Ce ions).  
The pseudo fermion  Green's functions for $f^1$  $j=5/2$, $7/2$ angular
momentum  multiplets are
\begin{equation}
G_{jm}(z)=\frac{1}{z-\epsilon_{fj}-\Sigma_{jm}(z)}
\end{equation}
and the pseudo boson Green's function for the $f^0$  is
\begin{equation}
D(z)=\frac{1}{z-\Sigma_0(z)}.
\end{equation}
Then we insert a self-energy into the propagators
of pseudo particles. This gives coupled integral equations for the ionic
propagator self energies, $\Sigma_0(z)$, $\Sigma_{jm}(z)$.
>From the leading order diagrams of Fig.~\ref{self.dia}, the coupled equations
for the self energies are
\begin{eqnarray}
\Sigma_0(z) &=&V^2\sum_j N_j \sum_k\frac{f_k}{z+\epsilon_k-\epsilon_{fjm}-
\Sigma_{jm}(z+\epsilon_k)} \nonumber \\
  &=& V^2\sum_j N_j \sum_kf_kG_{jm}(z+\epsilon_k)\\
\Sigma_{jm}(z) &=& V^2 \sum_k \frac{1-f_k}{z-\epsilon_k-\Sigma_0(z-\epsilon_k)}
\nonumber \\
&=& V^2 \sum_k(1-f_k)D(z-\epsilon_k),
\end{eqnarray}
where $V$ is the hybridization
strength between the conduction band and the atomic orbitals and $N_j$ is the
degeneracy of the spin-orbit multiplet $j$.
It is convenient to introduce the spectral functions $A_{jm}(\omega)$,and
$B(\omega)$ for pseudo-particle  Green's functions.
\begin{eqnarray}
G_{jm}(i\omega)& =& \int \frac{d\rho}{\pi}\frac{A_{jm}(\rho)}{i\omega-\rho}
\nonumber\\
A_{jm}(\omega) &=&-\mbox{Im}\;G_{jm}(\omega)\\
D(i\nu)& =& \int \frac{d\zeta}{\pi}\frac{B(\zeta)}{i\nu-\zeta}\nonumber\\
B(\nu) &=& -\mbox{Im}\;D(\nu)
\end{eqnarray}
In addition to spectral functions $A_{jm}(\rho)$ and $B(\zeta)$, it is necessary
 to
 introduce negative frequency spectral functions $A_{jm}^{(-)}(\rho)$ and
$B^{(-)}(\zeta)$.  These spectra are given by
\begin{eqnarray}
A_{jm}^{(-)}(\rho, T) & = & A_{jm}(\rho,T)e^{-\beta(\rho-E_0)}\\
B^{(-)}(\zeta, T) &=&  B(\zeta,T)e^{-\beta(\zeta-E_0)},
\end{eqnarray}
 where $E_0$ is the ground state energy relative to the Fermi energy.
The impurity partition function $Z_{4f}$ is given by
\begin{eqnarray}
Z_{4f}(T)& = &\int\frac{d\zeta}{\pi}[\sum_{jm}A_{jm}(\zeta,T)+B(\zeta,T)]
    e^{-\beta\zeta}.\nonumber \\
& =& e^{-\beta E_0}\int\frac{d\zeta}{\pi}[\sum_{jm}A_{jm}^{(-)}(\zeta,T)
+B^{(-)}(\zeta,T)].
\end{eqnarray}
At $T\rightarrow 0$, $Z_{4f}(T)$ becomes
\begin{equation}
Z_{4f}(0) = e^{-\beta E_0}.
\end{equation}

The iteration of these coupled equations for the self energies  generates a set
of diagrams which includes all
non-crossing diagrams, but does not correspond to any specific order in the $1/N
$
 expansion by treating $NV^2$ as ${\it O}(1)) $ where $V$ is the hybridization
strength between the conduction electron and the atomic orbitals.
The set of diagrams summed by these equations includes all the
terms of order ${\it O}(1)$ and ${\it O}(1/N)$ and a subset of contributions
from the
higher order terms.
The lowest order skeleton diagrams which are  not included  are
 of order ${\it O}(1/N^2)$.
All the diagrams that enter at ${\it O}(1)$ and ${\it O}(1/N)$ have non-crossing
conduction lines.
Specifically, the leading order vertex corrections, which are ${\it O}(1/N^2)$,  
are not included in the NCA.

\begin{figure}
\vspace*{7pt}         
\begin{center}       
\leavevmode           
\epsfxsize=4.0in
\epsfbox{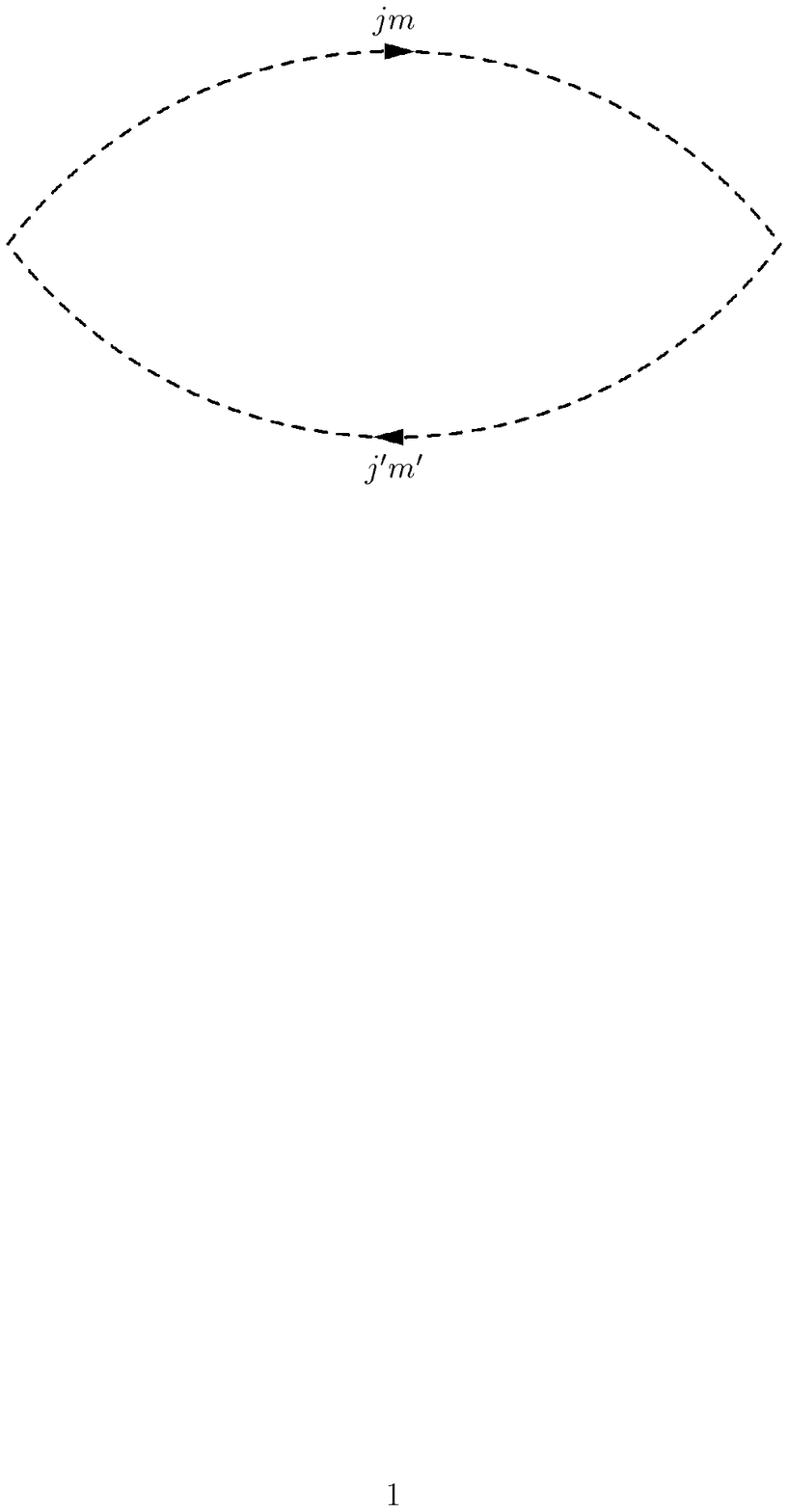}
\end{center}
\vspace{0.5cm}
\caption
{ Leading order Feynman diagram for static magnetic susceptibility.
In this diagram, only $f$ electrons are coupled to the field.}
\label{sus.dia}
\end{figure}

These self-consistent integral equations are solved  to second order in the
hybridization for the ionic propagator self energies.
Then   physical properties, such as the resistivity and  magnetic susceptibility
$\chi$,  are calculated as
convolutions of these propagators.
Fig.~\ref{sus.dia} shows a leading order Feynman diagram for the static
magnetic susceptibility and its convolution integral is given by Eq.
(\ref{eq:chi}). This is discussed more in section ~\ref{sus.ssec}.

The NCA shows a  pathological behavior (due to the
truncation of the
diagrammatic expansion) for a temperature scale $T_p \ll T_0$ in this
conventional Anderson
model. However,  provided the $f^1$ occupancy $n_f \ge 0.7$, 
and  $N\ge 4$,  this is not a problem,  as shown in Ref.~\cite{bcw}, in that comparison of NCA
results with exact
thermodynamics from the Bethe-Ansatz shows agreement at the few percent
level above $T_p$.
Hence, this is a reliable method for our purposes.

\begin{figure}
\vspace*{7pt}         
\begin{center}       
\leavevmode           
\epsfxsize=4.0in
\epsfbox{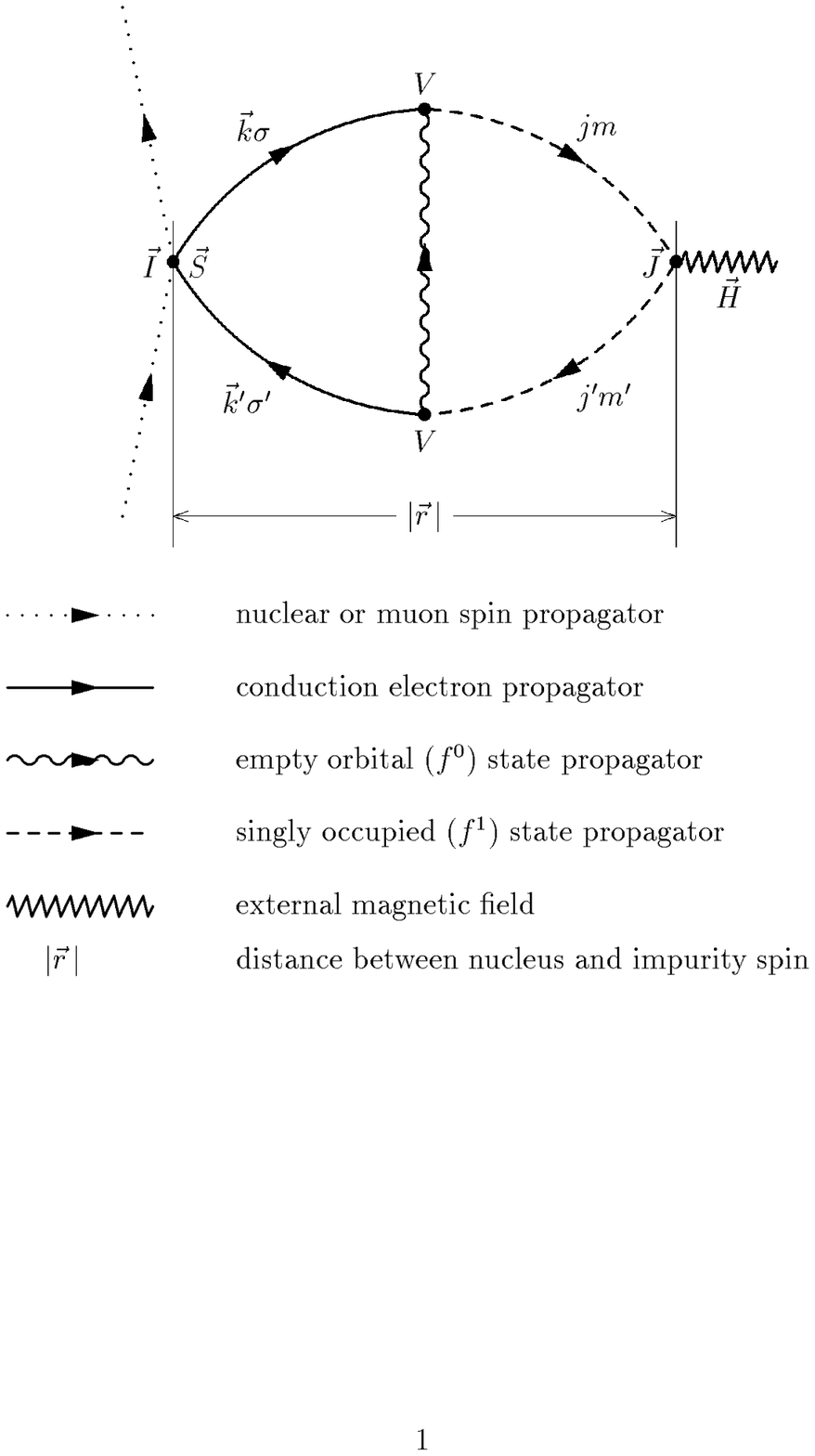}
\end{center}
\vspace{0.5cm}
\caption
{ Feynman Diagram for the Knight shift calculation for Ce ions. 
This is the lowest order
diagram of coupling between Ce local moment and nuclear spin in the infinite
$U$ Anderson model. All the propagators are explained in the figure. 
For the incoherent calculation the conduction electron propagator is a
bare electron propagator and when the the coherence effect for conduction 
electron is included ( multiple scattering off of $f$-sites) , 
it becomes a dressed propagator.}
\label{feyn}
\end{figure}

Our numerical procedure, briefly, consists of solving the NCA integral
equations for the
Anderson Hamiltonian specified above on a logarithmic mesh with order
600 points chosen
to be centered about the singular structures near the ground state
energy $E_0\sim
\epsilon_{f5/2}$.  We then feed the self-consistent propagators for the
empty and singly
occupied orbitals into the convolution integrals obtained from the
diagram  of Fig.~ \ref{feyn},
which allows for evaluation of the Knight shift at arbitrary angle and
distance from
the nuclear site. This  will be explained more in  section~\ref{shift.ssec}.
  It is convenient to take the nuclear
site as the origin in this case leading to phase factors $e^{-i\vec
k\cdot R}$
in the hybridization Hamiltonian ${\cal H}_{cf}$, where $\vec r$ is the
nuclear-Ce site separation.
These factors give the oscillations and position space angular
dependence in the Knight shift $K$.

\subsection{Magnetic Susceptibility}
\label{sus.ssec}
The static magnetic susceptibility is a direct indicator of the nature of the
ground state for the Kondo and Anderson model. Near room temperature, the 
susceptibility $\chi(T)$ displays a Curie-Weiss temperature dependence for most
heavy-electron materials. $\chi$ is linearly  related to the Knight shift
$K(T)$.
At low temperature, $\chi(T)$ does not follow the Curie-Weiss law and for
some systems, a linear relation of the Knight shift $K(T)$  to the magnetic 
susceptibility breaks down and hence shows a  Knight shift anomaly. 

In general the magnetic susceptibility comes from 
self-correlation of the conduction band magnetization($\chi_{cc}$), 
the self-correlation of the $f$ magnetization($\chi_{ff}$), and the mutual 
correlation of $f$ and band components($\chi_{cf}$).
The leading diagram comes from the second term where the only $f$ electrons are
 coupled to the field and
this gives a good approximation to the overall susceptibility. 
This diagram for magnetic susceptibility is
in  Fig.~\ref{sus.dia}.
Then, in the NCA, the magnetic susceptibility in the zero field
limit  can be written~\cite{hewson} 
\begin{equation}
\chi(T)=\sum_{j} \frac{\mu_j^2N_j}{Z_{4f}(T)} 2\int \frac{d\zeta}{\pi}A_{jm}
^{(-)}(\zeta,T)\mbox{Re}G_{jm}(\zeta,T)\label{eq:chi}
\end{equation}
where $\mu_j$ is the effective magnetic moment which is  defined  as $\mu_j^2=
(g_j\mu_B)^2j(j+1)/3$ where $g_j$ is the Land\'{e} $g$ factor for the $j$ 
multiplet and $Z_{4f}(T)$ is the impurity partition function.

Also we can get the van Vleck magnetic susceptibility between $j$ and $j'$ 
 angular momentum multiplet.
\begin{eqnarray}
\chi^{''}_{m_jm_{j'}}(\omega)
&=&\mu_B^2\frac{|\langle jm_j|J_z+S_z|j'm_{j'} \rangle |^2}{Z_{4f}(T)} (1-e^{-\beta\omega})\nonumber \\
&& \times \int \frac{d\zeta}{\pi} A_{jm_j}(\zeta+\omega)A_{j'm_{j'}}^{(-)}(\zeta)\nonumber \\
& & 
\label{eq:chijj'}
\end{eqnarray}
With $\langle \frac{5}{2},m|J_z+S_z|\frac{7}{2},m' \rangle  = -(\sqrt{49-4m^2}/14)
\delta_{mm'} $, the total van Vleck susceptibility for $j=5/2$ and $j'=7/2$ is
\begin{eqnarray}
\chi_{vv}^{''}(\omega) &=& \sum_{m_j,m_{j'}}\chi_{m_jm_{j'}}^{''}(\omega) \nonumber \\
 & =& \frac{8}{7}\mu_B^2\frac{(1-e^{-\beta\omega})}{Z_{4f}(T)} \int
\frac{d\zeta}{\pi} A_{jm}(\zeta+\omega)A_{j'm}^{(-)}(\zeta).
\end{eqnarray}
The susceptibility sum rule is derived from the zero frequency limit of the
Hilbert transform of $\chi_{vv}^{''}(\omega,T)$,
\begin{equation}
\chi_{vv}(T)=\chi'_{vv}(0,T) =\int \frac{d\omega}{\pi}\frac{\chi^{''}_{vv}(\omega,T)}
            {\omega}
\end{equation}

In our calculation,  both the $j=5/2$, $7/2$ contributions to the static magnetic susceptibility and the 
 van Vleck
susceptibility are considered for Ce ions. For Yb ions, only the static susceptibility of 
the $j=7/2$ ground spin-orbit multiplet is calculated because that the energy 
gap between two spin orbit multiplets ($j=7/2$ and $j=5/2$)  is large (
about $1.2eV$) For Y$_{1-x}$U$_x$Pd$_3$, only the van Vleck 
susceptibility between different $\Gamma$ states is  calculated because
the assumed $\Gamma_3$ ground state is a non-magnetic doublet.

\subsection{Knight Shift}
\label{shift.ssec}
Knight shift measurements on the  nuclear spins of  non-$f$ ions in Kondo or
heavy electron materials
 can probe the local induced magnetic fields. The additional fields come from 
all the possible polarization sources, such as  conduction electron spin polarization. 
For $f$ electrons, the radius wave function 
is small and  they are well screened, so there is little possibility
of direct overlap interactions between the nuclear spins and local moments.
In particular, the polarization of conduction
 electrons by the polarized local Kondo impurities, {\it i.e.} the transferred
$f$-electron polarization, is usually expected to have the most 
significant temperature dependent contribution.

The Knight shift of Heavy electron materials is induced by the
indirect interactions of the magnetic impurity and host nuclear spin
mediated by the conduction electrons.
Without the charge fluctuations introduced by the hybridization
interaction between conduction electron and $f$ electron, this
indirect interaction has the RKKY interaction form.
So at high temperatures, the Knight shift follows the RKKY interaction and
at low temperatures where the Kondo effect appears, the Knight shift
can show deviation from the RKKY form.

To calculate the Knight shift we need to evaluate the Feynman diagram in
Fig~\ref{feyn} which is the lowest order diagram coupling the nuclear spin to,
say,  Ce
magnetic moments, ignoring direct Coulomb exchange coupling.
 All the propagator symbols are  explained in the figure.
For the incoherent calculation, the conduction electrons are assumed to belong
to a broad, featureless, and symmetric band of half width $D$.
The conduction electron propagator is taken to be a
bare electron propagator, {\it i.e.} it includes no self
energy effects reflecting multiple scattering off the $f$-sites.
 When the the lattice coherence effects of 
conduction electrons are included, the self energy arising from 
the  scattering of conduction
electron at every $f$  site is
included in the conduction electron propagator in an approximate
way.

The Knight shift for a nuclear spin or muon at $\vec r$ is approximated as
\begin{eqnarray}
  K(r,\theta)& =&-\frac{8\pi V^2}{3N_s}\frac{g_e\mu^2_B}{\Omega}\sum_{}
 f_{jj'}(\alpha,\theta) \int k^2\,dk\nonumber \\
 & & \times \int k'^2\, dk' I_{jj'}(\epsilon_k, \epsilon_{k'}),
\label{eq:knight}
\end{eqnarray}
where $g_e$ is the electron $g$-factor and $\mu_B$ is the Bohr magneton. 
Here $\theta$ is the angle between $z$ axis and 
the bond direction $\vec r$ which connects
the nucleus or muon to a given $f$ ion in the crystal and $\alpha$ is the angle
between the field axis and bond axis. 
$I_{jj'}(\epsilon_k, \epsilon_{k'})$ is given by
\begin{eqnarray}
I_{jj'}(\epsilon_k, \epsilon_{k'})&  =& j_3(kr)j_3(k'r) \frac{1}{Z_{4f}}
\sum_{\omega,\nu} G_0(k\sigma,i\omega)G_0(k'\sigma',i\omega)\nonumber \\
&& \times G_{jm}(i(\omega+\nu))G_{j'm'}(i(\omega+\nu))D(i\nu) \nonumber \\
 & =&2\;\frac{j_3(kr)j_3(k'r)}{\epsilon_k- \epsilon_{k'}}(f(\epsilon_k)
I_{1jj'}(\epsilon_k) \nonumber \\
     & & + (1-f(\epsilon_k)) I_{2jj'}(\epsilon_k)),
\label{eq:ikk'}
\end{eqnarray}
where
\begin{eqnarray}
I_{1jj'}(\epsilon_k) & = &\frac{2}{Z_{4f}}\!\int\!\frac{d\zeta}{\pi} \int \!
\frac{d\rho}{\pi}\frac{(B^{(-)}
 (\zeta)A_{jm}(\rho)\mbox{Re}G_{j'm'}(\rho)}{\epsilon_k+\zeta-\rho}\nonumber \\
& = & \frac{1}{Z_{4f}}\!\int\!\frac{d\zeta}{\pi}B^{(-)}(\zeta)
    \mbox{Re}G_{jm}(\zeta+\epsilon_k)\nonumber \\
 && \;\;\;\;\times\mbox{Re}G_{j'm'}(\zeta+\epsilon_k)\\
I_{2jj'}(\epsilon_k) & = &\frac{-2}{Z_{4f}}\! \int\!\frac{d\rho}{\pi} \int\! \frac{d\zeta}{\pi}
\frac{B(\zeta)A_{jm}^{(-)}(\rho)\mbox{Re}G_{j'm'}(\rho))}{\epsilon_k+\zeta-\rho}\nonumber\\
 & = &  \frac{2}{Z_{4f}} \!\int\! \frac{d\rho}{\pi}A_{jm}^{(-)}(\rho)\mbox{Re}G_{j'm'}(\rho)
                         \mbox{Re}D(\rho-\epsilon_k).
\end{eqnarray}
The derivation is explained in detail in Appendix~\ref{k_sum.app}.

We can analytically evaluate  the inner $k'$ integral as 
\begin{equation}
\int_{0}^{\infty}k'^2\;dk'\frac{j_3(k'r)}{k^2-k'^2},  
\end{equation}
and the results are presented in   Appendix~\ref{j3int.app}.

For a magnetic field in the $z$ direction ({\it i.e.}, $\theta=\alpha)$, 
and for $I=1/2$, 
the angular dependent function in  the Knight shift, $f_{jj'}(\alpha,\theta)$, is 
given by 
\begin{equation}
f_{jj'}(\theta) =\cos^2 \theta f_{jj'}^z+\frac{\sin^2\theta }{4} (f^{1-}_{jj'}
+f^{2-}_{jj'} +f^{1+}_{jj'} + f^{2+}_{jj'}).
\end{equation}
Here $f^z_{jj'}(\theta)$ is 
\begin{eqnarray}
f_{jj'}^z(\theta) &=& \sum_{m_j, m_{j'}}\langle jm_j|J_z+S_z|j'm_{j'} \rangle \sigma^z_{\alpha \alpha}
\langle j'm_{j'}|\hat{r} \rangle \langle \hat{r}|jm_j \rangle  \nonumber \\
       &=& \sum_{m_j,\alpha,m_3} \alpha \langle jm_j|J_z+S_z|j'm_{j'} \rangle 
      \langle j'm_{j'}|3m_3; \frac{1}{2}\alpha \rangle \nonumber \\
    && \times \langle 3m_3;\frac{1}{2}\alpha|jm_j \rangle |Y_{3m_3}(\hat{r})|^2.
\end{eqnarray}
and $f^{1\pm}_{jj'}(\theta)$ and $f^{2\pm}_{jj'}(\theta)$ are
\begin{eqnarray}
f^{1\pm}_{jj'}(\theta) &=& \sum_{m_jm_{j'}\alpha,\beta}\langle j,m_j|J_{\pm}|j',
m_{j'}\rangle \langle j'm_{j'} |\hat{r}\rangle \sigma^{\mp}_{\alpha\beta}
\langle \hat{r}| j,m_j \rangle \nonumber \\
&=& \sum_{m_j,m_{j'},m_3, m'_3,\alpha,\beta}\!\!\!\langle j,m_j|J_{\pm}|j',m_{j'}
\rangle  \langle j'm_{j'}|3m'_3; \frac{1}{2}\alpha\rangle \nonumber \\
 & & \times\langle 3m'_3; \frac{1}
{2}\alpha|\sigma^{\mp}_{\alpha,\beta}|3m_3;\frac{1}{2}\beta \rangle
\langle 3m_3;\frac{1}{2}\beta |jm_j\rangle \nonumber \\
     & & \times Y^*_{3m_3}(\hat{r})Y_{3m'_3}(\hat{r})\\
f^{2\pm}_{jj'}(\theta) &=& \sum_{m_jm_{j'}\alpha\beta}\langle jm_j|S_{\pm}|j',
m_{j'}\rangle \langle j'm_{j'} |\hat{r}\rangle \sigma^{\mp}_{\alpha\beta}
\langle \hat{r}| j,m_j \rangle \nonumber \\
&=& \sum_{m_j,m_{j'},m_3, m'_3,\alpha,\beta}\!\!\!\langle j,m_j|S_{\pm}|j',m_{j'}
\rangle  \langle j'm_{j'}|3m'_3; \frac{1}{2}\alpha\rangle \nonumber \\
 & & \times\langle 3m'_3; \frac{1}
{2}\alpha|\sigma^{\mp}_{\alpha,\beta}|3m_3;\frac{1}{2}\beta \rangle
\langle 3m_3;\frac{1}{2}\beta |jm_j\rangle \nonumber \\
   & & \times Y^*_{3m_3}(\hat{r})Y_{3m'_3}(\hat{r})
\end{eqnarray}
 The explicit values for $f_{jj'}(\theta)$ are 
\begin{eqnarray}
f_{\frac{5}{2}\frac{5}{2}}(\theta)&=&\frac{9}{28\pi}(1-8\sin^2\theta +6\sin^4\theta)\\
f_{\frac{7}{2}\frac{7}{2}}(\theta)&=&\frac{2}{7\pi}[2+12\sin^2\theta-9\sin^4\theta]\\
f_{\frac{5}{2}\frac{7}{2}}(\theta)&=&\frac{3}{28\pi}[4-4\sin^2\theta+3\sin^4\theta].
\end{eqnarray}

To assess the relevance of this single site physics to the periodic
compound CeSn$_3$, 
we also carried out incoherent lattice sums over a few hundred radial
shells of
 Ce atoms around the Sn nucleus. For each atom in a shell the distance $r$ and angle
$\theta$ of Ce ion is calculated and the Knight shift of each ion
is evaluated for given position.
 Then the contribution from each ion is added to
get the total Knight shift. 
For the Y$_{1-x}$U$_x$Pd$_3$, impurity configuration averaging is also carried out.
This single site physics is known to be a good approximation at high
temperatures where the ions are incoherent with one other, and known
to provide a very accurate description of the thermodynamics in
many cases.
For  CeSn$_3$, given the tetragonal symmetry at the Sn site, 
we fixed the field in the $z$ direction and averaged over the $xz$,$xy$ and
$yz$ host planes for the Sn nucleus.  
Note that $yz$ plane is equivalent to $xz$.

For YbCuAl, which has hexagonal symmetry, we have to consider three 
possible field directions, along the $x$, $y$ and $z$ axes. 
Al has an $I=5/2$ nuclear spin  and the NMR shift was obtained from derivative
spectra
 of the central ($ 1/2 \leftrightarrow -1/2$) NMR transition.
Then
\begin{equation}
f_{jj'}(\alpha,\theta)= A\cos\alpha f^z_{jj'}(\theta)+\frac{B}{4}\sin\alpha
[f^+_{jj'}(\theta)+f^-_{jj'}(\theta)].
\end{equation}
where $f^{\pm}_{jj'}=f^{1\pm}_{jj'}+f^{2\pm}_{jj'}$.
Also $A$ and $B$ are given by
\begin{eqnarray}
A &=& \frac{1}{4}(25\cos^5\theta-26\cos^3\theta+5\cos\theta)\nonumber \\
B &=& \frac{1}{4}\sin\theta(25\cos^4\theta-14\cos^2\theta+1).
\end{eqnarray}
Then, 
\begin{eqnarray}
f_{\frac{7}{2}\frac{7}{2}}(\alpha,\theta)
&=&\frac{2}{7\pi}[2A\cos\alpha(1+3\sin^2\theta) \nonumber \\
&&+B\sin\alpha(5+3\cos^2\theta)],\\
f_{\frac{5}{2}\frac{5}{2}}(\alpha,\theta)
&=&\frac{9}{28\pi}[A\cos\alpha(1-4\sin^2\theta)\nonumber \\
& &  -B\sin\alpha (3-2\sin^2\theta)].
\end{eqnarray}
For a detailed derivation, see  Appendix~\ref{angle.app}.

To the extent that the dynamics of the empty orbital
can be neglected, the Knight shift expression (Eq.~(\ref{eq:knight}))
 factorizes into a nearly temperature
independent RKKY interaction (modified due to
the spin-orbit coupling and anisotropic hybridization from the original
form) times
the $f$-electron susceptibility.  Thus, no anomaly results from the
diagram in this
limit. In this limit, the susceptibility in the diagram corresponds to
the
the leading order estimate used in Ref. \cite{bcw} to compare with
exact Bethe-Ansatz results.

We can gauge the effects of charge fluctuations with a simple approximation
~\cite{cox1,ph88}. For $T=0$, the empty orbital
propagator may be written in an approximate two-pole form, one with
amplitude $1-Z$,
$Z=\pi k_BT_0/N\Gamma$, centered near zero energy, and one with
amplitude $Z$ centered
at $\epsilon_g \sim \epsilon_{f}-k_BT_0$ which reflects the anomalous ground
state mixing due
to the Kondo effect. The singly occupied  propagator has a simple pole 
structure.
\begin{eqnarray}
B(\omega) &=& (1-Z)\delta(\omega) +Z\delta(\omega-\epsilon_g) \nonumber \\
A_m(\omega) &=& \delta(\omega-\epsilon_{f})\nonumber\\
B^{(-)}(\omega) &=& Z\delta(\omega-\epsilon_g) \nonumber \\
A_m^{(-)}(\omega) &=& \frac{1-Z}{2}\delta(\omega-\epsilon_g).
\end{eqnarray}
Then 
\begin{eqnarray}
I_1(\epsilon_k)&=&Z/(k_BT_0-\epsilon_k)^2 \nonumber \\
I_2(\epsilon_k)&=&\frac{1-Z}{k_BT_0}(\frac{1-Z}{\epsilon_k-\epsilon_{f}}+
\frac{Z}{\epsilon_k})
\end{eqnarray}
Now, we can perform the $k$ integral putting above equations in Eq.
(\ref{eq:ikk'}). At low temperatures only conduction electrons which
have momentum close to $k_F$ participate in the interaction.
We can rewrite the radial momentum $k$ as
\begin{equation}
k=k_F+(k-k_F) \approx k_F +\frac{\epsilon_k}{\hbar v_F}
\end{equation}
Then we can write
\begin{equation}
e^{ikr}=e^{ik_Fr}e^{i\epsilon_k r/\hbar v_F}~.
\end{equation}
>From this we see that 
small $\epsilon_k$ gives large contribution to the Knight shift.
The contribution from the integral 
$I_1$ depends on whether $r\gg \hbar v_f/k_BT_0$.
For  $r\gg \hbar v_f/k_BT_0$ only very small $\epsilon_k < k_BT_0$ 
contribute the $k$ integral and we can approximate
$1/(k_BT_0-\epsilon_k) \approx 1/k_BT_0 $ and this contribution has the
amplitude $Z/k_BT_0=\pi/N\Gamma$.
For  $r\ll \hbar v_f/k_BT_0$, the Knight shift
 has contributions from $\epsilon_k > k_BT_0$ 
and this term has  the  amplitude $Z/D$. The amplitude of the Knight shift 
outside of the
coherence length $\xi_k=\hbar v_F/k_BT_0$ is $D/k_BT_0$ times
bigger that that at inside the the Kondo screening cloud.
This term can give the Knight shift anomaly.
Because $\epsilon_{f}$ is much bigger than the $\epsilon_k$, 
 the first term of $I_2$  gives conventional RKKY
oscillations modulo the
anisotropy and altered range dependence induced by the $m,\hat k$
dependence of the hybridization.
The amplitude of the second term goes to zero above the Kondo
temperature.  This term also may contribute to the anticipated
anomalous Knight
shift, and within such a two-pole approximation may be seen to be
finite within $\xi_K$,
have a stronger distance dependence in that regime, but possess an
amplitude of order
$Z$ only within this distance regime.  Beyond $\xi_K$, the amplitude is
of order 1/$N$
and
the shape of the spin oscillations
is the same as that found from the high frequency pole of the empty
orbital propagator.

\begin{figure}
\vspace*{7pt}         
\begin{center}       
\leavevmode           
\epsfxsize=4.0in
\epsfbox{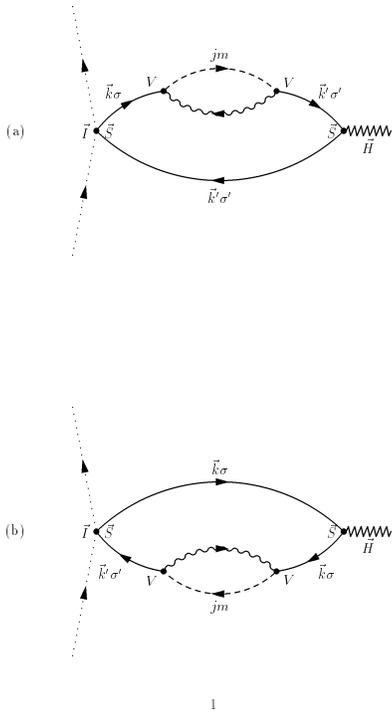}
\end{center}
\vspace{0.5cm}
\caption
{ Feynman Diagram for the additional contribution to the
 Knight shift which was noted by  S\o rensen and Affleck 
\protect\cite{soaf95,soaf96,baaf96}. This is significant in 
magnitude for distances beyond the
Kondo screening cloud radius and not included in our calculation.
 This contribution is of potential importance for
any dilute system, but  is not important in our
lattice calculation ( see, Sec.~\ref{shift.ssec} for discussion.). 
 All the propagators are explained in the Fig.~\ref{feyn}.} 
\label{soaf-diag}
\end{figure}

S\o rensen and Affleck~\cite{soaf95,soaf96,baaf96} have noted that for a
single impurity an
additional contribution is present below the Kondo screening cloud which
is not present in this calculation.  This corresponds to the diagrams
shown in Fig.~\ref{soaf-diag}, in which scattering occurs off of the 
impurity for one or other conduction legs.  This process will occur for
{\it any} impurity in a metal, and at a fundamental level corresponds, 
to the contribution induced by the field dependence of $k_F$ for
the two different spin branches.  The contribution then goes as $1/R^2$,
but only outside the Kondo coherence length where the low temperature
screening cloud can be regarded as a potential scatterer.  (The $1/R^2$
follows trivially from differentiating the Friedel oscillation with
respect to $k_F$.)  This contribution is of potential importance for 
any dilute system, but we argue that it is not important in our
lattice context (or, for that matter, for any system with an appreciable 
concentration of impurities).  The reason is that the $T$-matrix insertions of 
Fig.~\ref{soaf-diag} will go over to self-energy insertions in the 
lattice, as we sum over all possible sites.  These self-energy
insertions will simply provide the renormalized Pauli susceptibility
contribution to the Knight shift, which is not the dominant
contribution.

\subsection{Coherent Lattice Effects}
\label{cohe.ssec}
Coherent lattice effects are included within the local approximation ( $d=
\infty$ limit) to the lattice model. 
In this approximation a conduction electron self energy is included using the
average -T matrix approximation~\cite{ekl74} which assumes 
that conduction electron scatters off every $f$-electron site.
This corresponds to a first iteration of the local approximation.
In contrast,   these multiple scattering 
processes are ignored in the incoherent limit.
The NCA approximation treats intra-site interactions to all orders.
In this calculation we consider intersite coupling which involves simple
hopping process in perturbation theory (ATA) and ignore intersite interactions 
which involving transfer of particle-hole pairs between sites.
This coherent lattice effect may reduce the Kondo screeinig length~\cite{millee}.

The Anderson lattice model for spin 1/2 has a conduction electron Green's 
function given by~\cite{yy85}
\begin{eqnarray}
G_c(\vec k, \omega,T)&=&(\omega-\epsilon_k-\Sigma_c(\omega,T))^{-1}\nonumber\\
&=&(\omega-\epsilon_k-\frac{V^2}{\omega-\epsilon_f
-\Sigma_f^{int}(\vec k, \omega, T)})^{-1} \label{eq:gc1}.
\end{eqnarray}
Where $\Sigma_f^{int}(\vec k, \omega, T)$ is the $f$-electron self energy 
arising from $f-f$ interactions. 
 The same results follow for the N-fold  degenerate model.
We remove wave-vector dependence of the self energy  
by neglecting the  intersite interactions. 
The $f$ electron self energy comes from the $f-f$ interaction 
$(\Sigma^{int}_{f})$ 
 and hybridization with the conduction electrons.
This hybridization energy is given by
\begin{equation}
V^2D(\omega)=\frac{V^2}{N_s}\sum_{\vec k}\frac{1}{\omega-\epsilon_k}.
\label{eq:d}
\end{equation}
For a  featureless symmetric  band $D(\omega)V^2$ becomes, for $\omega 
\rightarrow 0$, $i\Gamma$,
where $\Gamma=\pi N(0)V^2$ is the single particle hybridization width($N(0)$ the
density of state at the Fermi energy).
Here the conduction
electron can not be scattered at the same $f$ electron site twice and this 
site restriction gives the cancellation of the hybridization self
energy term ($-i\Gamma$) in the full $4f$ Green's function~\cite{coxgre}.

Specifically, the band electron self energy is written within this approximation
as ~\cite{grewe84,grewe82}
\begin{eqnarray}
\Sigma_c(\omega, T)& =& V^2[(G_{4f}(\omega,T))^{(-1)} + V^2D(\omega)]^{(-1)}\nonumber\\
& =& V^2G_{4f}^{(int)}(\omega,T),
\label{eq:sigmac}
\end{eqnarray}
where $G_{4f}$ is the the full $4f$ on-site Green's function   given  by~\cite{bcw}
\begin{eqnarray}
G_{4fjm}(\omega)& =& \frac{1}{Z_{4f}}\int \frac{d\xi}{\pi} [\;B^{(-)}(\xi)
G_{jm}(\omega+\xi+i\delta)\nonumber\\
&&-A_{jm}^{(-)}(\xi)D(\xi-\omega-i\delta)\;]
\end{eqnarray}
within the NCA.

 Since   vertex corrections are neglected in the
NCA, the conduction electron Green's
function is determined completely through $\Sigma_f$ by hybridization and 
 any   resistance solely arises
from the damping of band states due to the imaginary part of $f$ electron
self energy.
A realistic estimate of $\Sigma_f$ is  important to study 
lattice coherence effects. 
By the standard Fermi-liquid phase space argument~\cite{hewson}, the imaginary part
 of the exact, on-site f-electron self 
energy  is given , for low frequencies and temperatures, by
\begin{equation}
-\mbox{Im} \Sigma_{4f}(\omega,T) \sim \omega^2+(\pi k_BT)^2+\Gamma\label{eq:imf}
\end{equation}
In the NCA calculation, due to the approximation involved, the minimum value in
 $-\mbox{Im}\Sigma_{4f}( \omega, T) $ does not occur precisely at the Fermi energy
(though it differs only by a small fraction of $T_0$) and is not equal to 
$\Gamma$~\cite{coxgre}.  So in our numerical calculations, 
 $\mbox{Im}\Sigma_{4f}( \omega, T) $ is extrapolated to
$\omega, T \rightarrow 0$ and  $-\mbox{Im}\Sigma_{4f}(0,0)$  is 
replaced by $\Gamma$.

In our calculation, the $f$-electrons have $j=5/2$ and $j=7/2$ states by 
spin-orbit coupling. Thus the conduction electron self energy has two terms 
from each $j$ state, viz.
\begin{eqnarray}
\Sigma_c(\omega,\alpha,T) &=& \sum_{j,m_j} |\langle j,m_j|V(k)|\vec k, \alpha
\rangle|^2G_{4fj}
^{(int)}(\omega)\nonumber \\
&=&4\pi V^2 \sum_{jm_j}|\langle j,m_j|3m_3;\frac{1}{2}\alpha\rangle|^2\nonumber\\
 && \;\;\; \times |Y_{3m_3}(\hat{k})|^2 G_{4fj}^{(int)}(\omega)\nonumber \\
& =&  V^2(3G_{4f5/2}^{(int)} +4G_{4f7/2}^{(int)}).
\end{eqnarray}
where the one particle hybridization strength $V(\vec K)  = V$ is, taken to
be independent of $|\vec k|$ in this calculation, and we used
\begin{eqnarray}
\sum_m|\langle\frac{5}{2}m|3m_3;\frac{1}{2}\alpha\rangle|^2|Y_{3m_3}(\hat{k})|^2 & = & 
\frac{3}{4\pi} \nonumber \\
\sum_m|\langle\frac{7}{2},m|3m_3;\frac{1}{2}\alpha\rangle|^2|Y_{3m_3}(\hat{k})|^2 &=&
\frac{1}{\pi}. 
\end{eqnarray}

With the inclusion of lattice coherence effects, the 
$I_{jj'}(\epsilon_k,\epsilon_{k'})$ term in the Knight shift calculation is 
changed from the incoherent form, Eq. (\ref{eq:ikk'}), to
\begin{eqnarray}
\lefteqn{\int k^2dk \int k'^2dk'I_{jj'}(\epsilon_k,\epsilon_{k'})=}\nonumber \\
& &-\!\int\!\frac{d\xi}{\pi}
\mbox{Im}J(\xi)\mbox{Re}J(\xi) [f(\xi)I_{1jj'}(\xi) + (1-f(\xi))I_{2jj'}(\xi)]\nonumber \\
&&
\end{eqnarray}

where
\begin{eqnarray}
J(\xi)& =& \int dk\frac{k^2j_3(kr)}{\xi-\epsilon_k- \Sigma_c(\xi)},\\
I_{1jj'}(\xi) & = & \frac{2}{Z_{4f}}\!\int\!\frac{d\zeta}{\pi}\! \int\! \frac{d\rho}{\pi}\frac{(B^{(-)}
(\zeta)A_{jm}(\rho)\mbox{Re}G_{j'm'}(\rho)}{\xi+\zeta-\rho}\nonumber \\
& = &\!\frac{1 }{Z_{4f}}\!\int\!\frac{d\zeta}{\pi}B^{(-)}\!(\zeta)
\mbox{Re}G_{jm}\!(\zeta+\xi) \mbox{Re}G_{j'm'}\!(\zeta+\xi),\nonumber\\
  & & \\
I_{2jj'}(\xi) & = &\frac{-2}{Z_{4f}}\!\int\!\frac{d\zeta}{\pi}\! \int\! \frac{d\rho}{\pi}
\frac{B(\zeta)A_{jm}^{(-)}(\rho)\mbox{Re}G_{j'm'}(\rho))}{\xi+\zeta-\rho}\nonumber\\
        & = & \! \frac{2}{Z_{4f}}\! \int\! \frac{d\rho}{\pi}A_{jm}^{(-)}\!(\rho)
\mbox{Re}G_{j'm'}(\rho) \mbox{Re}D(\rho-\xi).
\end{eqnarray}

\section{Results}
\label{result.sec}
In this chapter, the numerically calculated  results for  
the   Knight shift $K(T)$ and magnetic susceptibility $\chi(T)$ of heavy
electron materials such as CeSn$_3$, YbCuAl, and U$_{0.2}$Y$_{0.8}$Pd$_3$ will
be presented.
CeSn$_3$ and YbCuAl are  concentrated heavy electron materials and 
 U$_{0.2}$Y$_{0.8}$Pd$_3$ is a proposed two channel quadrupolar
Kondo heavy electron alloy.

\begin{figure}
\vspace*{7pt}          
\begin{center}         
\leavevmode            
\epsfxsize=3.0in
\epsfbox{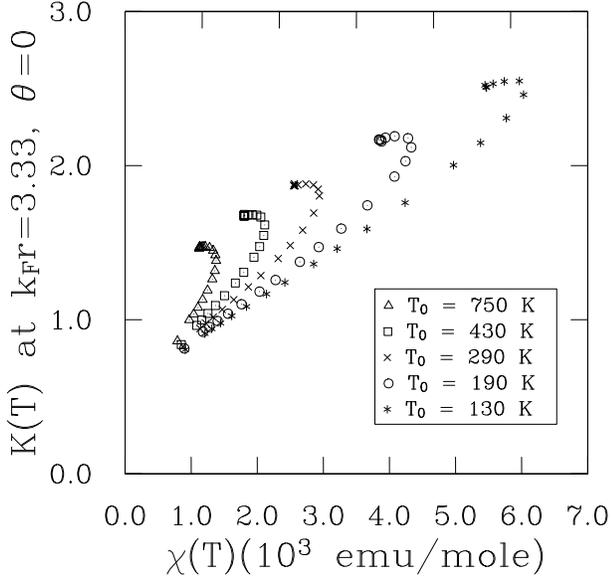}
\end{center}
\caption{ Calculated Knight shift $K(T)$ vs. susceptibility $\chi(T)$
for a single Ce site at $k_Fr=3.33$ from a nuclear moment and angle
$\Theta =0$.   Fixing the $f$-level energy $\epsilon_{5/2}=-2$eV, and
the spin orbit splitting $\Delta_{so}=0.29eV$, the hybridization
is varied to illustrate the dependence of the
nonlinearity on the magnitude of $T_0$ (which ranges from 750 K to 130
K in these
calculations as $\Gamma$ varies from $0.165eV$ to $0.130eV$).  The diagram 
of Fig.~\ref{feyn} is used to calculate $K(T)$.
The magnitude of the non-linearity diminishes as $T_0$ is reduced.
The theoretical Knight shift has been shifted by a common offset and
scale factor  to match the susceptibility.}
\label{Ce-temp}
\end{figure}

First, the Knight shifts $K(\vec{r}, T)$   are systematically 
calculated  for different values of the Kondo temperature $T_0$ which 
is controlled by the
hybridization width $\Gamma = \pi N(0)V^2$, where $N(0)$ is the conduction
electron density of states at the Fermi energy and $V$ is the one particle
hybridization strength,  for Ce and Yb compounds. 
These results  show 
that the magnitude of the Knight shift anomaly   depends
upon the distance $ \vec r$ between the local magnetic moment and the  
nucleus and the Kondo 
temperature $T_0$.  There is an anomaly even for the small distance $\vec r$.
The magnitude  of deviation between a linear $K$ $.vs$  $\chi$
 relation is systematically increased when the distances are increased and 
the Kondo temperatures are increased. 
These results are shown in Fig.~\ref{Ce-dist} and Fig.~\ref{Ce-temp}. 
These calculations support  Ishii's idea of an anomalous 
conduction electron spin density cloud~\cite{i76} which sets in beyond the 
Kondo screening length $\xi_K=\hbar v_F/k_BT_0$, where $v_F$ is the Fermi velocity.

The lattice sum is carried out over a few hundred shells. 
 In subsection~\ref{cesn3.ssec} the results for CeSn$_3$ 
are  compared with  experiments. Both the 
$^{119}$Sn NMR Knight shift and $\mu$sr Knight shift are
studied. Also the influence of lattice coherence  of the conduction electrons on
  both the 
NMR and positive muon Knight shift is investigated
 using the average T-matrix approximation
and the numerically calculated results for CeSn$_3$ are 
compared with the experiments and also the calculated incoherent Knight shift.
The calculated 
$^{119}$Sn NMR Knight shift agrees well with the experiment. The incoherent
$\mu$sr
Knight shift shows an anomaly but has opposite sign. For coherent case, the Knight 
shift from different muon site gives an anomaly with opposite sign. We may fit
the experimental results by averaging out two $\mu$sr Knight shifts.

For YbCuAl, because of its complicated crystal structure, the incoherent 
lattice sum is carried out  over several thousand atoms. 
The calculated $^{27}$Al NMR Knight shift   
 results are  mentioned in subsection~\ref{ybcual.ssec}.
 These results show excellent agreement with the experiments.

In the last subsection, the Knight shift and the magnetic susceptibility of
U$_{0.2}$Y$_{0.8}$Pd$_3$ is discussed. 
We do a full incoherent lattice sum and impurity configuration averaging for 
U$_{0.2}$Y$_{0.8}$Pd$_3$.
Our study doesn't show the low temperature Knight shift anomaly like experiment.
 
\subsection{Systematic Calculations}
\label{sys.ssec}
To see the systematic behavior of the magnetic susceptibility and the Knight
shift, we have calculated $\chi(T)$ and $K(\vec r, T)$  for different 
 Kondo temperatures $T_0$. For spin-orbit coupling and zero crystal field 
splitting, $T_0$   is given  by
\begin{equation}
  k_BT_0= D(\frac{\Gamma}{\pi|\epsilon_f|})^{1/N_g}(\frac{D}
{\Delta_{so}})^{N_{ex}/N_g} \exp(\frac{\pi \epsilon_f}{N_g\Gamma}),
\label{eq:T0}
\end{equation}
where  $N_g$ is the  degeneracy of the ground spin orbit multiplet, $N_{ex}$
is the degeneracy of the excited multiplet,  $\epsilon_f$ is the energy level 
position of the ground multiplet,  $\Delta_{so}$ is the energy gap between
two spin orbit multiplets, and  $\Gamma = \pi N(0)V^2$
is the  hybridization width
($D$ is the physical Lorentzian bandwidth of the conduction electron). 
The values of the  parameters  we used for
Ce and Yb ions are listed in Table~\ref{ceyb}.

For several different  hybridization widths 
$\Gamma$, the Knight shift was studied as a function of temperatures and
distance $\vec r$ between the local impurity spin and nucleus at fixed angle
$\theta$ between $\vec r$ and  quantization axis 
$\vec k -\vec k'$.
In these calculations all other variables such as $\epsilon_f$, 
and $\Delta_{so}$
were fixed to the values which give the best magnetic susceptibility $\chi(T)$ fit 
to  experimental results of CeSn$_3$ for Ce 
ions and YbCuAl for Yb ions(see subsection~\ref{cesn3.ssec} and subsection
~\ref{ybcual.ssec} for the parameter values.). The Knight shift is scaled to the
susceptibility by matching at high temperatures.
For all the calculations the conduction electron band width $D$ was assumed 
to be $3eV$.
Because of the small gap between $j=5/2$ and $j=7/2$ states of Ce compound
($\Delta_{so}=0.29eV$ for Ce ions and $\Delta_{so}=1.30eV$ for Yb ions) ,
the van Vleck term  is included for only Ce compound studies.

\begin{figure}
\vspace*{7pt}         
\begin{center}       
\leavevmode           
\epsfxsize=3.0in
\epsfbox{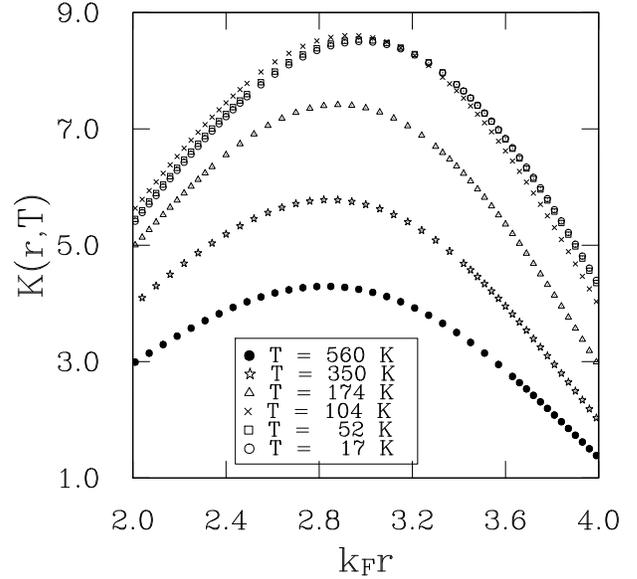}
\end{center}
\caption
{ Knight shift $K( r, T)$ at fixed angle $\theta =0$ and Kondo
temperature $T_0=430K$ for Ce ions on a fine scale. 
We use the dimensionless
variable $k_Fr$ with  the  Fermi wave vector $k_F=0.65\AA^{-1}$ instead of $r$.
Around $k_Fr=3.2$ the lines of $T=104K$ and $T=52K$ are crossed.
This can be explained that the Knight shift summation 
converges faster at higher 
temperatures and it has  longer range at lower temperatures as 
mentioned by E.S. S\o rensen and I. Affleck \protect\cite{soaf95,soaf96,baaf96}.
}
\label{K-t}
\end{figure}

\begin{figure}
\vspace*{7pt}          
\begin{center}         
\leavevmode            
\epsfxsize=3.0in
\epsfbox{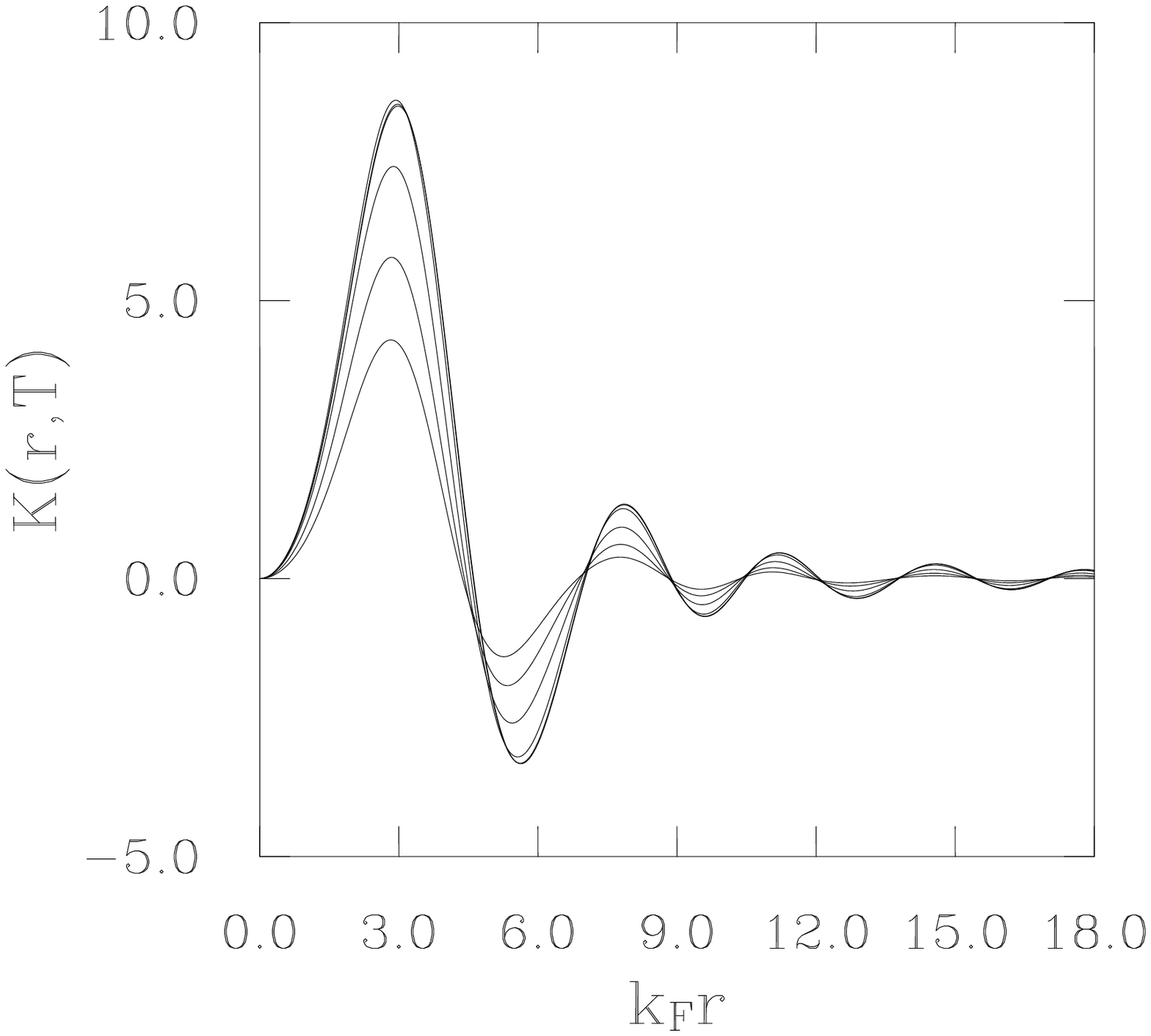}
\end{center}
\caption{ Knight shift $K( r, T)$ at fixed angle $\theta =0$ and Kondo
temperature $T_0=430K$ for Ce ions. 
We use dimensionless
variable $k_Fr$ with  the  Fermi wave vector $k_F=0.65\AA^{-1}$ instead of $r$. 
This shows an oscillatory RKKY like behavior.
The amplitude is decreased as the temperature is increased and the
distance $r$ is increased. The calculations are done at the same temperatures 
in Fig~\ref{K-t}.} 
\label{K-rt}
\end{figure}

Fig.~\ref{K-rt} shows the calculated Knight shift $K(\vec r, T)$ as a function of 
separation $r$ and temperatures $T$ at fixed angle $\theta=0$ and hybridization 
width
$\Gamma=0.050667D$ ($T_0=430K$) for Ce ions.
We use a dimensionless 
variable $k_Fr$ with  the  Fermi wave vector $k_F=0.65\AA^{-1}$ instead of $r$.
The Knight shift on a fine scale is 
shown  in Fig.~\ref{K-t}.    The Knight shift shows an oscillatory RKKY-like
behavior and the the total magnitude 
decreases when distance is increased. 
$|K(T)|$ at fixed distance $r$  is first increased  and then decreased
(it becomes almost constant),
when the temperature is lowered. 
In the Fig.~\ref{K-t}, we can see that the curves for $T=104K$ and $T=52K$ 
 cross around $k_Fr=3.2$. The temperature where
$K(\vec r, T)$ has the maximum value is a function of separation $r$ and has lower
 value with larger separation $r$. The Knight shift has different $r$ dependence
whether temperature is above the Kondo temperature or below the Kondo
temperature as   
mentioned by E.S. S\o rensen and I. Affleck~\cite{soaf95,soaf96,baaf96}.
The Knight shift converges faster at higher 
temperatures (above the Kondo temperature) and it has  longer range at lower temperatures. 
The Knight shifts of Yb ions  at fixed angles show similar behavior.

\begin{figure}
\vspace*{7pt}          
\begin{center}       
\leavevmode            
\epsfxsize=3.0in
\epsfbox{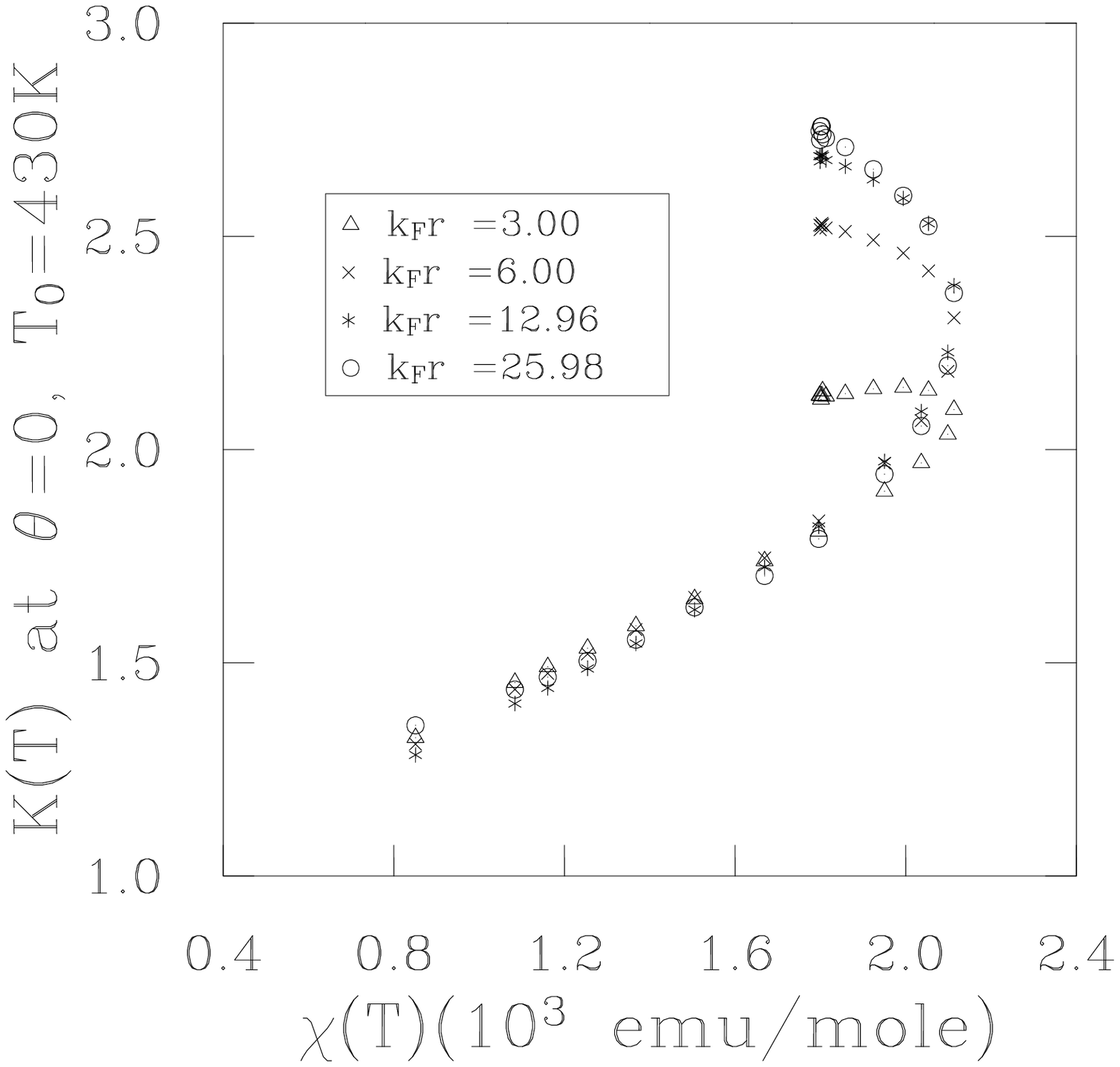}
\end{center}
\caption
{  Calculated Knight shift $K(T)$ for a single Ce site vs.
susceptibility $\chi(T)$
for varying separation with the Kondo scale used to fit the CeSn$_3$
$\chi(T)$(see Fig.~\ref{CeSn3}). For each plot the angle is held at $\theta=0$ 
with respect
to the nuclear moment-Ce axis. The diagram
of Fig.~\ref{feyn} is used to calculate $K(T)$.
 The magnitude of the non-linearity
diminishes as $k_Fr$ is
reduced.  The theoretical Knight shifts have been shifted by offset
and scale factors to match the susceptibility; this does not affect
the relative size of the anomaly.}
\label{Ce-dist}
\end{figure}

For a fixed Kondo temperature $T_0$ and distance $r$, 
the knight shift $K(T)$ shows a 
linear relation with the magnetic susceptibility $\chi(T)$ at high temperature
 and it starts to deviate from the linear relation  and shows an anomaly 
when temperature is lower than $T_{max}$ where $\chi(T)$ reaches its maximum
value.
Fig.~\ref{Ce-dist} shows  the Knight shift at $\theta = 0$, $T_0=380K$ as a 
function of the magnetic susceptibility for different separations, with temperature 
as an implicit variable for Ce ions.
These curves show a  linear $K$ $vs.$ $\chi$  relation at high temperatures
(in this figure, high 
temperatures correspond to the lower left corner) and show non-linear $K-vs.-
\chi$ relations at low temperatures.
The anomaly, {\it i.e.} 
the magnitude of the non-linearity, diminishes as the separation
$r$ is decreased.
In Fig.~\ref{Ce-temp},  the  Knight shift is investigated at $k_Fr=3.3$, and 
$\theta=0$  for different Kondo temperatures.  
As in Fig.~\ref{Ce-dist}, temperature is an implicit variable and high 
temperatures correspond to the lower left corner.
The magnitude of deviation from the linear relation is decreased when the Kondo 
temperature is reduced.
For CeSn$_3$, $k_F\xi_k\approx 25$ where  $\xi_K=\hbar v_F/k_BT_0$ is the 
Kondo screening length with the Fermi velocity $v_F$.
So our calculation is done well 
inside the conduction
electron screening spin cloud. 
Very similar results are obtained for Yb compounds.

This results qualitatively confirm the Ishii's argument~\cite{i76}
 that when the 
radius $r$ is bigger than Kondo screening  length $\xi_K=\hbar v_F/k_BT_0$,
where $v_F$ is the Fermi velocity, the anomalous conduction spin density
cloud sets in. 
Far inside this radius, conventional
temperature independent RKKY oscillations dominate of the kind observed
by Boyce and Slichter.
Outside the screening length, at $T=0$,
the anomalous term will dominate also with an RKKY form
but an amplitude of order $D$, the conduction bandwidth, compared with
$D(N(0)J)^2$ for the Ruderman-Kittel term, where $N(0)$ is the conduction
electron density of states at the Fermi energy and $J$ the conduction
 electron- local moment exchange coupling.  Ishii did not calculate the 
explicit temperature dependence of this structure, but did anticipate that 
it would vanish above the Kondo scale.
Scaling analysis confirmed the asymptotic factorization of the
Knight shift for short distance and low temperature~\cite{cjk87}.
A possible understanding of the
Boyce and Slichter results, then, is that the Cu nuclei they sampled were
at distances $r<<\xi_K$ from the Fe ions.
But, in our calculation, the anomaly is still present  and the
magnitude is surprisingly large  for short distance provided $T_0$ is large. 
A heuristic basis for understanding this is the  two-pole approximation, as 
discussed in subsection~\ref{shift.ssec}.

\subsection{CeSn$_3$}
\label{cesn3.ssec}
The compound CeSn$_3$ has the f.c.c. AuCu$_3$  crystal  structure.
The local symmetry at every cerium site is cubic. With decreasing temperature, the 
magnetic susceptibility $\chi(T)$ of CeSn$_3$
 shows first typical
Curie-Weiss-like behavior, followed by a maximum at $T_{max} \approx
 40 K $ and tending to a constant value at $T=0$~\cite{mv75,m85}.  
CeSn$_3$ has positive amplitude of Knight shift for the $^{119}$Sn nuclei.
However, our calculated amplitude of the Knight shift before  scaling it to
$\chi$ is
actually negative. This implies that the fit is sensible only if the 
assumed contact coupling between conduction and nuclear spins is negative. 
This actually makes sense because the Sn nucleus 
should dominantly couple through core polarization, which produces a negative
effective contact coupling. 
The NMR Knight shift $K(T)$ and  $\chi(T)$ are related  linearly 
 above  $T_{max}$. But the Knight shift doesn't follow the magnetic
susceptibility below  $T_{max}$ and  has weaker temperature dependence . 
The $^{119}$Sn NMR Knight shift has a positive anomaly, {\it i.e.} it has larger
magnitude than that of magnetic susceptibility below the $T_{max}$.  
The  positive muon spin rotation($\mu$sr) measurement shows that the
positive muon Knight shift also 
exhibits an anomaly below $T_{max}$, but it has a different sign with respect
to the susceptibility as compared to
 the NMR Knight shift~\cite{m85}. 

\subsection{NMR Knight Shift}
\label{nmr.subsec}
To study the Knight shift for CeSn$_3$, first the  hybridization width $\Gamma$
for CeSn$_3$ is decided by comparing the calculated magnetic 
susceptibility and experimental magnetic susceptibility, assuming a conduction
electron density of states half width $D=3eV$.
As in the systematic calculation, a  fixed value for parameter  
$\epsilon_{f5/2}=-2eV=-0.6667D$ is assumed, and $\Delta{so}=0.29eV=0.096667D$  is used.
Then $D$ is varied  a little to  fine tune the results. 
 The Knight shift is calculated with the fixed Kondo
temperature (specified by $\Gamma=0.050666D$) and conduction band width $D=2.655eV$. 
The best fit 
Kondo temperature  which is given  by Eq. (\ref{eq:T0}) is $T_0=0.0123$,
 in units of $D$ {\em i.e.}  $T_0=380K$.

For the Knight shift, an incoherent lattice sum is carried over several hundred shells
of surrounding Ce ions about a given $^{119}$Sn nucleus. We assume the Knight shift
contribution of each ion to be described by this single impurity model,
known to be a good approximation at high temperatures where the ions are
incoherent with one another, and known to provide a very accurate description
of the thermodynamics in many cases.

Because Sn is on the face of  each unit cell,  we also average the
Knight shift over the
reference frame determined by whether $^{119}$Sn is on the $xy$ plane, or the $xz$ plane(for $\vec H 
\parallel \hat{z}$). Note that $yz$ is equivalent to $xz$.
Like the systematic calculation, a Van-Vleck term is included for both
magnetic susceptibility $\chi(T)$ and the Knight shift $K(T)$ calculation.
The Knight shift is scaled by an intermediate range temperature match to 
the susceptibility, because the experimental data is measured only up to
room temperatures.

\begin{figure}
\vspace*{7pt}         
\begin{center}        
\leavevmode           
\epsfxsize=3.0in
\epsfbox{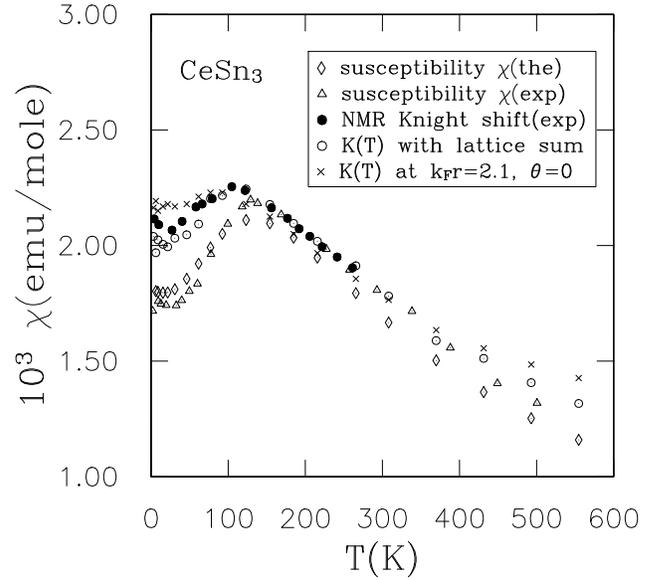}
\end{center}
\caption{  Temperature dependence of $^{119}$Sn Knight Shift $K(T)$ and Ce
$\chi(T)$ (both
calculated and experimental results \protect \cite{mv75,m85}) for CeSn$_3$. The
theoretical $K(T)$ is
calculated using the diagram of Fig.\ref{feyn} and with $T_0$ chosen to fit the
experimental
$\chi(T)$ data. A full (incoherent) lattice sum is carried
out over several hundred shells of atoms.}
\label{CeSn3}
\end{figure}

As shown in Fig.~\ref{CeSn3} the results agree well with the experiments
in spite of the oversimplified conduction electron band structure.
Experimentally  the $^{119}$Sn  NMR Knight shift has a 
positive sign and is linearly 
related to Ce magnetic susceptibility $\chi(T)$ at high temperatures.
$K(T)$ shows an anomalous deviation from Ce $\chi(T)$ at low temperatures
and the magnitude is bigger than $\chi$.  At high temperatures above
$400K$, $K(T)$ is reduced slowly with increasing temperature and shows
 deviation from $\chi$.  Experiment measures the Knight
shift only up to  $300K$, so we are not sure whether  this behavior appears
in experiment.
The magnitude of the anomalous contribution goes down with distance from the 
Sn nucleus and the theoretical data at a fixed distance 
which most closely match those of experiment
 are taken $k_Fr=2.1 $ and $\theta =0$.
 Note that this distance is an order of magnitude smaller than the Kondo screening length 
$\xi_K = \hbar v_F/k_BT_0 \approx 25/k_F$.

\subsection{Lattice Coherence  Effects}
\label{cohe.subsec}
We study lattice coherence effects within a
local approximation($d=\infty$ expansion) to the lattice model.
The conduction electron propagator in Fig.~\ref{feyn} becomes dressed
and the conduction electron self energy is calculated using the average T-matrix
approximation, which is exact for a Lorentzian density of states within the
local approximation~\cite{ata1,ata2}, and otherwise corresponds to the first iteration
 of the $d=\infty$ self consistency.
 The same parameters for the incoherent Knight shift calculation  are
used for the coherent lattice calculation.

\begin{figure}
\vspace*{7pt}          
\begin{center}         
\leavevmode            
\epsfxsize=3.0in
\epsfbox{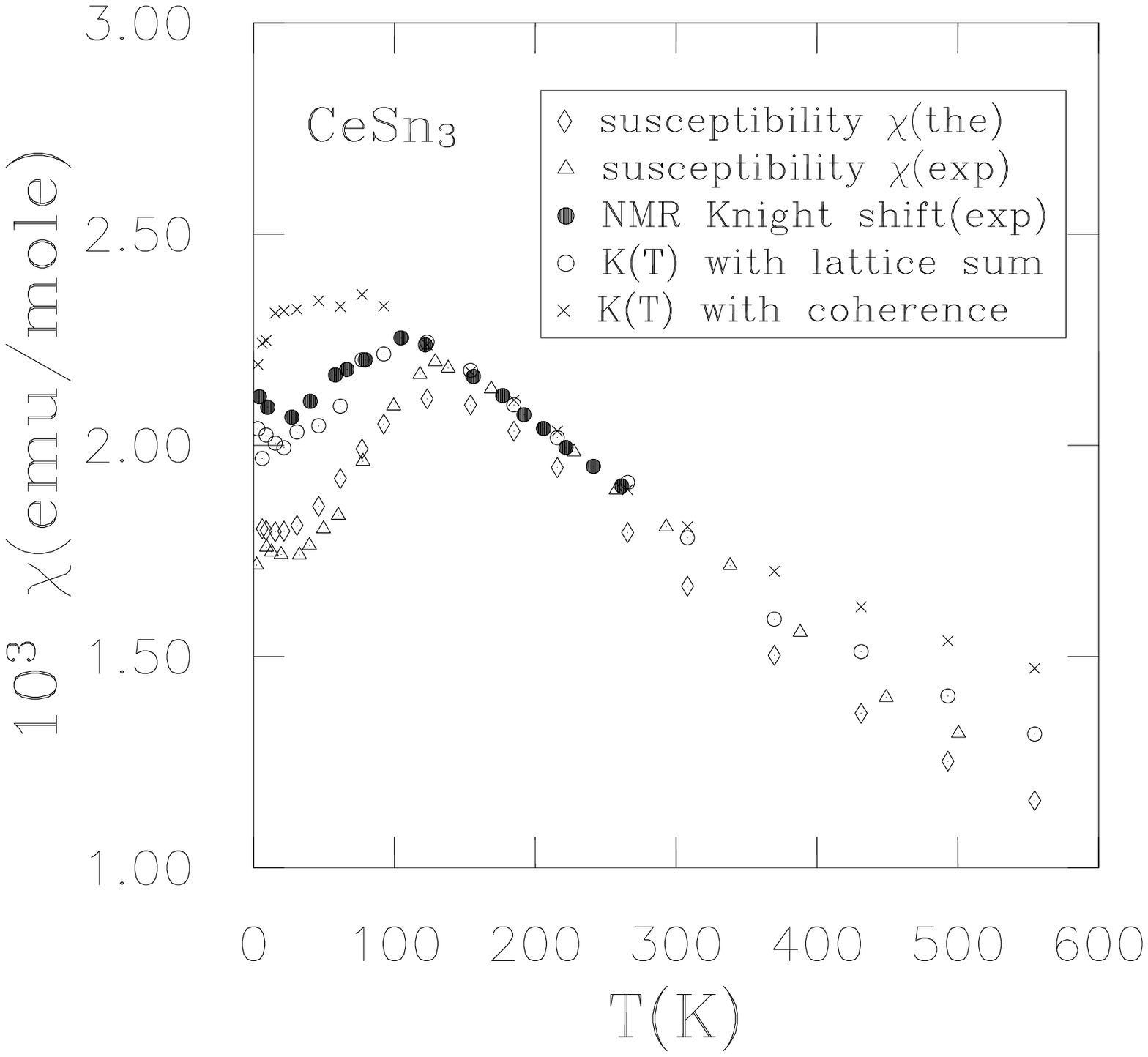}
\end{center}
\caption{  Temperature dependence of $^{119}$Sn Knight Shift $K(T)$ with coherence 
effect and Ce $\chi(T)$ (both
calculated and experimental results \protect \cite{mv75,m85}) for CeSn$_3$. The
theoretical $K(T)$ is
calculated using the diagram of Fig.~\ref{feyn} and with $T_0$ chosen to fit the
experimental
$\chi(T)$ data. A full coherent lattice sum is carried
out over several hundred shells of atoms
  using  the average T-matrix approximation to approximate the coherent
conduction self energy.}
\label{cohe}
\end{figure}

The calculated results fit experimental data well in
the region of temperatures between
$100K$ and $250K$ where the experimental data exist. 
At low temperature, the Knight shift with coherent
lattice effects also shows an anomaly, with a little discrepancy with the experiment.
At higher temperatures it shows tails which has bigger values than the 
experimental susceptibility(see Fig.~\ref{cohe}).
If we fit the results to this high temperature magnetic susceptibility, then
the Knight shift has bigger values at the maximum temperature and
shows sharp changes with temperature changes.
We are not sure which is the best way to fit the result with the
 experiment.
The calculated magnitudes(before the scaling to fit the experimental results)
of both incoherent and coherent Knight shift have similar values at high
temperatures where coherent lattice effects are small. 
Explicit values are shown in Table~\ref{kvalue}.

\begin{figure}
\vspace*{7pt}         
\begin{center}         
\leavevmode            
\epsfxsize=3.0in
\epsfbox{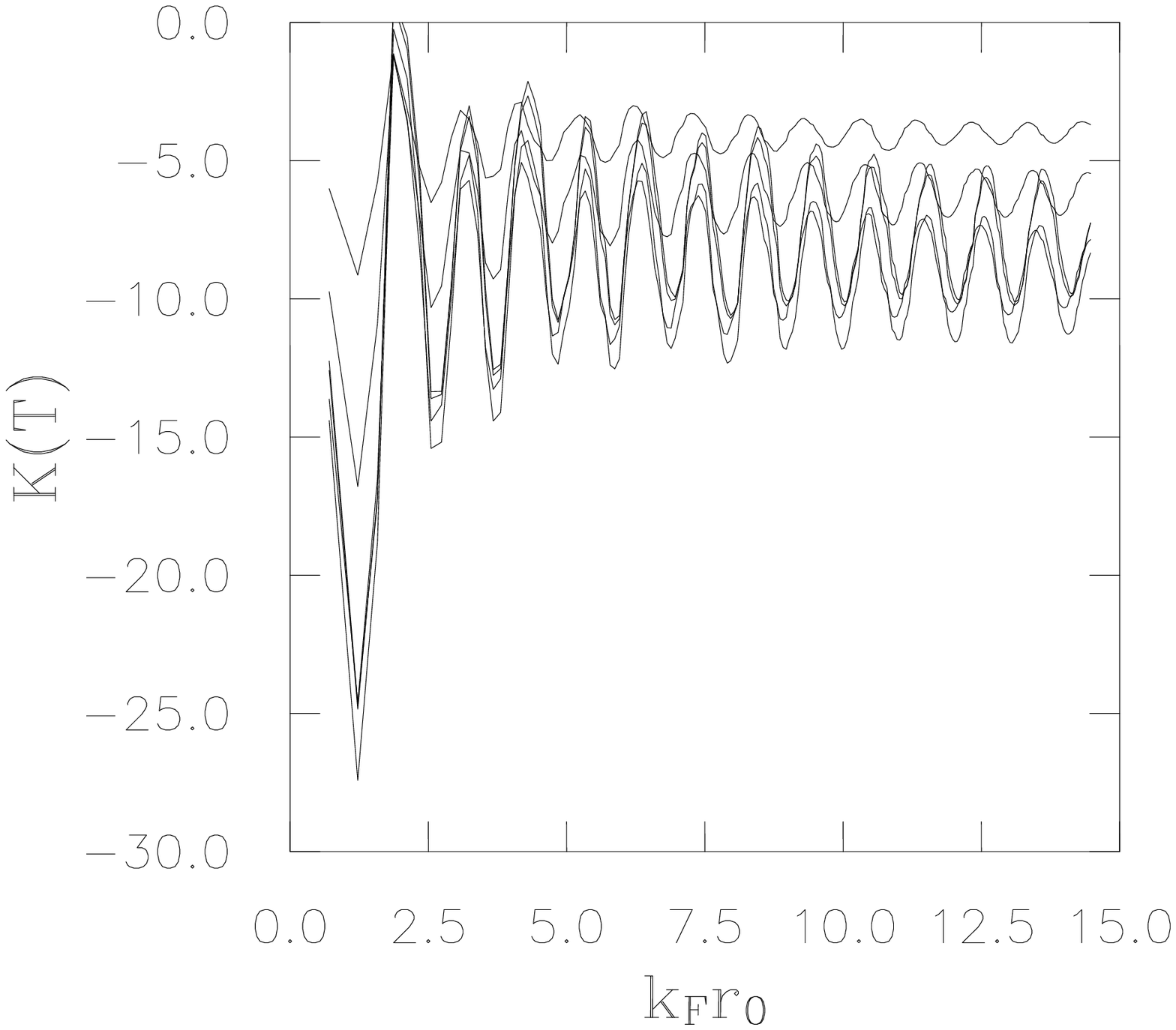}
\end{center}
\caption{ Temperature dependence of  Knight Shift $K(T)$ with small constant
damping $-\mbox{Im}\Sigma_f=0.01$ in incoherent lattice sum. 
In this figure $K(T)=\sum_{k_Fr < k_Fr_0}K(r,T)$. The calculations are
done at the same temperatures  in Fig.~\ref{K-t}. 
The small absolute magnitude corresponds to the high temperature and it
increases as the temperatures goes down. 
The convergence is slower than the coherence lattice sum.}
\label{sgamma}
\end{figure}

\begin{figure}
\vspace*{7pt}          
\begin{center}         
\leavevmode            
\epsfxsize=3.0in
\epsfbox{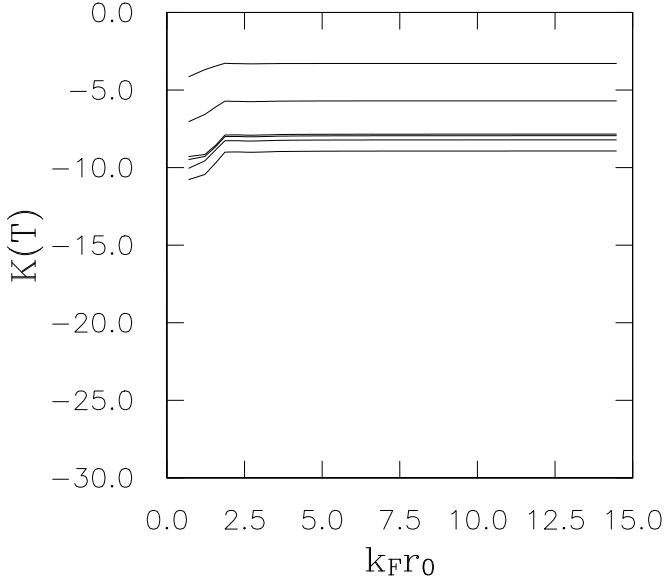}
\end{center}
\caption{ Temperature dependence of  Knight Shift $K(T)$ with large constant
damping $-\mbox{Im}\Sigma_f=1$ in incoherent lattice sum. 
In this figure $K(T)=\sum_{k_Fr < k_Fr_0}K(r,T)$. The calculations are
done at the same temperatures  in Fig.~\ref{K-t}. 
The small absolute magnitude corresponds to the high temperature and it
increases as the temperatures goes down. 
They  converge very fast. an impurity at large distance  doesn't contribute
to the total Knight shift because of the large damping, {\it i.e.} short
life time of the conduction electrons.}
\label{lgamma}
\end{figure}

\begin{figure}
\vspace*{7pt}         
\begin{center}         
\leavevmode            
\epsfxsize=3.0in
\epsfbox{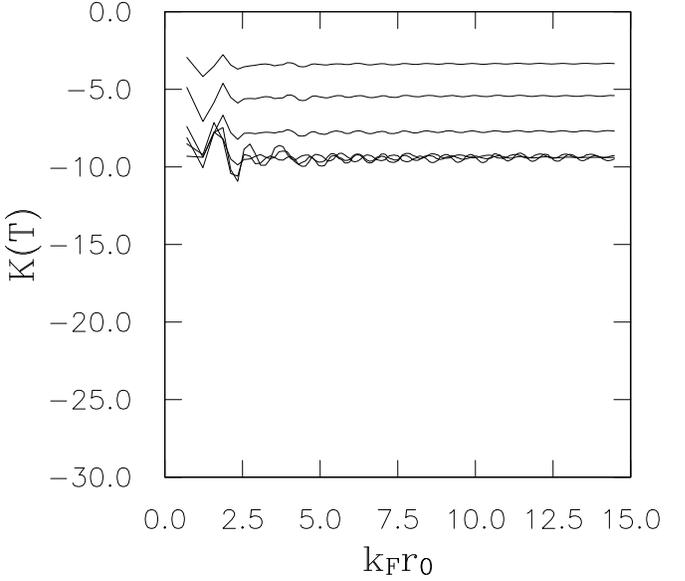}
\end{center}
\caption{ Temperature dependence of  Knight Shift $K(T)$ with coherent lattice 
sum.
In this figure $K(T)=\sum_{k_Fr < k_Fr_0}K(r,T)$. The calculations are
done at the same temperatures  in Fig.~\ref{K-t}. 
The small absolute magnitude corresponds to the high temperature and it
increases as the temperatures goes down. 
The convergence is faster than the small damping case but slower than
larger damping case.}
\label{shift}
\end{figure}

The average magnitude and amplitude of the oscillation of the Knight shift
 is decreased from the 
that of the incoherent lattice sum; we believe  
because of the damping effects brought in
by the imaginary part of conduction electron self energy. 
To test this idea, we added a 
phenomenological constant damping to the incoherent 
lattice sum. Fig.~\ref{sgamma} shows the result for small damping ,
$-\mbox{Im}\Sigma_f=0.01D$ and Fig.~\ref{lgamma} shows the results for large 
damping, $-\mbox{Im}\Sigma_f=1D$. In these
 figures  $K(T)=\sum_{k_Fr < k_Fr_0}K(r,T)$.
This study shows that the amplitude of oscillation is
reduced  when the damping is increased. And the amplitude of the
Knight shift converges the faster
when the damping is the bigger when distance is increased. 
An impurity at  large distance  doesn't contribute
to the total Knight shift because of the large damping, {\it i.e.} short
life time of the conduction electrons.
Also Fig.~\ref{shift} shows the Knight shift calculated 
with the coherent lattice sum.  
It shows a  behavior intermediate 
between small and large damping calculations.

\subsection{$\mu$sr Knight Shift}
\label{usr.subsec}
 Positive muons, stopped in a solid, come to rest at interstitial sites where 
the muon spin performs a Larmor  precession in the local magnetic field.
The muon Knight shift is then a measure of the local magnetic susceptibility.
For CeSn$_3$,
from volume considerations it is most likely that the muon preferentially
occupies the octahedral interstices of the AuCu$_3$ structure  (as it does in
metals with the closely related fcc structure), rather than the tetrahedral
sites. There are, however, two inequivalent octahedral sites, one at  the
center of the cubic unit cell and the other at the middle of the Ce atoms
which has  a non-cubic symmetry. So in principle two resonances are expected.
In the experiment by Wehr and {\it et al.}~\cite{wehr84}, only one resonance was
observed because either the muon performs a site average by fast diffusion
or the frequency difference is small  with respect to the apparent width 
of the signal which is given by the muon lifetime and the intrinsic width
$\gamma$.

In the Curie-Weiss regime of CeSn$_3$ above 200K, the temperature dependence 
of the positive muon Knight shift  is linearly related to the bulk
magnetic susceptibility.  In the intermediate valence regime of CeSn$_3$
the local magnetization as experienced by the muon decreases more strongly
below  200K than the magnetization of the $4f$ state as deduced from the bulk
susceptibility. This behavior reflects either a modification of the 
transferred hyperfine fields between the $4f$ moments and the muon or 
signals the influence of an additional negative Knight shift contribution
which    was absent or small in the  high temperature ranges.
Comparing with the NMR Knight shift, the  
  $\mu$sr Knight shift has a maximum at higher temperature and the sign
of anomaly is opposite.
This anomalous reduction of positive muon Knight shift might be regarded
as an indication for an additional negative d-electron Knight shift 
contribution\cite{wehr84}.
In terms of a band picture, the increase of d character at the Fermi level
can be understood as a $4f-5d$ hybridization effect. It is supported by
de Haas-van Alphen measurements~\cite{jc81}. Positive muons sitting between 
Ce atoms  should be particularly sensitive to variations 
in the $5d$ states from the  symmetry of these orbitals.

\begin{figure}
\vspace*{7pt}          
\begin{center}         
\leavevmode            
\epsfxsize=3.0in
\epsfbox{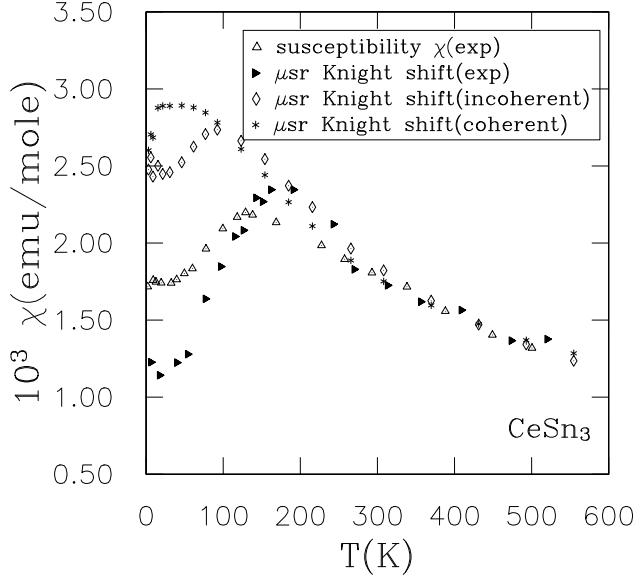}
\end{center}
\caption{  Temperature dependence of positive muon Knight Shift $K(T)$ and Ce
$\chi(T)$ (both
calculated and experimental results \protect\cite{m85,wehr84}) for CeSn$_3$. The
theoretical $K(T)$ is
calculated  assuming that muon sits at the center of cubic unit cell.
Other parameters are same as for the $^{119}$Sn NMR Knight shift calculation.}
\label{ce-usr1}
\end{figure}

We have calculated the $\mu^{+}$ Knight shift both possible muon
sites. Both incoherent and coherent lattice sums are carried out over  several 
hundred shells of Ce atoms.  All other parameters for the calculations are 
same as for the $^{119}$Sn NMR Knight shift calculations. 
Fig.~\ref{ce-usr1} shows the results assuming that muon sits at the center of 
the cubic unit cell. At high temperatures, the calculated $\mu$sr Knight shift 
agrees well with the experimental data and shows a linear relation with the
bulk magnetic susceptibility. Both incoherent and coherent lattice sum studies
give the correct magnitude for the low temperature Knight shift anomaly but
the wrong sign. Also the maxima occur at lower temperature than the data 
and the magnitude is
bigger than the experiment.

\begin{figure}
\vspace*{7pt}          
\begin{center}         
\leavevmode            
\epsfxsize=3.0in
\epsfbox{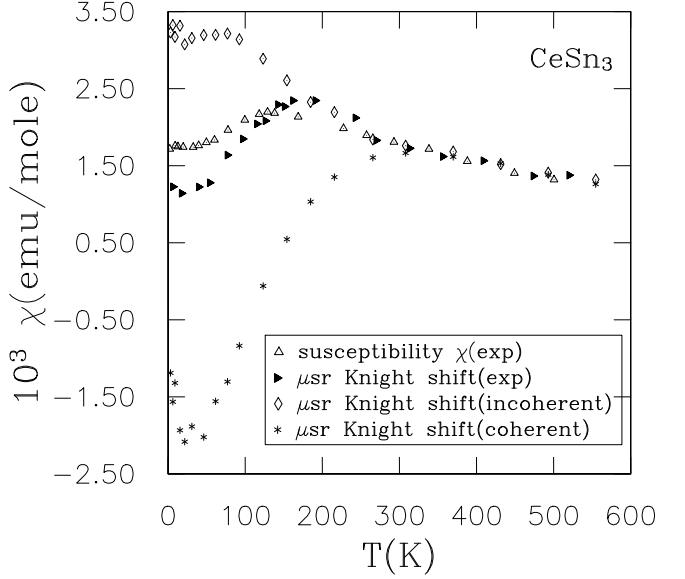}
\end{center}
\caption{  Temperature dependence of positive muon Knight Shift $K(T)$ and Ce
$\chi(T)$ (both
calculated and experimental results \protect\cite{m85,wehr84}) for CeSn$_3$. The
theoretical $K(T)$ is
calculated  assuming that muon sits at the middle of the axis between the
Ce atoms.
Other parameters are  same as for the $^{119}$Sn NMR Knight shift calculation.}
\label{ce-usr2}
\end{figure}

\begin{figure}
\vspace*{7pt}          
\begin{center}         
\leavevmode            
\epsfxsize=3.0in
\epsfbox{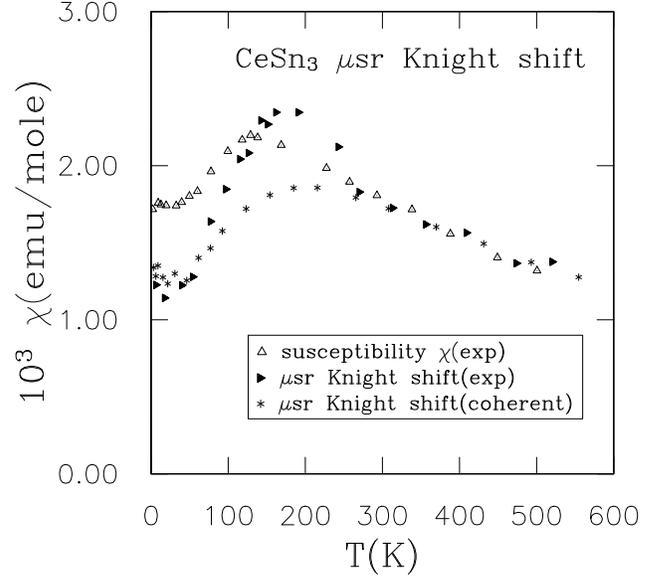}
\end{center}
\caption{  Temperature dependence of positive muon Knight Shift $K(T)$ and Ce
$\chi(T)$ (both
calculated and experimental results \protect\cite{m85,wehr84}) for CeSn$_3$. The
theoretical $K(T)$ is
calculated  averaging the $\mu$sr Knight
 shift from  both positions with fractional occupancy ratio $f=2/3$ ( 2/3 of 
the  muon is at the center of CeSn$_3$ unit cell.). }
\label{usr-add}
\end{figure}

Results are shown in Fig.~\ref{ce-usr2} for the case  where that muon
sits at  the middle of the Ce-Ce band axis.
In this case, results with the incoherent lattice sum show similar behavior 
to the previous calculations(assuming the muon sits at the center of the
unit cell). 
But the result of the calculation   shows very interesting behavior even 
though it does not agree with experiment. 
The Knight shift starts to deviate from the linear relation at higher
temperature and the magnitude of the anomaly is larger than the 
experimental result.
Note that the sign of the anomaly is agrees with experiment.
There is a possibility to fit the experimental data by 
 averaging the $\mu$sr Knight
 shift from  both positions. 
Because there is no experimental data which gives
information about the fractional site occupancy we just averaged two Knight
shifts with several fractional occupancy ratio $f$ which is defined as
\begin{equation}
K_{\mbox {total}}=fK_{\mbox {center}}+(1-f)K_{\mbox {bond-axis}}.
\end{equation}
Fig.~\ref{usr-add} shows the result for  $f=2/3$. Our calculation misses
the maximum point, but the  low temperature anomaly agrees with the
experimental result.

Also, if the muon 
is not situated at a site of cubic symmetry, dipolar fields from the induced
local moments may give the dominant contribution  to the positive muon 
Knight shift. The magnetic dipole which is inversely  related to the mass, can
give comparable contribution to the positive muon Knight shift contrast
to the $^{119}$Sn NMR Knight shift~\cite{cox1}.
The direct dipolar interaction energy of two magnetic dipoles $\vec m_1$ and
$\vec m_2$, separated by $\vec r$ is given
\begin{equation}
U=\frac{1}{r^3}[\vec m_1\cdot\vec m_2 -3(\vec m_1\cdot \hat{r})(\vec m_2
\cdot\hat{r})]
\end{equation}
The 
dipolar part of the Knight shift tracks directly the atomic susceptibility
of the nearest neighbor f-electron ions while the contact hyperfine
contribution is what we are interested in. 
In our calculation the dipolar field effect is of course exactly  cancelled out
when we  average over  the reference frame.

\subsection{YbCuAl} 
\label{ybcual.ssec}

The ternary inter-metallic compound YbCuAl has the hexagonal 
Fe$_2$P type  crystal structure~\cite{oe73,dmc68}, in which each Yb atom 
has the same local environment.

\begin{figure}
\vspace*{7pt}         
\begin{center}         
\leavevmode            
\epsfxsize=3.0in
\epsfbox{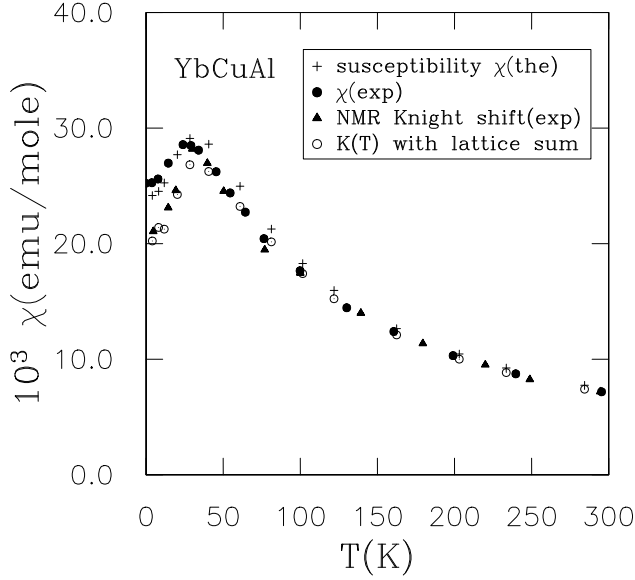}
\end{center}
\caption{  Temperature dependence of $^{27}$Al Knight Shift $K(T)$ and Yb 
$\chi(T)$ (both
calculated and experimental results~\protect\cite{mb79,m85,me77}) for YbCuAl.
 The theoretical $K(T)$ is
calculated using the diagram of Fig.~\ref{feyn} and with $T_0$ chosen to fit the
experimental
$\chi(T)$ data. A full (incoherent) lattice sum is carried
out over twenty four thousand atoms.}
\label{ybcual}
\end{figure}

At low temperatures, the magnetic susceptibility  $\chi(T)$ has a large 
constant value($\chi(0)=25.5 \times 10^{-3} e.m.u./mole$ Yb atoms) and 
a maximum value at 
  $T_{max} \approx 27K$. There is a Curie-Weiss like behavior above 
$T_{max}$~\cite{mb79,m85,me77}.
$^{27}$Al NMR shift data were obtained from derivative spectra of the central ($ 1/2 
\leftrightarrow -1/2$) NMR transition.
  Above  $T_{max}$, $\chi$ and $K$ track each other, as expected if only one 
mechanism is appreciably temperature dependent. Here the Yb magnetization 
is the obvious candidate for the temperature-dependent contributions to
both $\chi(T)$ and $K(T)$. This linear $K-vs.-\chi$ relation has been used 
to determine the relative scales of the $\chi(T)$ and shift coordinate axes
in Fig.~\ref{ybcual}. The linear relation breaks down below $T_{max}$.

The  $^{27}$Al NMR Knight shift has negative sign and the absolute magnitude of
the low temperature Knight shift is smaller than the magnetic susceptibility,
opposite to the case of CeSn$_3$. 
 For YbCuAl, the ground state energy for f state is taken as 
$\epsilon_{f7/2}= -1eV=
-0.333D$ and $\Delta_{so}=1.3eV=0.4333D$. Because of the large value of  
$\Delta_{so}$, we can neglect the interaction between $j=7/2$
ground state and $j=5/2$ excited state.
Without this  Van-Vleck term, we can estimate the 
conduction electron band half width
 $D$ using the zero temperature  magnetic susceptibility $\chi(0)$ value.
\begin{equation}
 \chi(0)= \frac{g_{j}^{2}j(j+1)u_{B}^{2}N_0}{3k_BT_0} 
\end{equation} 
With $\chi(0)=25.5 \times 10^{-3}$, we get $k_BT_0=8.7974\times 10^{-3}eV$.
Also with the relation
\begin{equation}
\left(\frac{T_0}{D} \right)_{exp}=\left(\frac{T_0}{D} \right)_{the}
\end{equation}
where $D_{the}=3eV$, and $(T_0/D)_{the}=0.00184$, we get $D_{exp}=4.7812eV$.
But, we get the best magnetic susceptibility fit with $D=3.5eV$ and $\Gamma=
0.02033D$. 
The incoherent lattice sum is carried out for over 8000
shells of atoms, larger than for CeSn$_3$. Because of the complicated crystal structure, each shell
only includes a few atoms.

Calculated results are shown in Fig.~\ref{ybcual}. 
 Both the magnetic susceptibility and the Knight shift agree well  with
experimental results and can explain both the magnitude and sign of
Knight shift anomaly in spite of the oversimplified conduction electron
band structure.  We note that the sign of the anomaly is opposite to
that of Ce in this case, and indeed we find that these contributions go
in opposite directions numerically.

\subsection{Y$_{0.8}$U$_{0.2}$Pd$_{3}$}
\label{yupd3.ssec}

The Y$_{1-x}$U$_x$Pd$_3$$^3$ system has aroused great interest, following the
discovery of non-Fermi liquid behaviour for uranium concetrations around 
$x=0.2$~\cite{yupd,sm92,sm94}.  This descrepancy has been interpreted as
possibly 
arising from a two channel quadrupolar Kondo effect~\cite{c87,yupd} or from critical
effects of a new kind of second-order phase transition at zero temperature~\cite{at91}.
We will mainly discuss the composition $x=0.2$.

Y$_{0.8}$U$_{0.2}$Pd$_{3}$ has a cubic AuCu$_3$  crystal structure.
The  $j=4$ ground state of
U$^{4+}$ is split to $\Gamma_1$ singlet, $\Gamma_3$ nonmagnetic doublet,
and $\Gamma_4$ and $\Gamma_5$ magnetic triplet.
Knowledge of the crystal field ground state is a
crucial test for the validity of the quadrupolar Kondo model and there are
several neutron scattering experiments to decide the ground state of 
Y$_{0.8}$U$_{0.2}$Pd$_{3}$~\cite{mooketal,eb95,dm95}.
Mook {\it et. al.}~\cite{mooketal}
interpreted their results in terms of a crystal field level scheme with a
$\Gamma_3$ doublet ground state and $\Gamma_5$ and $\Gamma_4$ excited triplet
states at 5 and $16meV$, respectively and thus support the two channel 
quadrupolar Kondo effect interpertation.
In this case, $\chi(T)$ originates in the Van-Vleck susceptibility associated
with transition from a $\Gamma_3$ nonmagnetic ground doublet into excited-state
 $\Gamma_5$ and $\Gamma_4$ triplets 
 The $\Gamma_3$ is described by a quadrupolar pseudospin.
This couples to pseudospin variables of a conduction $\Gamma_8$ quartet in
time-reversed channels with the antiferromagnetic pseudospin coupling  mediated
by virtual charge fluctuations to magnetic doublets in excited-state configurations.

McEwen {\it et al.}~\cite{eb95} saw  a peak of magnetic origin  at $36meV$ and another
 peak around $4meV$ and explained this with the transition  between $\Gamma_3$
ground state with excited states $\Gamma_4$, $\Gamma_1$ and $\Gamma_5$.
They couldn't find a peak at $16meV$.

Dai {\it et al.}~\cite{dm95} reported a $\Gamma_5$ magnetic ground state with polarized
inelastic neutron scattering experiment and $\Gamma_3$ and $\Gamma_4$ excited
states at 5, 39 $meV$. We note, however, in contrast to this 
interpretation, that there is no quasielastic scattering around
$\hbar\omega=3.6meV(\sim 42K=k_bT_0)$, expected for a conventional magnetic
Kondo effect.

\begin{figure}
\vspace*{7pt}          
\begin{center}         
\leavevmode            
\epsfxsize=3.0in
\epsfbox{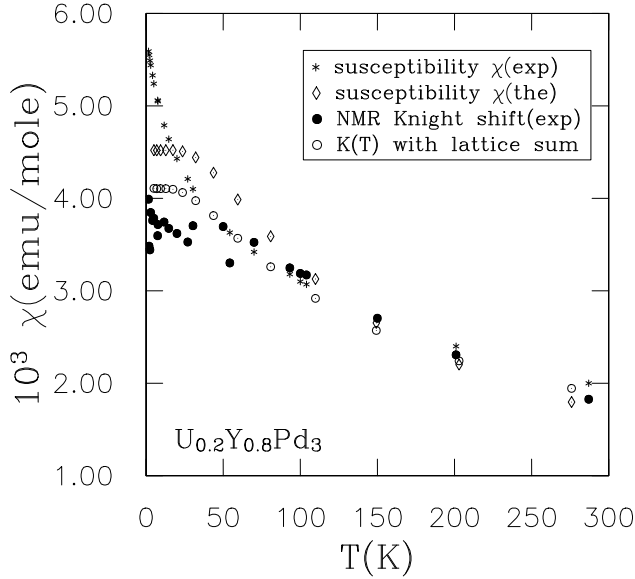}
\end{center}
\caption{  Temperature dependence of $^{89}$Y Knight Shift $K(T)$ and  U
$\chi(T)$ (both
calculated and experimental results~\protect\cite{lm94,lb95}) for Y$_{0.8}$U$_{0.2}$Pd$_{3}$. The
theoretical $K(T)$ is
calculated using the diagram of Fig.~\ref{feyn} and with $T_0$ chosen to fit the
experimental
$\chi(T)$ data. A full (incoherent) lattice sum and impurity configuration
averaging is carried
out over a several shells.}
\label{yupd3.fig}
\end{figure}

For $x=0.1$ and 0.2, a breakdown of the expected linearity between the 
NMR Knight 
shift and the bulk susceptibility $\chi$ is found below $\sim 50K$~\cite{lm94}.
The magnetic susceptibility exhibited an upturn at low temperatures(Curie tail),
indicating the presence of magnetic impurities(see Fig.\ref{yupd3.fig}). 
The impurity magnetic
susceptibility was subtracted~\cite{lb95}.
 The temperature dependence of $\chi(T)$~\cite{mooketal}
 suggests that the mechanism for the
two channel behavior is the quadrupolar Kondo effect~\cite{c87}.
\begin{equation}
\frac{\chi(T)}{\chi(0)} \approx 1-A(\frac{T}{T_0})^{(1/2)}
\end{equation}

In our study,
we consider  $f^2$ and $f^3$  configurations for U ions and  only the Hund's
rule ground states, {\it i.e.} $f^2$, $j=4$ and $f^3$, $j=9/2$ 
spin-orbit states are kept for the calculation. 
 $j=4$ states is split to $\Gamma_3$  ground doublet, 
$\Gamma_5$, $\Gamma_4$ and $\Gamma_1$ excited states and $j=9/2$ multiplet is
split to $\Gamma_6$ doublet and two $\Gamma_8$ quartets. 
The conduction electron band width $D$ is assumed to be $3eV$. 
All parameter values are listed in  Table~\ref{uvalue} in the unit of $D$.
The incoherent lattice sum is carried out over three  hundred shells 
with impurity 
configuration averaging.

Fig.~\ref{yupd3.fig} shows both the experimental and calculated  magnetic 
susceptibility and $^{89}$Y NMR Knight shift.
In our calculation, both the bulk magnetic susceptibility and the Knight shift
become constant when the temperature goes to zero and thus a Knight shift 
anomaly doesn't arise.
Our calculated  magnetic susceptibility saturates when the 
temperature goes to zero
and doesn't show the low temperature singularity like experiment. This may arise  
from the  numerical calculation  or from 
intrinsic properties such that real ground 
state may be magnetic as discussed earlier.
A separate possibility is that the weak admixture of excited $f^3$
magnetic states contributes a weakly log divergent contribution to
$\chi$; this possibility may be explored elsewhere.  
Regardless, 
the  calculated Knight shift agrees well with the experimental data in 
magnitude and temperature dependence.

\section{Summary }
\label{end.sec}

In this paper, we have calculated the magnetic susceptibility and
the Knight shift for the 
heavy electron materials within the infinite-$U$ single impurity Anderson model
using NCA method.

In our calculations we can explain that the Knight shift anomaly in 
heavy electron materials with the simplified single impurity Kondo effect. 
There exists a Knight shift anomaly at short distance $r<\xi_K$, with 
amplitude proportional to $T_0/\Gamma$.

Our calculations show generally good agreement with experimental results
in spite of the oversimplified
band structure. Especially, the short distance Knight shift depends on the
detailed structure on conduction electron band and our calculation shows 
large contributions  from the short distance Knight shift.
For  future work, we can include more 
realistic conduction electron band structure which can be calculated with
LMTO (linearized Muffin-Tin Orbital) method.

\acknowledgments
This research was supported by a grant from the U.S. Dept. of Energy, Office
of Basic Energy Science Division of  Material Research.  We thank 
H.R. Krishna-murthy, D.E. MacLaughlin, M. Steiner, and J.W. Wilkins for
useful conversations on this work and related topics.  

\appendix
\section{U$^{4+}$ Ions}
\label{u.app}

For the U compound ,  the crystal field effect (CEF),  
which  lift the angular momentum degeneracy of U  ions and 
their spin-orbit multiplet decomposes into irreducible representation of
the cubic field need to be included.
The distinction from the CeSn$_3$ and YbCuAl cases is that the apparent
crystal field splitting $\Delta_{CEF}\,\stackrel{ >}{\tiny \sim}\,T_0$
for CeSn$_3$ and YbCuAl.
The $f^2, j=4$ Hund's rule ground state of the U ion 
is split to a $\Gamma_3$ nonmagnetic
doublet, $\Gamma_5$ and $\Gamma_4$ magnetic triplet  and $\Gamma_1$
singlet states\cite{mooketal}. The $f^3, j=9/2$ spin-orbit multiplet is split
to $\Gamma_7$ doublet and two $\Gamma_8$ quartets.
For an explicit derivation, see Appendix~\ref{cef.app} and
the articles  by K.R. Lea,
M.J. Leask and W.P. Wolf~\cite{llw} and by M.T. Hutchings~\cite{hutch}.
The eigenstates of $\Gamma_i$ CEF states for $j=4$ and $j=9/2$ multiplets are in 
Table~\ref{table3} and  Table~\ref{table4}~\cite{llw,hutch}.
In  cubic symmetry, the coefficients of CEF states  depend upon 
the parameter $x$
 which is fixed by the ratio of the fourth  and sixth
degree terms in a  short distance  expansion of the cubic field in the 
Hamiltonian of the crystal electric field, and upon the 
parameter $W$ which is an overall  scale
factor fixed by the crystal field strength. 
In this calculation, we use $x_3=0.3693$ and $W_3= 2.746\times 10^{-4}$ to have 
$j=9/2$ $\Gamma_6$ for the ground state of the  $f^3$ configuration and
$x_2=-0.648$ and $W_2=-3.95\times10^{-4}$ to have $j=4$ $\Gamma_3$ for 
the ground
 state for $f^2$ configuration.  For further details, see Appendix~\ref{cef.app}.
The choice of the overall phase is arbitrary in defining the CEF eigenstates.

A brief  descriptions of the different irrep labels of the cubic group
is as follows:\\
(1) $\Gamma_1$ and $\Gamma_2$  are orbital singlets.\\
(2) $\Gamma_3$ is an orbital (non-magnetic, or non-Kramers') doublet and
usually labeled by $\pm$.  \\
(3) $\Gamma_4$ and $\Gamma_5$ are magnetic triplets and
labeled by $0, \pm 1$.\\
(4) $\Gamma_6$ and $\Gamma_7$ are magnetic Kramers' doublets and 
labeled with
pseudo-spin $\uparrow$ or $\downarrow$. That is, the $\Gamma_6$ and $\Gamma_7$
CEF states are similar to the $j=1/2$ angular momentum manifold.\\
(5) $\Gamma_8$ is a magnetic quartet ($\Gamma_8 = \Gamma_3\otimes \Gamma_7$).
and  labeled by $\pm \uparrow/\downarrow$.  

In  the Anderson model picture, the conduction electrons can hop on
and off the atomic orbitals at the impurity site. 
the $l=3$ conduction electron partial waves are most strongly coupled 
to the $f$ electrons in U ions (for the isotropic hybridization, only the $l=3$ 
components can hybridize with the $f$ orbitals.). 
In the presence of the spin-orbit coupling, the $l=3$ conduction electron state
splits into the  $j=5/2$, $7/2$ 
conduction electron states.   These $j$ multiplets further split into the
CEF irreducible representations, $\Gamma_6$, $\Gamma_7$ doublets and $\Gamma_8$
quartet in crystal environment. 
The CEF eigenstates for $j=5/2$, and $j=7/2$ 
multiplets are listed in Table~\ref{J2.5} and Table~\ref{J3.5}.

The hybridization  Hamiltonian of our model U compound in the absence of the CEF is
given by
\begin{eqnarray}
{\cal H}_{cf}& =& {\sum_{kj_cm_c}}\sum_{m_2m_3}[\;V_kc^{\dagger}_{kj_cm_c}\Lambda(j_cm_c;
f^2j_2=4m_2|f^3j_3=\frac{9}{2}m_3)\nonumber \\
& &|f^2j_2=4m_2\rangle \langle f^3j_3= \frac{9}{2} m_3|  + h.c.\:].
\end{eqnarray}
Here the reduced matrix elements are
\begin{eqnarray}
\lefteqn{\Lambda(j_cm_c;f^2j_2=4m_2|f^3j_3 = \frac{9}{2}m_3)} \nonumber \\
&\equiv& \langle f^2j_2=4m_2|
f_{j_cm_c}|f^3j_3=\frac{9}{2}m_3\rangle \nonumber \\
&=&K(j_cj_2|j_3)\langle j_cm_c;j_2m_2|j_3m_3\rangle
\end{eqnarray}
where $j_c=5/2,7/2$. The Wigner-Eckart theorem is used for 
the last line and the 
prefactor $K(j_cj_2|j_3)$ is the fractional parentage coefficient.
In our calculation where  $j_2=4$ and $j_3=9/2$,  
$K(j_cj_2|j_3)=\sqrt{7/9}$  for both $j_c=5/2$ and $j_c=7/2$. 
$\langle j_cm_c;j_2m_2|j_3m_3\rangle$  are
the Clebsch-Gordan coefficients. 
If there is a crystal electric field, the Anderson hybridization between the 
CEF states $\Gamma$s  needs to be re-evaluated in the basis appropriate
to the crystal field split states. We have
\begin{eqnarray}
{\cal H}_{cf}&=&{\sum_{kj_c\Gamma_c\alpha_c}}\sum_{\Gamma\alpha\Gamma\beta}
[\;V_kc^{\dagger}_{kj_c\Gamma_c\alpha_c}\Lambda(j_c\Gamma_c\alpha_c;
4\Gamma\alpha|\frac{9}{2}\Gamma\beta)\nonumber \\
 & &\;\;|f^24\Gamma\alpha\rangle \langle f^3\frac{9}
{2}\Gamma\beta| + h.c.\:]\nonumber \\
&=&{\sum_{kj_c\Gamma_c\alpha_c}}\sum_{\Gamma\alpha\Gamma\beta}[\;V_k
K(j_c;4|\frac{9}{2})\langle j_c\Gamma_c\alpha_c;4\Gamma\alpha|\frac{9}{2}
\Gamma\beta\rangle\nonumber \\
 & &\;\;c^{\dagger}_{kj_c\Gamma_c\alpha_c}|f^24\Gamma\alpha\rangle \langle
f^3\frac{9}{2}\Gamma\beta| + h.c.\:].
\end{eqnarray}
The reduced matrix $\Lambda(j_c\Gamma_c\alpha_c;
4\Gamma\alpha|\frac{9}{2}\Gamma\beta)$ is implicitly defined in above equation.
All the possible selection rules for the hybridization are listed in 
Table~\ref{table8}.

\section{Crystal Electric Field  Effect}
\label{cef.app}
In the crystal lattice, magnetic ions feel  an electrostatic field  
produced by the neighboring ions. This crystal field lifts the
degeneracy of the angular momentum of the magnetic ions.
The most common method to calculate the effect of  the crystal electric field 
is the operator equivalent techniques~\cite{s52}
 which exploits the Wigner-Eckart theorem to replace the electrostatic
potential terms in the Hamiltonian by operators in the angular momentum
space of the ground multiplet.
It depends on the symmetry of  the crystal and the orbital angular momentum
$j$ of the magnetic electrons.
The most general Hamiltonian with cubic symmetry~\cite{llw,hutch}
 can be written as
\begin {equation}
{\cal H}_{cef} = B_4(O^0_4+5O^4_4)+ B_6(O^0_6-21O^4_6)
\end{equation}
where 
\begin{eqnarray}
O^0_4 &=& 35J_z^4-[30J(J+1)-25]J_z^2\nonumber\\
&& -6J(J+1)+3J^2(J+1)^2 \nonumber\\
O^4_4 &=& \frac{1}{2}(J_+^4+J_-^4) \nonumber \\
O^0_6 &=& 231J_z^6-105[3J(J+1)-7]J_z^4 -5J^3(J+1)^3  \nonumber \\
   &&+[105J^2(J+1)^2-525J(J+1)+294]J_z^2\nonumber \\
   & &      -5J^3(J+1)^3 + 40J^2(J+1)^2-60J(J+1) \nonumber \\
O^4_6 &=& \frac{1}{4}[11J_z^2-J(J+1)-38](J_+^4+J_-^4)\nonumber\\
&& + \frac{1}{4}(J_+^4+J_-^4)[11J_z^2-J(J+1)-38].
\end{eqnarray}

The coefficients $B_4$ and $B_6$ are the factors which determine the scale of the crystal 
field splittings. In a simple point charge model, 
they are linear functions of $<r^4>$ and $<r^6>$, the
mean fourth and sixth powers of the radii of the magnetic electrons, and thus
depend on the detailed nature of the magnetic ion wave functions.
We treat these as phenomenological  parameters because these are very difficult to
calculate quantitatively.

Following Ref~\cite{llw},
we rewrite the Hamiltonian as
\begin{equation}
{\cal H}_{cef} = B_4F(4)\frac{O_4}{F(4)} + B_6F(6)\frac{O_6}{F(6)},
\label{eq:cef}
\end{equation}
where\\
$O_4=[(O^0_4+5O^4_4]$, and $O_6=[O^0_6-21O^4_6]$ and $F(4)$ and $F(6)$ is the 
common factors to all the matrix elements of fourth and sixth degree terms.
In order to cover all the possible values of the ratio between the fourth and
sixth degree terms,
we put
\begin{eqnarray}
B_4F(4) &=& Wx \label{eq:b4}\\
B_6F(6) &=& W(1-|x|)\label{eq:b6}
\end{eqnarray}
where $-1<x<+1$.\\
It follows that 
\begin{equation}
\frac{B_4}{B_6} =\frac{x}{1-|x|}\frac{F(6)}{F(4)}\label{ratio},
\end{equation}
so that $B_4/B_6=0$ for $x=0$ and $B_4/B_6=\pm \infty$ for $x=\pm 1$.

Rewriting Eq. (\ref{eq:cef}) we have
\begin{equation}
{\cal H}_{cef} =W[x(\frac{0_4}{F(4)}) +(1-|x|)(\frac{O_6}{F(6)})].
\end{equation}
 Here, the eigenvalue is related the crystal electric energy by the
scale factor $W$.
 
In our calculations, we need the relations between parameters $x_2$ and $W_2$
for $f^2$ configuration and $x_3$ and $W_3$ for $f^3$ configuration of U ions.
To get $j=4$ $\Gamma_3$ and $j=9/2$ $\Gamma_6$ doublets as 
the ground states of $f^2$ and
$f^3$ configuration, we choose $x_2<0$ and $x_3 > 0$ and $W_2 <0 $.
$W_2$ is decided by the energy splitting between $\Gamma_3$ and $\Gamma_5$
states, $\Delta_{35}$.
\begin{equation}
W_2 = -\Delta_{35}/(84+114x_2)
\end{equation}
>From the  Eq. (\ref{eq:b4}), Eq. (\ref{eq:b6}) and Eq. (\ref{ratio}),
and factors which is given in  Table 1 of reference ~\cite{llw}, 
we get the relations of parameters between $f^2$ and $f^3$ configurations.
\begin{eqnarray}
-3.143x_3(1-|x_2|) & = & x_2(1-|x_3|) \\
W_3 & =& 0.3963 W_2\frac{x_2}{x_3}
\end{eqnarray}
>From the above equations we get  $x_3=0.3693$ and   $W_3= 2.746\times10^{-4}$
for $x_2=-0.648$ and $W_2=-3.95\times10^{-4}$ .
 
\section{Evaluation of Knight shift k Sum}
\label{k_sum.app}
In this appendix, the form of $I(\epsilon_k, \epsilon_{k'})$ expressed in
Eq. (\ref{eq:knight}) is derived. We define
\begin{eqnarray}
I_{jj'}(\epsilon_k, \epsilon_{k'})&  =& j_3(kr)j_3(k'r)\frac{1}{\beta^2}
 \sum_{\omega,\nu} G_0(k\sigma,i\omega)G_0(k'\sigma',i\omega)\nonumber \\
 & & \times G_{jm}(i(\omega+\nu))
G_{j'm'}(i(\omega+\nu))D(i\nu) \label{eq:kk'sum}
\end{eqnarray}
where $\omega[\nu]=(2n+1)\pi i/\beta[2n\pi i/\beta]$ and the usual 
fermion(boson) Matsubara frequencies.
 $G_0$ is the bare conduction electron propagator, $G_{jm}$ is a pseudofermion 
 propagator for spin-orbit multiplet $j$, and $D$ is a pseudoboson propagator.
For the coherent calculation the bare conduction 
propagator is replaced by the dressed electron
propagator.  This coherence effect will be considered in Appendix \ref{cohe.app}.

Then  Eq. (\ref{eq:kk'sum}) can be  rewritten as
\begin{eqnarray}
I_{jj'}(\epsilon_k , \epsilon_{k'})& =& \frac{1}{Z_{4f}}
         j_3(kr)j_3(k'r)\frac{1}{\beta^2}
\sum_{\omega,\nu} \frac{1}{i\omega-\epsilon_k} \frac{1}{i\omega-\epsilon_{k'}}
\nonumber \\
 & &\times \int \frac{d\zeta}{\pi}\frac{B(\zeta)}{i\nu-\zeta}
\int\frac{d\rho}{\pi}\frac{A_{jm}(\rho)}{i(\omega+\nu)-\rho}\nonumber \\
& & \times     \int\frac{d\rho'}{\pi}\frac{A_{j'm'}(\rho')}{i(\omega+\nu)-\rho'}
      \label{eq:k_sum}.
\end{eqnarray}
 Let's first do the $\nu $ summation.
\begin{eqnarray}
 (A) & = & \frac{1}{\beta}\sum_{\nu}\frac{1}{i\nu-\zeta}\frac{1}{i(\omega+\nu)-\rho}
        \frac{1}{i(\omega+\nu)-\rho'} \nonumber \\
     & = & \frac{ 2f(\rho) -2N(\zeta)}{(\rho-\rho')(\rho-i\omega-\zeta)}
        \label{eq:nu_sum}
\end{eqnarray} 
where $f(\rho)[N(\zeta)]=(e^{\beta\rho}+1)^{-1}[(e^{\beta\zeta}-1)^{-1}]$.
 In this Eq. (\ref{eq:nu_sum}), only $(\rho - \zeta - i\omega)$ term is depend on $\omega$.
 Let's do the $\omega$ sum. We get
\begin{eqnarray}
 (B) & = &\frac{1}{\beta}\sum_{\omega}\frac{1}{i\omega-\epsilon_k}\frac{1}{i\omega-\epsilon_{k'}}
 \frac{1}{(\rho-i\omega-\zeta)}\nonumber \\
   & = & \frac{2f(\epsilon_k)-2f(\rho-\zeta)}{(\epsilon_k - \epsilon_{k'})
        (\epsilon_k+\zeta-\rho)}. \label{eq:w_sum}
\end{eqnarray} 
Now taking  the $\lambda \rightarrow -\infty$ projection
\begin{eqnarray}
 e^{-\beta\lambda}(N(\zeta)-f(\rho))
&  =&e^{-\beta\lambda}(\frac{1}{e^{\beta(\zeta-\lambda)}
-1} + \frac{1}{e^{\beta(\rho-\lambda)}+1})\nonumber \\
&= & e^{-\beta\zeta}+e^{-\beta\rho}.
\label{eq:projection}
\end{eqnarray}
Also  with the relations
\begin{equation}
\int\frac{d\rho'} {\pi}\frac{A_{j'm'}(\rho')}{\rho-\rho'} = ReG_{j'm'}(\rho),
\label{eq:regf}
\end{equation}
 and $f(\rho-\zeta)(e^{-\beta\zeta}+e^{-\beta\rho})= e^{-\beta\rho} $
\begin{eqnarray}
I_{jj'}(\epsilon_k,\epsilon_{k'})&=&\frac{4}{Z_{4f}}\frac{j_3(kr)j_3(k'r)}
{\epsilon_k- \epsilon_{k'}} \int\frac{d\zeta}{\pi} \int \frac{d\rho}{\pi}\nonumber\\
&&\times [\frac{(B^{(-)}(\zeta)A_{jm}(\rho)ReG_{j'm'}(\rho)f(\epsilon_k)}{\epsilon_k+\zeta-\rho}\nonumber \\
 & &+\frac{B(\zeta)A_{jm}^{(-)}(\rho)ReG_{j'm'}(\rho)(f(\epsilon_k)-1)}{\epsilon_k+\zeta-\rho}].
\end{eqnarray}
The above equation, can be rewritten 
\begin{eqnarray}
I_{jj'}(\epsilon_k,\epsilon_{k'})&=&2\frac{j_3(kr)j_3(k'r)}
{\epsilon_k- \epsilon_{k'}}\nonumber \\
& & \times (f(\epsilon_k)I_{1jj'}(\epsilon_k) + (1-f(\epsilon_k))
I_{2jj'}(\epsilon_k))
\end{eqnarray}
where
\begin{eqnarray}
I_{1jj'}(\epsilon_k) & = &\! \frac{2}{Z_{4f}}\!\int\!\frac{d\zeta}{\pi} \int\! 
\frac{d\rho}{\pi}\frac{(B^{(-)}
(\zeta)A_{jm}(\rho)ReG_{j'm'}(\rho)}{\epsilon_k+\zeta-\rho}\nonumber \\
& = & \!\frac{1}{Z_{4f}}\!\int\!\frac{d\zeta}{\pi}B^{(-)}\!(\zeta)
ReG_{jm}(\zeta+\epsilon_k)ReG_{j'm'}(\zeta+\epsilon_k)\nonumber\\
&& \\
I_{2jj'}(\epsilon_k) & = &\!\frac{-2}{Z_{4f}}\! \int\!\frac{d\zeta}{\pi}
\! \int\! \frac{d\rho}{\pi}
\frac{B(\zeta)A_{jm}^{(-)}(\rho)ReG_{j'm'}(\rho))}{\epsilon_k+\zeta-\rho}\nonumber\\
        & = &\! \frac{2}{Z_{4f}}\! \int \!\frac{d\rho}{\pi}A_{jm}^{(-)}(\rho)ReG_{j'm'}(\rho)
                         ReD(\rho-\epsilon_k).\nonumber \\
&& 
\end{eqnarray}

\section{Analytic Calculations of Inner $k'$ Integral in the Knight shift}
\label{j3int.app}

 We can calculate the inner $k'$ integral  analytically. This is given by
\begin{equation}
\int_{0}^{\infty}k'^2dk'\frac{j_3(k'r)}{k^2-k'^2}\label{eq:k'int}
\end{equation}
 where $j_n$ is the spherical Bessel function.
The above integration is broken into 4 terms which may be expressed in
terms of  Sine and Cosine integrals.\\
The first term  is
\begin{eqnarray}
\lefteqn{ \int_{0}^{\infty}dk'\frac{k'^2}{k^2-k'^2} \frac{\cos k'r}{k'r}} \nonumber \\
    & = & \frac{1}{r}[Ci(kr)\cos kr + si(kr)\sin kr+ \frac{\pi}{2}\sin kr] 
\label{eq:k'1}
\end{eqnarray}
  where the Sine  and Cosine integrals are 
\begin{eqnarray}
  Si(z) & = &\int_{0}^{z} \frac{\sin t}{t} dt\\
  si(z) & = & Si(z)- \frac{\pi}{2}\\
  Ci(z) & = & \gamma + \ln z + \int_{0}^{z} \frac{\cos t -1}{t} dt .
\end{eqnarray}
For more explicit formalism of Sine and Cosine integrals see the
Reference~\cite{as70}.
The second term in our integration is
\begin{eqnarray}
\lefteqn{ \int_{0}^{\infty}dk'\frac{k'^2}{k^2-k'^2} \frac{-6\sin k'r}{(k'r)^2}} \nonumber \\
    & = & \frac{-6}{kr^2}[Ci(kr)\sin kr -si(kr)\cos kr+\frac{\pi}{2}\cos kr].
\label{eq:k'2}
\end{eqnarray}
The third term is
\begin{eqnarray}
\lefteqn{ \int_{0}^{\infty}dk'\frac{k'^2}{k^2-k'^2} \frac{-15\cos k'r}{(k'r)^3}}
 \nonumber \\ 
    & = & \frac{-15}{k^2r^3}[-Ci(0)+Ci(kr)\cos kr +si(kr)\sin kr\nonumber\\
 & & + \frac{\pi}{2}\sin kr] .  \label{eq:k'3}
\end{eqnarray}
$Ci(0)$ is divergent logarithmically, but is  exactly canceled by a term
 from next integral.\\
Our fourth term in the integral is given by
\begin{eqnarray}
\lefteqn{ \int_{0}^{\infty}dk'\frac{k'^2}{k^2-k'^2} \frac{15\sin k'r}{(k'r)^4}}
 \nonumber \\
   & = &  \frac{15}{k^3r^4}[Ci(kr)\sin kr -si(kr)\cos kr+\frac{\pi}{2}\cos kr]\nonumber \\
& & + \frac{15}{k^2r^3}[1-Ci(0)].
\label{eq:k'4}
\end{eqnarray}
 Note the explicit cancellation of $Ci(0)$ between Eq. (\ref{eq:k'3}) and 
Eq. (\ref{eq:k'4}). 
Putting Eq.(\ref{eq:k'1})-Eq.(\ref{eq:k'4}) together, 
we have our inner momentum integral in our
Knight shift expression define in Eq. (\ref{eq:k'int}) given by 
\begin{eqnarray}
\lefteqn{\int_{0}^{\infty}k'^2dk'\frac{j_3(k'r)}{k^2-k'^2}}\nonumber \\
& =& \frac{1}{r}[Ci(kr)\cos kr + si(kr)\sin kr+ \frac{\pi}{2}\sin kr] \nonumber\\
  & &- \frac{6}{kr^2}[Ci(kr)\sin kr -si(kr)\cos kr+\frac{\pi}{2}\cos kr]\nonumber\\
  & &- \frac{15}{k^2r^3}[Ci(kr)\cos kr +si(kr)\sin kr+\frac{\pi}{2}\sin kr -1]\nonumber\\ 
 &  &+  \frac{15}{k^3r^4}[Ci(kr)\sin kr -si(kr)\cos kr+\frac{\pi}{2}\cos kr] .
\end{eqnarray}

\section{Angular Dependence of the Knight Shift}
\label{angle.app}
In this appendix the derivation of $f_{jj'}(\alpha,\theta)$, the angular 
dependence of the Knight shift, will be discussed. 
 
First,  we need to  calculate the expectation value of the magnetic
moment operator in
the $z$ direction 
\begin{equation}
 \langle jm_j|J_z+S_z|j'm_{j'}\rangle =m_j\delta_{jj'}+ 
\langle jm_j|S_z|j'm_{j'}\rangle.
\end{equation}
 Then for $ j=j'$, by Wigner-Eckart theorem gives
\begin{equation}
    \langle jm_j|J_z+S_z|jm_{j'}\rangle=\delta_{m_jm_{j'}}g_jm_j.
\end{equation}
 where $g_j$ is the Land\'{e} $g$-factor and $g_{5/2}=6/7$ and $g_{7/2}=8/7$
for $l=3$ and $s=1/2$ which is the case for $4f^1$ and $4f^{13}$ configuration.
And for $j \neq j'$ 
\begin{eqnarray}
\lefteqn{ \langle jm_j|J_z+S_z|j'm_{j'}\rangle  =  \langle jm_j|S_z|j'm_{j'}\rangle} \nonumber \\
     & = & \delta_{m_jm_{j'}} \sum_{m_l\alpha} \alpha \langle jm_j|lm_l;s\alpha\rangle 
        \langle lm_l;s\alpha|j'm_{j'}\rangle 
\end{eqnarray}
where $\alpha=\pm 1/2$.
  For $l=3$, $s=1/2$,  $j=5/2$ and $j'=7/2$ 
\begin{eqnarray}
   \langle \frac{5}{2} m_j|S_z|\frac{7}{2} m_{j'}\rangle& = &\delta_{m_jm_{j'}}
\! \sum_{m_3\alpha} \alpha \langle \frac{5}{2}m_j|3m_3;s\alpha\rangle
   \langle 3m_3;s\alpha|\frac{7}{2}m_j\rangle \nonumber\\
                     & = & -\frac{\sqrt{49-4m_j^2}}{14}\delta_{m_jm_{j'}}.\!\!
\end{eqnarray}
given
\begin{eqnarray}
  \langle j_1 m-\alpha; \frac{1}{2}\alpha|j,m\rangle & = & (\frac{j_1\pm m/2}{2j_1+1})^{1/2} \; 
   \mbox{for}\; j=j_1+1/2 \\
            & = & \mp(\frac{j_1\mp m/2}{2j_1+1})^{1/2} \; \mbox{for}\;
              j=j_1-1/2  
\end{eqnarray}
where 
\begin{eqnarray}
  \langle 3m_3;\frac{1}{2}\pm\frac{1}{2}|\frac{7}{2}m\rangle & =&
(\frac{7\pm2m}{14})^{1/2}\; \mbox{for}\; j=7/2 \label{eq:we1}\\
  \langle 3m_3;\frac{1}{2}\pm\frac{1}{2}|\frac{5}{2}m\rangle&=&
\mp(\frac{7\mp2m}{14})^{1/2} \;\mbox{for}\; j=5/2 \label{eq:we2}.
\end{eqnarray}
Then
\begin{eqnarray}
 \langle  \frac{5}{2} m_j|S_z|\frac{7}{2} m_{j'} \rangle & =  & 
   -\frac{\sqrt{12}}{7} \;\mbox{for}\; m_j=\pm 1/2\nonumber\\
     & =  &    -\frac{\sqrt{10}}{7} \;\mbox{for}\; m_j=\pm 3/2 \nonumber\\
     & =  &    -\frac{\sqrt{6}}{7} \;\;\mbox{for}\; m_j=\pm 5/2
\end{eqnarray}

In the Knight shift calculation the angular dependence comes from the Zeeman
term in the Hamiltonian.
\begin{equation}
{\cal H}_z = -\mu_BH_z(L_z+2S_z)=-\mu_B H_z(J_z+S_z),
\end{equation}
where external magnetic field $\vec H = H_z \hat{z} $.

Now let's calculate the angular part of the Knight shift. 
To do the lattice sum, we have to consider the difference of  the field axis and
 the bond
 direction  $\vec r$ which connects the  nucleus or muon to a given $f$ ion.
Let the angle between the field axis and $\vec r$
 is $\alpha$,  and the angle between the $z$ axis and bond axis is $\theta$.
First when the field is along the $z$ direction, {\it i.e.} $\theta=\alpha$ and
nuclear spin $I=1/2$, the angular
momentum operator $J_z$, which is quantized in the bond direction 
  becomes
$\cos\theta J_z -\sin\theta J_x=\cos\theta J_z -\sin\theta(J_+ + J_-)/2$ in the
new reference frame when the material has the cubic symmetry.
Also, the nuclear spin operator $\sigma_z$ becomes $\cos\theta\sigma
_z
-\sin\theta\sigma_x=\cos\theta\sigma_z-\sin\theta(\sigma_+ +\sigma_-)/2$.
Then the surviving terms in the Knight shift calculation are
\begin{eqnarray}
\lefteqn{\cos^2\theta\sigma_z(J_z+S_z) +\sin^2\theta\sigma_x(J_x+S_x) =}
\nonumber \\
& & \cos^2\theta\sigma_z(J_z+S_z)+\frac{\sin^2\theta}{4}(\sigma_+(J_-+S_-)
+\sigma_-(J_+S_+)).\nonumber \\
&&
\label{eq:ftheta}
\end{eqnarray}
Then the total angular part is
\begin{equation}
f_{jj'}(\theta) =\cos^2 \theta f_{jj'}^z+\frac{\sin^2\theta }{4} (f^{1-}_{jj'}
+f^{2-}_{jj'} +f^{1+}_{jj'} + f^{2+}_{jj'}).
\end{equation}
Where $f_{jj'}^z$ is 
\begin{eqnarray}
f^z_{jj'}(\theta) &=& \sum_{m_jm_{j'}\alpha} \langle jm_j|J_Z+s_Z|j'm_{j'}\rangle\sigma^z_{\alpha \alpha}
 \langle j'm_{j'}|\hat{r}\rangle \langle \hat{r}|jm_j\rangle \nonumber \\
       &=& \sum_{m_j\alpha m_3} \alpha  \langle jm_j|J_z+S_z|j'm_{j'}\rangle
      \langle j'm_{j'}|3m_3; \frac{1}{2}\alpha\rangle \nonumber \\
    & &\times \langle 3m_3;\frac{1}{2}\alpha|jm_j\rangle|Y_{3m_3}(\hat{r})|^2
\end{eqnarray}
 where $\alpha=\pm 1/2$ and $Y_{lm}(\hat{r})$ is the spherical harmonics.

Then for
1)$ j =j'$ 
\begin{eqnarray}
f_{jj}^z(\theta)& = &\sum_{m_jm'_j\alpha m_3} \alpha  \langle jm_j|J_z+S_z|jm'_j\rangle \langle jm'_j|3m_3;
\frac{1}{2}\alpha\rangle \nonumber\\
 &&\times \langle 3m_3;\frac{1}{2}\alpha|jm_j\rangle|Y_{3m_3}(\hat{r})|^2 \nonumber\\
& = &\sum_{m_j\alpha}g_jm_j \alpha | \langle jm_j|3m_j-\alpha;\frac{1}{2}\alpha
\rangle|^2 Y_{3m_j-\alpha}(\hat{r})|^2  \nonumber \\
 &&
\end{eqnarray}
With the above equation and Eq. (\ref{eq:we1}) and Eq. (\ref{eq:we2})
\begin{eqnarray}
f_{\frac{5}{2}\frac{5}{2}}^z(\theta) &=&\frac{9}{28\pi}(1-4\sin^2 \theta) \\
f_{\frac{7}{2}\frac{7}{2}}^z(\theta) &=& \frac{4}{7\pi}(1+3\sin^2 \theta) .
\end{eqnarray}
2) $j \neq j'$, {\it i.e.} $j=5/2$ and $j'=7/2$

This term gives the van Vleck Knight shift 
\begin{eqnarray}
f_{\frac{5}{2}\frac{7}{2}}^z(\theta) &=& \sum_{m_j\alpha m_3} \alpha  \langle jm_j|J_z+S_z|j'm_{j'}\rangle
 \langle j'm_{j'}|3m_3; \frac{1}{2}\alpha\rangle \nonumber\\
&&\times \langle 3m_3;\frac{1}{2}\alpha|jm_j\rangle
    |Y_{3m_3}(\hat{r})|^2 \nonumber\\
 &=& \frac{3}{14\pi}(2-\sin^2 \theta).
\end{eqnarray}

Now let's derive  $f^{i\pm}_{jj'}$, the angular part of Knight shift from the 
second term of Eq. (\ref{eq:ftheta}), where $j=j'$
\begin{eqnarray}
f^{1\pm}_{jj'}(\theta) &=&\! \sum_{m_jm_{j'}\alpha\beta}\!\!\langle j,m_j|J_{\pm}|j',
m_{j'}\rangle \langle j'm_{j'} |\hat{r}\rangle \sigma^{\mp}_{\alpha\beta}
\langle \hat{r}| j,m_j \rangle \nonumber \\
&=&\!\! \sum_{m_jm_{j'}m_3 m'_3\alpha\beta}\!\!\!\!\langle j,m_j|J_{\pm}|j',m_{j'}
\rangle  \langle j'm_{j'}|3m'_3; \frac{1}{2}\alpha\rangle \nonumber \\
 & & \times\langle 3m'_3; \frac{1}
{2}\alpha|\sigma^{\mp}_{\alpha,\beta}|3m_3;\frac{1}{2}\beta \rangle
\langle 3m_3;\frac{1}{2}\beta |jm_j\rangle \nonumber \\
&& \times Y^*_{3m_3}(\hat{r})Y_{3m'_3}(\hat{r})
\end{eqnarray}
with $\langle j,m_j|J_{\pm}|j',m_{j'}\rangle =\sqrt{(j\pm m_j)(j\mp m_j+1)}
\delta_{jj'}\delta_{m_{j'}m_j\mp1} $  
and $\langle 3m'_3; \frac{1} {2}\alpha|\sigma^{\mp}_{\alpha,\beta}
|3m_3;\frac{1}{2}\beta \rangle = \sqrt{\frac{3}{4}-\beta(\beta\mp 1)}
\delta_{\alpha \beta\mp 1} \delta{m_3m'_3}$ 
$=\delta_{m_3m'_3}\delta_{\alpha,-\beta}$. 

 The above equation gives
\begin{eqnarray}
f^{1\pm}_{jj}(\theta)&=&\sum_{m_j} \sqrt{(j\pm m_j)(j\mp m_j\!\mp\! 1)} \langle jm_j\!\mp\!1|3m'_3
; \frac{1}{2}\mp\!\frac{1}{2}\rangle\nonumber \\
&&\times\langle 3m_3:\frac{1}{2}\pm\!\frac{1}{2}|jm_j
\rangle
|Y_{3m_j\mp \frac{1}{2}}(\hat{r})|^2.
\end{eqnarray}

Hence
\begin{eqnarray}
f^{2\pm}_{jj'}(\theta) &=& \sum_{m_jm_{j'}\alpha\beta}\langle jm_j|S_{\pm}|j',
m_{j'}\rangle \langle j'm_{j'} |\hat{r}\rangle \sigma^{\mp}_{\alpha\beta}
\langle \hat{r}| j,m_j \rangle \nonumber \\
&=&\! \sum_{m_jm_{j'}m_3 m'_3\alpha\beta}\!\!\langle j,m_j|S_{\pm}|j',m_{j'}
\rangle  \langle j'm_{j'}|3m'_3; \frac{1}{2}\alpha\rangle \nonumber \\
 & & \times\langle 3m'_3; \frac{1}
{2}\alpha|\sigma^{\mp}_{\alpha,\beta}|3m_3;\frac{1}{2}\beta \rangle
\langle 3m_3;\frac{1}{2}\beta |jm_j\rangle \nonumber \\
&& \times Y^*_{3m_3}(\hat{r})Y_{3m'_3}(\hat{r})
\end{eqnarray}
where
\begin{eqnarray}
\langle j m_j|S_{\pm}|jm'_j\rangle &=&\! \sum_{m_3m'_3\alpha\beta}\!\langle jm_j|
3m_3;\frac{1}{2}\alpha\rangle
\langle 3m'_3;\frac{1}{2}\beta|jm'_j\rangle \nonumber \\
&&\;\;\times \langle 3m_3;\frac{1}{2}\alpha |S_{\pm}|3m'_3;\frac
{1}{2}\beta\rangle\nonumber\\
 \mbox{  for  } j=5/2;& =& -\frac{\sqrt{(7\mp 2m)(5\pm 2m)}}{14}\delta_{m'_jm_j\mp1}
  \nonumber \\
 \mbox{  for  } j=7/2;& =& \frac{\sqrt{(7 \pm2m)(9 \mp 2m)}}{14}\delta_{m'_jm_j\mp1}.
  \nonumber \\
&&
\end{eqnarray}
Then
\begin{eqnarray}
f^{2\pm}_{jj}(\theta)&=&\sum_{m_j}\langle j m_j|S_{\pm}|jm_j\mp 1\rangle\langle 
jm_j\mp1| 3m'_3 ; \frac{1}{2}\mp\frac{1}{2}\rangle\nonumber\\
&&\;\;\times \langle 3m_3:\frac{1}{2}\pm
\frac{1}{2}|jm_j \rangle |Y_{3m_j\mp \frac{1}{2}}(\hat{r})|^2\nonumber \\
&=& |\langle jm_j\mp1| 3m'_3 ; \frac{1}{2}\mp\frac{1}{2}\rangle|^2
|\langle 3m_3:\frac{1}{2}\pm\frac{1}{2}|jm_j \rangle|^2\nonumber\\
&&\times |Y_{3m_j\mp \frac{1}{2}}(\hat{r})|^2.
\end{eqnarray}

1) contribution from $j=5/2$. Defines
\begin{eqnarray}
f^{\pm}_{\frac{5}{2}\frac{5}{2}}(\theta)&=&f^{1\pm}_{\frac{5}{2}\frac{5}{2}}
+f^{2\pm}_{\frac{5}{2}\frac{5}{2}}\nonumber \\
&=& \sum_{m_j} \sqrt{(\frac{5}{2}\pm m_j)(\frac{5}{2}\mp m_j+1)}
( \mp)\sqrt{\frac{7\mp 2m_j} {14}}\nonumber\\
&&\;\;\times (\pm)\sqrt{\frac{7\pm 2(m_j\mp1)}{14}}
|Y_{3m_j\mp\frac{1}{2}}(\hat{r})|^2\nonumber \\
&&+ \sum_{m_j}\frac{7\pm 2(m_j\mp1)} {14}\frac{7\mp 2m_j}{14}
|Y_{3m_j\mp\frac{1}{2}}(\hat{r})|^2\nonumber \\
&=&-\frac{9}{14\pi}[1+2\cos^2\theta]= g_{j=\frac{5}{2}}f^{1\pm}_{\frac{5}{2}\frac{5}{2}}
\end{eqnarray}
Then
\begin{equation}
f_{\frac{5}{2}\frac{5}{2}}(\theta) 
=\frac{9}{28\pi}(1-8\sin^2\theta +6\sin^4\theta)
\end{equation}

2)Contribution from $j=7/2$. Define
\begin{eqnarray}
f^{\pm}_{\frac{7}{2}\frac{7}{2}}(\theta)&=&f^{1\pm}_{\frac{7}{2}\frac{7}{2}}
+f^{2\pm}_{\frac{7}{2}\frac{7}{2}}\nonumber \\
&=&\frac{8}{14\pi}[5+3\cos^2\theta]=g_{j=\frac{7}{2}}f^{1\pm}_{\frac{7}{2}\frac{7}{2}}
\end{eqnarray}
Then
\begin{eqnarray}
f_{\frac{7}{2}\frac{7}{2}}(\theta) 
&=& \frac{2}{7\pi}[2+12\sin^2\theta-9\sin^4\theta]
\end{eqnarray}
$3)$ Contribution from Van-Vleck terms. Define
\begin{eqnarray}
f^{\pm}_{\frac{5}{2}\frac{7}{2}}(\theta) &=&\! \sum_{m_jm_{j'}\alpha\beta}\!\!\langle j,m_j|S_{\pm}|j'
m_{j'}|\rangle \langle j'm_{j'} |\hat{r}\rangle \sigma^{\mp}_{\alpha,\beta}
\langle \hat{r}| j,m_j \rangle \nonumber \\
&=&\!\! \sum_{m_jm_{j'}m_3 m'_3\alpha\beta}\!\!\!\!\langle j,m_j|S_{\pm}|j',m_{j'}
\rangle  \langle j'm_{j'}|3m'_3; \frac{1}{2}\alpha\rangle\nonumber \\
  & &\times \langle 3m'_3; \frac{1}{2}\alpha|\sigma^{\mp}_{\alpha,\beta}
|3m_3;\frac{1}{2}\beta \rangle
\langle 3m_3:\frac{1}{2}\beta |jm_j\rangle \nonumber\\
 &&\times Y^*_{3m_3}(\hat{r})Y_{3m'_3}(\hat{r})
\label{eq:vvf1}
\end{eqnarray}
with 
\begin{eqnarray}
\lefteqn{\langle j=\frac{5}{2},m_j|S_{\pm}|j'=\frac{7}{2} ,m_{j'}\rangle   }\nonumber \\
&=&\sum_{m_3m'_3\alpha\beta}\langle\frac{5}{2}m'|3m_3;\frac{1}{2}\alpha\rangle 
 \langle 3m_3;\frac{1}{2}\alpha |S_{\pm}|3m'_3;\frac{1}{2}\beta\rangle\nonumber \\
 && \; \times \langle 3m'_3;\frac{1}{2}\beta|\frac{7}{2} m\rangle \nonumber \\
&=& \mp\frac{1}{14}\sqrt{(5\mp 2m)(7\mp 2m)}
\label{eq:fvv2}
\end{eqnarray}
where $m'=m\pm 1$.

Insert in, the   Eq. (\ref{eq:fvv2}) into the Eq. (\ref{eq:vvf1}) gives
\begin{eqnarray}
f^{\pm}_{\frac{5}{2}\frac{7}{2}}(\theta) &=& \sum_m \mp\frac{\sqrt{(7+2m)(5+2m)}}{14}
\langle\frac{7}{2}
m|3m\pm\frac{1}{2};\frac{1}{2}\mp\frac{1}{2}\rangle \nonumber \\
 & & \times \langle 3 m\pm\frac{1}{2}; \frac{1}{2}\pm\frac{1}{2} |\frac{5}{2} 
m\pm 1\rangle |Y_{3m\pm\frac{1}{2}} (\hat{r})|^2 \nonumber \\
&=& \frac{3}{14\pi}[2+\sin^2\theta]
\end{eqnarray}
Then
\begin{equation}
f_{\frac{5}{2}\frac{7}{2}}(\theta) =\frac{3}{28\pi}[4-4\sin^2\theta+3\sin^4\theta]
\end{equation}
For  cubic crystal, CeSn$_3$, 
we fix the field in the $z$ direction and do the reference frame averaging
over the cases of the 
Sn atom  in the $xz$ plane or the $xy$ plane. Note that $yz$ plane
is equivalent to the $xz$ plane.

For the YbCuAl case which has the hexagonal symmetry, we have to consider three
possible field directions, along the $x$, $y$ and $z$ axis.
Then $J_z=\cos\alpha J_z-\sin\alpha J_x$ and $\sigma_z=A\sigma_z-B\sigma_x$. 
where 
\begin{eqnarray}
A &=& \frac{1}{4}(25\cos^5\theta-26\cos^3\theta+5\cos\theta)\nonumber \\
B &=& \frac{1}{4}\sin\theta(25\cos^4\theta-14\cos^2\theta+1)
\end{eqnarray}
 which is derived below.
 The explicit form of the rotation matrices is given by 
\begin{eqnarray}
\lefteqn{d_{m'm}^{(j)}(\theta) =} \nonumber \\
&& \sum_k \frac{(-1)^{k-m+m'}\sqrt{(j+m)!(j-m)!(j+m')!
(j-m')!}}{(j+m-k)!(j-k-m')!(k-m+m')!k!}\nonumber\\
&&\;\;\times (\cos \frac{\theta}{2})^{2j-2k+m-m'}
(\sin \frac{\theta}{2})^{2k-m+m'}
\end{eqnarray}
where we take the sum over $k$ whenever none of the arguments of factorials in 
the denominator are negative.
Al has an $I=5/2$ nuclear spin  and the NMR shift was obtained from derivative 
spectra
 of the central ($ 1/2 \leftrightarrow -1/2$) NMR transition.
\begin{eqnarray}
\lefteqn{d_{\frac{1}{2}\frac{1}{2}}^{(\frac{5}{2})}(\theta) 
= d_{-\frac{1}{2}-\frac{1}{2}}^{(\frac{5}{2})}(\theta)} \nonumber\\
&=& \sum_k\frac{(-1)^k\,3!\,2!}{(3-k)!(2-k)!k!k!}(\cos \frac{\theta}{2})^{5-2k}
(\sin \frac{\theta}{2})^{2k}\nonumber \\
 &=& \frac{1}{2}\cos \frac{\theta}{2}[5\cos^2\theta -2\cos\theta -1] =\alpha\\
\lefteqn{d_{\frac{1}{2}-\frac{1}{2}}^{(\frac{5}{2})}(\theta) 
= -d_{-\frac{1}{2}\frac{1}{2}}^{(\frac{5}{2})}(\theta)}\nonumber\\
&=& \sum_k\frac{(-1)^{k+1}\,2!\,3!}{(2-k)!(2-k)!(k+1)!k!}(\cos \frac{\theta}{2})^
{4-2k} (\sin \frac{\theta}{2})^{2k+1}\nonumber \\
 &=&-\frac{1}{2} \sin \frac{\theta}{2}[5\cos^2\theta+2\cos \theta-1]=-\beta
\end{eqnarray}
Then $A$ and $B$ are defined as
\begin{eqnarray}
A&=& \alpha^2-\beta^2 = \frac{1}{4}(25\cos^5\theta-26\cos^3\theta+5\cos\theta)\nonumber \\
B&=&2\alpha\beta = \frac{1}{4}\sin\theta(25\cos^4\theta-14\cos^2\theta+1)
\end{eqnarray}
Then
\begin{equation}
f_{jj'}(\alpha,\theta)= A\cos\alpha f^z_{jj'}(\theta)+\frac{B}{4}\sin\alpha
[f^+_{jj'}(\theta)+f^-_{jj'}(\theta)].
\end{equation}
The explicit values for $j=5/2$ and $j=7/2$ are given by 
\begin{eqnarray}
f_{\frac{5}{2}\frac{5}{2}}(\alpha,\theta)
&=&\frac{9}{28\pi}[A\cos\alpha(1-4\sin^2\theta)\nonumber\\
 && -B\sin\alpha (3-2\sin^2\theta)]\\
f_{\frac{7}{2}\frac{7}{2}}(\alpha,\theta)
&=&\frac{2}{7\pi}[2A\cos\alpha(1+3\sin^2\theta)\nonumber\\
&& +B\sin\alpha(5+3\cos^2\theta)].
\end{eqnarray}
For YbCuAl,  we don't include the  van Vleck contribution because of
the large spin orbit splitting as discussed in section ~\ref{ybcual.ssec}.

\section{Inclusion of Coherence}
\label{cohe.app}
When multiple scattering  is accounting for, the conduction electron Green's function in the Eq. (\ref{eq:kk'sum}) become 
dressed by self energy corrections which account for multiple scattering off
at $4f$ sites. As a results, $I(\epsilon_k , \epsilon_{k'})$ in  
Eq.~\ref{eq:k_sum} is generalized to 
\begin{eqnarray}
I_{jj'}(\epsilon_k , \epsilon_{k'}) &=& \frac{1}{Z_{4f}}
                  j_3(kr)j_3(k'r)\frac{1}{\beta^2}
\sum_{\omega,\nu}\int \frac{d\xi}{\pi} \frac{A_c(\epsilon_k,\xi)}{i\omega-\xi}\nonumber\\
&&\times \int \frac{d\xi'}{\pi} \frac{A_c(\epsilon_{k'},\xi')}{i\omega-\epsilon_{k'}}
 \int \frac{d\zeta}{\pi}\frac{B(\zeta)} {i\nu-\zeta}\nonumber \\
& &\times    \int\frac{d\rho}{\pi}\frac{A_{jm}(\rho)}{i(\omega+\nu)-\rho}
     \int\frac{d\rho'}{\pi}\frac{A_{j'm'}(\rho')}{i(\omega+\nu)-\rho'}.\nonumber \\
&& 
      \label{eq:cohe_sum}
\end{eqnarray}
We can define the conduction electron self energy $\Sigma_c$ by
\begin{equation}
G_c(k,\xi) = \frac{1}{\xi-\epsilon_k-\Sigma_c(\xi)}.
\end{equation}
Then
\begin{eqnarray}
A_c(\epsilon_k,\xi) &=& -\mbox{Im}G_c(k,\xi)\nonumber \\
&=& -\mbox{Im}\frac{1}{\xi-\epsilon_k-\Sigma_c(\xi)}.
\end{eqnarray}
The conduction electron self energy is  calculated using the average-T matrix
(ATA) approximation. \\
The $\nu $ summation is the same as Eq. (\ref{eq:nu_sum}).
 Let's do the $\omega$ sum,
\begin{eqnarray}
 (B) & = &\frac{1}{\beta}\sum_{\omega}\frac{1}{i\omega-\xi}\frac{1}{i\omega-\xi}
 \frac{1}{(\rho-i\omega-\zeta)}\nonumber \\
   & = & \frac{2f(\xi)-2f(\rho-\zeta)}{(\xi - \xi')
        (\xi+\zeta-\rho)}. \label{eq:w_sum2}
\end{eqnarray} 
With Eq.(\ref{eq:w_sum2}), the $ I_{jj'}(\epsilon_k , \epsilon_{k'})$ is written
\begin{eqnarray}
I_{jj'}(\epsilon_k , \epsilon_{k'})& =&j_3(kr)j_3(k'r)
\int\frac{d\xi}{\pi}A_c(\epsilon_k,\xi)\int \frac{d\xi'}{\pi}A_c
(\epsilon_{k'}, \xi') \nonumber\\
&&\!\!\times \int \frac{d\zeta}{\pi} B(\zeta)
 \int\frac{d\rho}{\pi}A_{jm}(\rho)
\int\frac{d\rho'} {\pi}A_{j'm'}(\rho')\nonumber\\
&&\!\!\times     \frac{( 2f(\rho) -2N(\zeta))(2f(\xi)-2f(\rho-\zeta))}
       {(\xi - \xi')(\rho-\rho')(\xi+\zeta-\rho)}.
\label{ew:cohe_sum3}
\end{eqnarray} 
In the above equation the $\xi'$ integration is given by
\begin{equation}
\int\frac{d\xi'}{\pi} \frac{A_c(\epsilon_{k'},\xi')}{\xi-\xi'} = \mbox{Re}
G_c(k'\sigma',\xi). \label{eq:regc}
\end{equation}
Putting,  Eq. (\ref{eq:regc}), Eq. (\ref{eq:projection}) and Eq. (\ref{eq:regf})
together 
 with $f(\rho-\zeta)(e^{-\beta\zeta}+e^{-\beta\rho})= e^{-\beta\rho} $,
$I_{jj'}(\epsilon_k , \epsilon_{k'})$ becomes
\begin{eqnarray}
I_{jj'}(\epsilon_k,\epsilon_{k'})&=&j_3(kr)j_3(k'r)\frac{4}{Z_{4f}}
\int\!\frac{d\xi}{\pi}A_c(\epsilon_k,\xi)\mbox{Re}G_c(k'\sigma',\xi)\nonumber\\
&&\!\!\!\!\times \int\!\frac{d\zeta}{\pi} \int\! \frac{d\rho}{\pi}
 [\frac{(B^{(-)}(\zeta)A_{jm}(\rho)\mbox{Re}G_{jm'}(\rho)f(\xi)}{\xi+\zeta-\rho}\nonumber\\
&&+\frac{B(\zeta)A_{jm}^{(-)}(\rho)\mbox{Re}G_{j'm'}(\rho)(f(\xi)-1)}
{\xi+\zeta-\rho}].
\end{eqnarray}
Where $J(\xi)$ is defined by
\begin{equation}
J(\xi) = \int dk\frac{k^2j_3(kr)}{\xi-\epsilon_k- \Sigma_c(\xi)}.
\end{equation}

 We can write 
\begin{eqnarray}
\lefteqn{ \int k^2dk\int k'^2dk'I_{jj'}(\epsilon_k,\epsilon_{k'})=}\nonumber\\
&&- \int\!\frac{d\xi}{\pi}\mbox{Im}J(\xi)\mbox{Re}J(\xi)
[f(\xi)I_{1jj'}(\xi) + (1-f(\xi))I_{2jj'}(\xi)],\nonumber\\
&&
\end{eqnarray}
where
\begin{eqnarray}
I_{1jj'}(\xi) & = & \frac{2}{Z_{4f}}\!\int\!\frac{d\zeta}{\pi} \int\! \frac{d\rho}{\pi}
\frac{(B^{(-)}(\zeta)A_{jm}(\rho)\mbox{Re}G_{j'm'}(\rho)}{\xi+\zeta-\rho}\nonumber \\
& = & \frac{1}{Z_{4f}}\!\int\!\frac{d\zeta}{\pi}B^{(-)}(\zeta)
\mbox{Re}G_{jm}(\zeta+\xi)\mbox{Re}G_{j'm'}(\zeta+\xi)\nonumber\\
&&\\
I_{2jj'}(\xi) & = &\frac{-2}{Z_{4f}}\! \int\!\frac{d\zeta}{\pi} \int\! \frac{d\rho}{\pi}
\frac{B(\zeta)A_{jm}^{(-)}(\rho)\mbox{Re}G_{j'm'}(\rho))}{\xi+\zeta-\rho}\nonumber\\
        & = & \frac{2}{Z_{4f}}\!\int\! \frac{d\rho}{\pi}A_{jm}^{(-)}(\rho)
\mbox{Re}G_{j'm'}(\rho)
                         \mbox{Re}D(\rho-\xi)\nonumber \\
&&
\end{eqnarray}

\begin{figure}
\vspace*{7pt}          
\begin{center}         
\leavevmode            
\epsfxsize=3.0in
\epsfbox{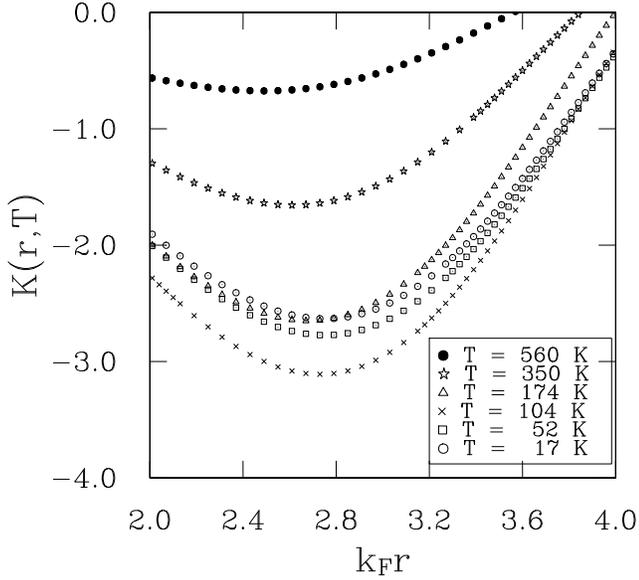}
\end{center}
\caption
{ Knight shift $K( r, T)$ at fixed angle $\theta =\pi/2$ and Kondo
temperature $T_0=430K$ for Ce ions on a fine scale. 
We use dimensionless
variable $k_Fr$ (with  the  Fermi wave vector $k_F=0.65\AA^{-1}$).
The Knight shift has maximum amplitude around $T=104K$.
Around $k_Fr=2.8$ the lines of $T=174K$ and $T=52K$ are crossed.}
\label{kt}
\end{figure}

\begin{figure}
\vspace*{7pt}          
\begin{center}         
\leavevmode           
\epsfxsize=3.0in
\epsfbox{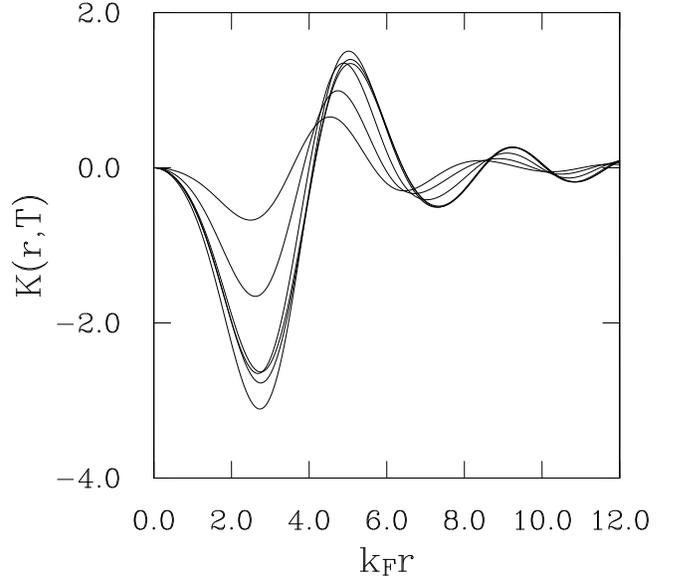}
\end{center}
\caption
{ Knight shift $K( r, T)$ at fixed angle $\theta =\pi/2$ and Kondo
temperature $T_0=430K$ for Ce ions. 
We use dimensionless
variable $k_Fr$ (with  the  Fermi wave vector $k_F=0.65\AA^{-1}$). 
This shows an oscillatory behavior like
the RKKY interaction form 
and the amplitude is decreased as the temperature is increased and the
distance $r$ is increased. The calculations are done at the temperatures
in Fig~\ref{K-t}. }
\label{K-rt2}
\end{figure}

\begin{table}
\caption{ Ground states and parameter values for, Ce and Yb ions model.
$j_g$ is the ground state spin-
orbit multiplet angular momentum and $j_{ex}$ is the excited state multiplet 
angular momentum. $N_g(N_{ex})$ is
the degeneracy of the ground (first excited) state multiplet. $\epsilon_f$  is
the energy of the ground configuration. and $\Delta_{so}$ is the energy difference 
between the $j_g$ and $j_{ex}$ multiplets}
\label{ceyb}
\begin{center}
\begin{tabular}{|@{\hspace{0.4in}} c|c|c@{\hspace{0.4in}}| } 
              &   Ce  &  Yb\\ 
\tableline
  Configuration & $4f^1$  & $4f^{13}$ \\ 
                & $4f^0$  & $4f^{14}$ \\  
     $j_g$        & 5/2   & 7/2   \\  
     $j_{ex}$     & 7/2   & 5/2    \\ 
    $N_g$        &  6    &  8     \\ 
    $N_{ex}$     &  8    &  6    \\   
      $\epsilon_f$       & $-2eV$   & $-1eV$   \\  
   $\Delta_{so}$  & $0.29eV$   & $1.3eV$   \\  
\end{tabular}
\end{center}
\end{table}

\begin{table}
\caption{ Parameter values for Y$_{0.8}$U$_{0.2}$Pd$_{3}$ in units of the
conduction electron band width $D=3eV$. $\epsilon_f$ is the energy of $j=4$ 
multiplet and $\Delta_{ij}$ is the energy difference between crystal field 
split $\Gamma_{i}$ and $\Gamma_{j}$ states.
$\Gamma$ is the single impurity hybridization width.
For the definition  of $x$ and
$W$, see  Appendix~\ref{cef.app}. $x_2$ and $W_2$ are
for $f^2$ configuration
and $x_3$ and $W_3$ are for $f^3$ configuration.}
\label{uvalue}
\begin{center}
\begin{tabular}{|@{\hspace{0.6in}}c@{\hspace{0.6in}}|c@{\hspace{0.6in}}|}
      $\epsilon_f$ & -0.333\\[0.05in]\hline
      $\Delta_{35}$ & 0.003\\[0.05in] 
      $\Delta_{34}$ & 0.008175\\[0.05in] 
      $\Delta_{31}$ & 0.019621\\[0.05in] 
      $\Delta_{68a}$ & 0.000136\\[0.05in] 
      $\Delta_{68b}$ & 0.013190\\[0.05in] 
      $\Gamma$      & 0.15 \\[0.05in]\hline
      $x_2$ & -0.648\\[0.05in] 
      $W_2 $& $ -3.95\times10^{-4}$ \\[0.05in]
      $x_3$ & 0.3693 \\[0.05in] 
      $ W_3$ & $ 2.746\times10^{-4}$\\[0.05in] 
\end{tabular}
\end{center}
\end{table}

\begin{table}
\caption
{Explicit values of  both incoherent and coherent Knight shift 
at high temperatures. They have similar values because coherent lattice
effects are small at high temperatures. These are unadjusted data.}
\label{kvalue}
\begin{center}
\begin{tabular}{@{\hspace{0.3in}} c    c  c @{\hspace{0.3in}}} 
 Temperature     &    \multicolumn{2}{c} {Knight Shift} \\ \cline{2-3}
      &   Incoherent  &  Coherent\\ \hline
925K&	-0.3579  &	-0.44855\\  
615K&	-2.4078  &	-2.42525\\  
555K&	-3.0157  &	-2.84562\\  
490K&	-3.7184  &	-3.34231\\  
430&	-4.5380  &	-3.95714\\  
\end{tabular}
\end{center}
\end{table}

\begin{table}
\caption{ Crystal electric field energy eigenstates for $j=4$ multiplet
in the cubic symmetry. The coefficient is independent of x} 
\label{table3}
\begin{center}
\begin{tabular}{@{\hspace{.4in}}cc@{\hspace{0.4in}}} 
  $j=4$ multiplet & states \\[0.05in] \hline
  $|\Gamma_1 \rangle $ & $\sqrt{\frac{5}{24}}|4\rangle +\sqrt{\frac{7}{12}}|0
\rangle +\sqrt{\frac{5}{24}}|-4\rangle$ \\[0.05in] 
  $|\Gamma_3 ;+ \rangle $ &$ \frac{1}{\sqrt{2}} (|2\rangle + |-2\rangle )$ 
         \\[0.05in] 
  $|\Gamma_3 ;-\rangle $ & $-\sqrt{\frac{5}{24}}|4\rangle +\sqrt{\frac{7}{12}}
|0 \rangle -\sqrt{\frac{5}{24}}|-4\rangle$ \\[0.05in] 
  $|\Gamma_4;0 \rangle $ & $\frac{1}{\sqrt{2}}(|4\rangle - |-4\rangle )$
\\[0.05in] 
  $|\Gamma_4;\pm 1 \rangle $ & $\frac{1}{\sqrt{8}}|\mp 3 \rangle +\sqrt{\frac{7}
{8}} |\pm 1 \rangle $ \\[0.05in] 
  $|\Gamma_5 ;0 \rangle $ &$ \frac{1}{\sqrt{2}} (|2\rangle - |-2\rangle )$
 \\[0.05in] 
  $|\Gamma_5 ;\pm 1\rangle $ &$\frac{1}{\sqrt{8}}|\mp 3 \rangle -\sqrt{\frac{7}
{8}} |\pm 1 \rangle $  \\[0.05in] 
\end{tabular}
\end{center}
\end{table}

\begin{table}
\caption{ Crystal electric field energy eigenstate for $j=9/2$ multiplet
in the cubic symmetry, the coefficient $a_i$'s and $b_j$'s depend on the $x$ 
which is the parameter which depends on the ratio of the fourth  and sixth
degree cubic field in the Hamiltonian of the crystal electric field. 
In this calculation, we use $x_3=0.3693$ and $W_3= 2.746\times10^{-4}$ to have
$j=9/2$ $\Gamma_6$ for the ground state of the  $f^3$ configuration and
$x_2=-0.648$ and $W_2=-3.95\times10^{-4}$ to have $j=4$ $\Gamma_3$ for the ground
 state for $f^2$ configuration.}
\label{table4}
\begin{center}
\begin{tabular}{@{\hspace{0.2in}} cc@{\hspace{0.2in}}  } 
  $j=9/2$ multiplet & states \\ \hline
  $|\Gamma_6;\pm \rangle $ & $\sqrt{\frac{9}{24}}|\pm \frac{9}{2} \rangle 
      +\sqrt{\frac{1}{24}}|\mp\frac{7}{2} \rangle +\sqrt{\frac{7}{12}}|\pm
\frac{1}{2}\rangle$ \\[0.05in]  
  $|\Gamma_{8a} ;1,\pm \rangle $ & $a_1\tablenotemark[1]|\pm\frac{9}{2} \rangle +a_2|\mp
\frac{7}{2} \rangle +a_3|\pm\frac{1}{2} \rangle )$ \\[0.05in]  
  $|\Gamma_{8a} ;2,\pm\rangle $ & $b_1|\pm\frac{5}{2}\rangle +b_2|\mp\frac{3}{2}
 \rangle$ \\[0.05in] 
  $|\Gamma_{8b} ;1,\pm \rangle $ & $a_6|\pm\frac{9}{2} \rangle +a_7|\mp
\frac{7}{2} \rangle +a_8|\pm\frac{1}{2} \rangle )$ \\[0.05in]  
  $|\Gamma_{8b} ;2,\pm\rangle $ & $b_3|\pm\frac{5}{2}\rangle +b_4|\mp\frac{3}{2}
 \rangle$ \\[0.05in] 
\end{tabular}
\tablenotetext[1]{For $x=0.3693$, 
$a_1=-0.1290$, $a_2=-0.1582$, $a_3=0.9789$, $b_1=0.7361$, $b_2=-0.6769$,
$a_6=0.7800$, $a_7=-0.6358$, $a_8=0.0016$, $b_3=0.6769$,  and $b_4=0.7361$.}
\end{center}
\end{table}

\begin{table}
\caption{ Crystal electric field energy eigenstates for $j=5/2$
multiplet in the cubic symmetry.}
\label{J2.5}
\begin{center}
\begin{tabular}{@{\hspace{0.3in}}cc@{\hspace{0.3in}}}
 $j=5/2$ Multiplet & States  \\[0.05in] \hline
$ |\Gamma_7^{(5/2)}; \uparrow / \downarrow \rangle $
 & $- \sqrt{1\over 6} ~ |\pm 5/2\rangle + \sqrt{5\over 6} ~|\mp 3/2\rangle$  \\[0.05in]

$ |\Gamma_8^{(5/2)}; +, \uparrow / \downarrow \rangle $
 & $ |\pm 1/2\rangle$ \\[0.05in] 
$ |\Gamma_8^{(5/2)}; -, \uparrow / \downarrow \rangle $
 & $\sqrt{5\over 6} ~|\pm 5/2\rangle + \sqrt{1\over 6} ~|\mp 3/2\rangle$ \\[0.05in] 
\end{tabular}
\end{center}
\end{table}

\begin{table}
\caption{Crystal electric field energy eigenstates for $j=7/2$
multiplet in the cubic symmetry.}\label{J3.5}
\begin{center}
\begin{tabular}{@{\hspace{0.3in}}cc@{\hspace{0.3in}}}
$j=7/2$ Multiplet & States \\[0.05in] \hline
$ |\Gamma_6^{(7/2)}; \uparrow / \downarrow \rangle $
 & $\pm \sqrt{5\over 12} ~|\mp 7/2\rangle \pm \sqrt{7\over 12}~|\pm 1/2\rangle$ \\[0.05in]
  
$ |\Gamma_7^{(7/2)}; \uparrow / \downarrow \rangle $
 & $ \pm {\sqrt{3} \over 2} ~|\pm 5/2\rangle \mp {1\over 2}~ |\mp 3/2\rangle$  \\[0.05in]
   
$ |\Gamma_8^{(7/2)}; +, \uparrow / \downarrow  \rangle $
 & $\pm \sqrt{7\over 12} ~|\mp 7/2\rangle \mp \sqrt{5\over 12}~|\pm 1/2\rangle$ \\[0.05in]
   
$ |\Gamma_8^{(7/2)}; -, \uparrow / \downarrow \rangle $
 & $\pm {1\over 2} ~|\pm 5/2\rangle \pm {\sqrt{3} \over 2} ~|\mp 3/2\rangle$ \\[0.05in]
\end{tabular}
\end{center}
\end{table}

\begin{table}
\caption{Selection rules for the Anderson hybridization between $f^2$ $\Gamma$
 and $f^3$ $\Gamma$ CEF states.  The meaning, for example, is that a
$\Gamma_6$ conduction electron doublet can combine with the $f_2$ $\Gamma_3$ doublet
to make an $f^3$ $\Gamma_8$ quartet.}
\label{table8}
\begin{center}
\begin{tabular}{@{\hspace{0.6in}} c  c@{\hspace{0.6in}}  } 
$\Gamma_c\otimes f^2\Gamma$ & $f^3\Gamma$ CEF states \\ \hline
$\Gamma_c \otimes \Gamma_1 $ & $\Gamma_c$\\[0.05in] 
$\Gamma_6 \otimes \Gamma_3 $ & $\Gamma_8 $ \\[0.05in] 
$\Gamma_7 \otimes \Gamma_3 $ & $\Gamma_8$ \\ 
$\Gamma_8 \otimes \Gamma_3 $ & $\Gamma_6 \oplus \Gamma_7 \oplus \Gamma_8$ \\[0.05in] 
$\Gamma_6 \otimes \Gamma_4 $ & $\Gamma_6 \oplus \Gamma_8 $ \\[0.05in] 
$\Gamma_7 \otimes \Gamma_4 $ & $\Gamma_7 \oplus \Gamma_8 $ \\[0.05in] 
$\Gamma_8 \otimes \Gamma_4 $ & $\Gamma_6 \oplus \Gamma_7 \oplus 2\Gamma_8$ \\[0.05in] 
$\Gamma_6 \otimes \Gamma_5 $ & $\Gamma_7 \oplus \Gamma_8 $ \\[0.05in] 
$\Gamma_7 \otimes \Gamma_5 $ & $\Gamma_6 \oplus \Gamma_8 $ \\[0.05in] 
$\Gamma_8 \otimes \Gamma_5 $ & $\Gamma_6 \oplus \Gamma_7 \oplus 2\Gamma_8$ \\[0.05in] 
\end{tabular}
\end{center}
\end{table}
 
\end{document}